# Vortex rings from notched nozzles

Trung Bao Le[1] and Fotis Sotiropoulos*[2]

## Abstract

The formation of a vortex ring from a piston/nozzle apparatus depends on the nozzle exit's shape. Here we study the formation mechanism and the evolution of asymmetrical rings from notched nozzles using direct numerical simulations. Three types of nozzle are used: i) V-notched; ii) A-notched and iii)W-notched nozzles. The results show that a distinct main vortex ring is formed following the shape of nozzle exit at initial formation time. However entrainment from the surrounding ambient fluid through the trough locations creates pairs of streamwise structure after the piston stops. This structure connects to the leading vortex ring at the trough position and propagates toward the nozzle's centerline. The special topology of the leading ring is augmented by the circumferential flow, which is split into smaller portions around the circumference. This flow drives the evolution of the ring at the later stage; excite instability growth during the flow evolution. The streamwise vortex tubes finally collide at the centerline of the nozzle and induce vortex reconnection at the end of the formation. This formation process is consistent in all examined cases, which suggests that it might be applicable for more complex nozzle cases.

---

[1] Division of Hydraulics, Thuy Loi University, 175 Tay Son, Hanoi, Vietnam

[2] Department of Civil Engineering, University of Minnesota, Minneapolis, MN 55414, USA

* corresponding author email: fotis@umn.edu

## 1. Introduction

Vortex ring is a fascinating phenomenon due to its importance in many physical processes. The topological structure of vortex ring and its evolution has drawn recently significant attention from researchers due to its importance in resolving fundamental questions of classical and quantum fluid (Arndt, 2003; Kleckner and Irvine, 2013).

The most studied problem is the formation of axi-symmetrical vortex ring structure, which has a torus structure (a perfect circular ring with a close vortex tube). The vorticity distribution is concentrated in a coherent region (or vortex core) with finite size (compact support). This problem is very much attractive since it can be described equivalently by a dipole of negative and positive vortex cores propagating on two dimensional plane (OFarrell, 2014). Much theoretical works have utilized this symmetric characteristic to estimate the stream functions, circulation strength, and propagation speed of the vortex ring (Batchelor, 2000). These analyses however are limited to fully formed vortex ring propagating in an infinite domain (Hill's spherical vortex, Norbury vortex rings, Pierrehumbert solutions).

A standard protocol to generate a vortex ring is the ejection of a column of fluid from a circular nozzle/orifice into an ambient fluid. The momentum transfer dynamics creates a vortex sheet on the interior side of the nozzle/orifice to form and rolls up to create a vortex ring. During this formation process, vorticity flux is continuously entrained from the vortex sheet into the main vortex cores. The vortex ring finally separates from the nozzle exit edge and maintains its circulation in a relatively long period of time. For a perfect circular ring, (Gharib et al., 1998) has shown that such entrainment process reaches a upper limit (maximum principle). Under this condition, the vortex ring core size cannot grow larger than a certain limit, which depends on the relative size of the displaced fluid column (L) over the diameter of the orifice (D). This limit is normally attained when the vortex ring is fully formed and "pinch-offs" from the original vortex sheet. Since the subsequent evolution of the vortex ring and its circulation depends strongly on how the initial momentum is entrained into the vortex core initially (Glezer, 1988), many works propose that the upper limit of vortex ring core size can be changed by adjusting the exit shape simultaneously with the starting flow (Dabiri, 2005).

The importance of nozzle exit's shape in vortex ring formation has been highlighted in recent works (Afanasyev, 2006). Afanasyev has demonstrated that it is possible to generate a dipole from a two-dimensional flow configuration where "pinch-off" phenomenon does not occur even at a large stroke length (L/D > 4). (Domenichini, 2011) advances the idea on the importance of the nozzle exit shape by hypothesizing that the curvature of the nozzle exit plays a critical role in the entrainment of vorticity into the vortex core during formation process.

In many natural (Dabiri, 2008; Le et al., November 2010; Le et al., 2012) and engineering processes (Belyakov et al., 2014; Meslem, 2010; Nastase and Meslem, 2010; Troolin and Longmire, 2010), vortex ring is widely formed under asymmetrically geometrical configuration. In biological flows, in particular, asymmetrical vortex ring has been found in pathologic human brain (Le et al., November 2010; Le et al., 2012), human heart (Le et al., 2012), attacking bubble of pistol shrimp (Hess et al., 2013) or jet-propelled swimmers (Katija and Dabiri, 2009). In engineering applications, non-circular nozzles have been used to generate asymmetrical vortex rings to enhance mixing (LONGMIRE et al., 1992b; Nastase and Meslem, 2010; New et al., 2005). Therefore, the understanding of asymmetrical vortex ring formation has important implications for practical applications, especially in biomedical and bioengineering.

A particular problem of interest is the formation of vortex ring from the mitral orifice of human heart. During diastolic phase of the heartbeat, a vortex ring is formed due to the sudden relaxation of the left ventricle, which creates a suction to draw blood from the left atrium to the left ventricle. Since the mitral orifice is not a perfect circle, the resulted vortex ring is not axi-symmetric (Le et al., 2012). In addition, other anatomical features of the mitral valve add further complexities to the vortex formation process. Traditionally, the vortex ring is treated equivalently as a perfect circular vortex ring (Gharib et al., 2006). However, (Le et al., 2012) pointed out that the asymmetry of the mitral vortex ring dominates its evolution. Therefore understanding the forming mechanism of vortex rings from complex nozzle could potentially provide important implications for the understanding of human heart physiology (Domenichini, 2011; O'Farrell and Dabiri, 2014).

In contrast to the perfect circular ring, the asymmetrical vortex ring has non-circular shape with non-uniform distribution of vorticity around its circumference (Le et al., 2011). It has fully three-dimensional structrure and cannot be represented by two-dimensional approximation (Domenichini, 2011). The understanding of asymmetrical vortex ring formation has only been known in details for certain cases when the non-circular exit plane is perpendicular to the nozzle exit.

The most well-known non-axisymmetric vortex ring is vortex ring from elliptical nozzles. As (O'Farrell and Dabiri, 2014) shows that the resulted vortex ring is found to have elliptical shape following the exit edge initially. Elliptical vortex ring has been found to exhibit oscillatory dynamics. The ring undergoes several stages including: i) axis-switching; ii) cross-links and iii) bifurcation. (ADHIKARI et al., 2009) reported that at low stroke ratio and low aspect ratio, elliptic vortex ring exhibits periodic oscillation and axis-switching phenomenon occurs as expected. However, at larger stroke L/D>2, the generating vortex sheet stays intact with the leading vortex ring and induces the ring structure to transient to three-dimensional wavy instability. The interaction of the vortex cores leads to instability and finally the breakdown of elliptic rings as small scale structures at later stages. Using detailed PIV measurements, (O'Farrell and Dabiri, 2014) shows that the trailing vortex sheet organizes into a crescent vortex core and interacts with the leading ring during the propagation process. At high aspect ratio, the elliptic ring deforms quickly so that its vortex core touches itself. This process leads to the separation of the elliptic ring into two separate smaller rings (bifurcation).

The vortex ring generation mechanism, however, has not been well understood in cases with complex nozzle exit shapes (Auerbach and Grimm, 2014). The most complex nozzle is classified as in-determinate origin nozzles where the nozzle exit has multiple notches and troughs (LONGMIRE et al., 1992b), lobed geometries (HuiHu et al., 2002; Nastase and Meslem, 2010) or arbitrary shapes. The simplest shape of in-determinate origin nozzle is inclined nozzles (Le et al., 2011; Troolin and Longmire, 2010; Webster and Longmire, 1998). In inclined nozzles, the exit plane is not perpendicular to the main nozzle axis. It aligns to the main nozzle axis with an inclined angle. The resulted inclined vortex ring exhibits a complex structure. The main vortex ring has been observed to follow the nozzle exit shape initially (Le et al., 2011) but quickly deforms under bending/rotating along the tranverse direction. In addition, experimental (Troolin and Longmire, 2010) and numerical works (Le et al., 2011) subsequently reveal a host of complex fluid mechanic phenomena during the formation of this inclined ring. The variation of vorticity flux around the nozzle circumference induces large entrainment of ambient fluid at the shortest lip and thus destabilizes the ring. As (Le et al., 2011) pointed out, the vortex-vortex and vortex-wall interactions are two most dominant processes. Vortex reconnection (S Kida and Takaoka, 1997) occurs at different stages of these interactions.

There are several unanswered questions regarding the formation and evolution of asymmetrical rings. Although the formation mechanism for inclined nozzles has been clarified by the previous work of (Le et al., 2011) it is not clear how such mechanism is applied for other in-determinate origin nozzles when multiple notched and troughs locations exist. The main questions need to be addressed in such complex nozzles: what are the role of peaks and troughs in generating streamwise structure? What is the role of the exit curvature in the vortex formation process? What is the role of the exit's shape on the propagation speed of the resulted vortex ring?

In this work, we carried out high resolution simulation of vortex ring from notched nozzles to clarify its formation mechanism and subsequent evolutionary process. Three types of nozzles are considered: i) V-notched; ii) A-notched and iii) W-notched nozzles. We demonstrate that vortex rings from notched nozzles have a complex formation process: its initial shape follows the nozzle exit's shape but due to the self-induced motion of the resulted vortex cores, the final topology includes a main vortex ring connecting to pairs of trailing vortex tubes. Our work is the extension of (Le et al., 2011) on the vortex ring formation from complex nozzles. Our numerical method is based on immersed boundary method, for the details of the numerical algorithms the reader is referred to (Gilmanov and Sotiropoulos, 2005).

Our paper is organized as follows. We first describe the generation condition of the vortex ring from notched nozzles. We then explain our numerical methods and computational setup for the case. Last, we report the formation process of the vortical structures from notched nozzles and their subsequent interactions.

## 2. Numerical method

We consider an incompressible flow of a Newtonian fluid with density ρ and kinematic viscosity υ driven impulsively through a cylindrical pipe with diameter D into a stagnant ambient fluid via a moving piston. The governing equations are the three-dimensional, unsteady incompressible continuity and Navier-Stokes equations. The governing equations read in Cartesian tensor notation as follows (i=1,2,3 and repeated indices imply summation):

$$\frac{\partial u_i}{\partial x_i} = 0$$

$$\frac{\partial u_i}{\partial t} + \frac{\partial (u_i u_j)}{\partial x_j} = -\frac{\partial p}{\partial x_i} + \upsilon \frac{\partial^2 u_i}{\partial x_i x_j}$$

In the above equations: $u_i$ is the i$^{th}$ component of the velocity vector $\vec{u}$; t is time; $x_i$ is the i$^{th}$ spatial coordinate; p is the pressure divided by ρ. The characteristic velocity scale is chosen as $U_0$ and the length scale is D.

The governing equations are solved using the sharp-interface curvilinear-immersed boundary method (CURVIB) (Borazjani et al., 2008; Ge and Sotiropoulos, 2007; Gilmanov and Sotiropoulos, 2005). The numerical method is described extensively in these publications and has also been applied to a wide range of engineering and biological flows, including flows with arbitrarily complex moving and flexible boundaries and fluid structure interactions (Le and Sotiropoulos, 2012). For that, only a brief description is provided here. The solid surfaces (the pipe walls, nozzle and piston surfaces in the present case) are treated as sharp-interface, immersed boundaries in a background Cartesian or curvilinear mesh and discretized with an unstructured triangular mesh (see Figure 1). Boundary conditions at background grid nodes in the immediate vicinity of an immersed boundary are reconstructed by interpolating along the local normal to the boundary such that the overall method is 2$^{nd}$ order accurate in space and time (Gilmanov and Sotiropoulos, 2005). The governing equations are discretized on a hybrid staggered/non-staggered grid using second-order accurate finite difference formulas and integrated in time using an efficient fractional step method (Ge and Sotiropoulos, 2007), which requires the solution of the non-linear momentum equations and a Poisson equation for the pressure variable. The implicit momentum equations are solved using Newton-Krylov methods while the GMRES solver with multigrid preconditioner is employed for the Poisson equation (Ge and Sotiropoulos, 2007). For more details about the numerical method the reader is referred to previous publications from our group (Borazjani et al., 2008; Ge and Sotiropoulos, 2007).

## 3. Computational setup

In order to easily compare with our previous works, the computational setup is identical to (Le et al., 2011) where the computational domain is a square tank of size *400mm (l) x 400 mm (w) x 760 mm (h)* in X, Y and Z direction respectively. The tank is consider closed at the bottom and has four side walls. The outlet is placed at the top where it is also the placement plane of the cylinders as shown in Figure 1 with the inner diameter of the cylinder is set as *D=72.8 mm* and the wall thickness of *A=1.7 mm*.

Grid sensitivity analysis has been carried out systematically as shown in (Le et al., 2011). The computational results have been compared with the experimental data of (Webster and Longmire, 1998), which shows that the computational grid provides sufficient resolution to capture the dynamics of the flow. Similarly, the grid here varies from the coarse grid (Grid 1-3M), finer grid (Grid 2-8M) and the finest grid (Grid 3-36M). The difference between them is the distribution of the grid points in the ROI. It is the rectangular box *[-0.75D:0.75D] x [-0.75D:0.75D] x [1.5D:3.25D]* in X, Y and Z respectively, indicated by thick line in the Figure 1. The grids are varied from 3 to 36 million grid points as a single block with structured mesh. The finest grid, Grid 3 (36M), is of size *301 x 301 x 413*. The grid spacing in the ROI is uniformly distributed $\Delta x = \Delta y = \Delta z = 0.728$ mm. The grid spacing is chosen to be uniform in the box (ROI) in both X and Y direction and Z direction. Other grid lines outside of the indicated ROI are stretched toward the tank walls. If otherwise noted, the results in this study are reported in the Grid 3.

| Case | I/D | L/D | S/D | T/D | C | Exit type |
|---|---|---|---|---|---|---|
| 1 | 0.25 | 1 | 0.5 | 2.25 | $135^0$ | V-notched |
| 2 | 0.25 | 1 | 0.5 | 2.25 | $45^0$ | A-notched |
| 3 | 0.25 | 1 | 0.5 | 2.25 | $45^0$ and $135^0$ | W-notched |

**Table 1: Simulation scenarios. I stands for initial position of the piston. S is the stopping distance to the shortest lip of the nozzle. T is the location of the common average lip. C is the angle of the cutting plane to the nozzle axis. See Figure 1 for the definition of each quantity.**

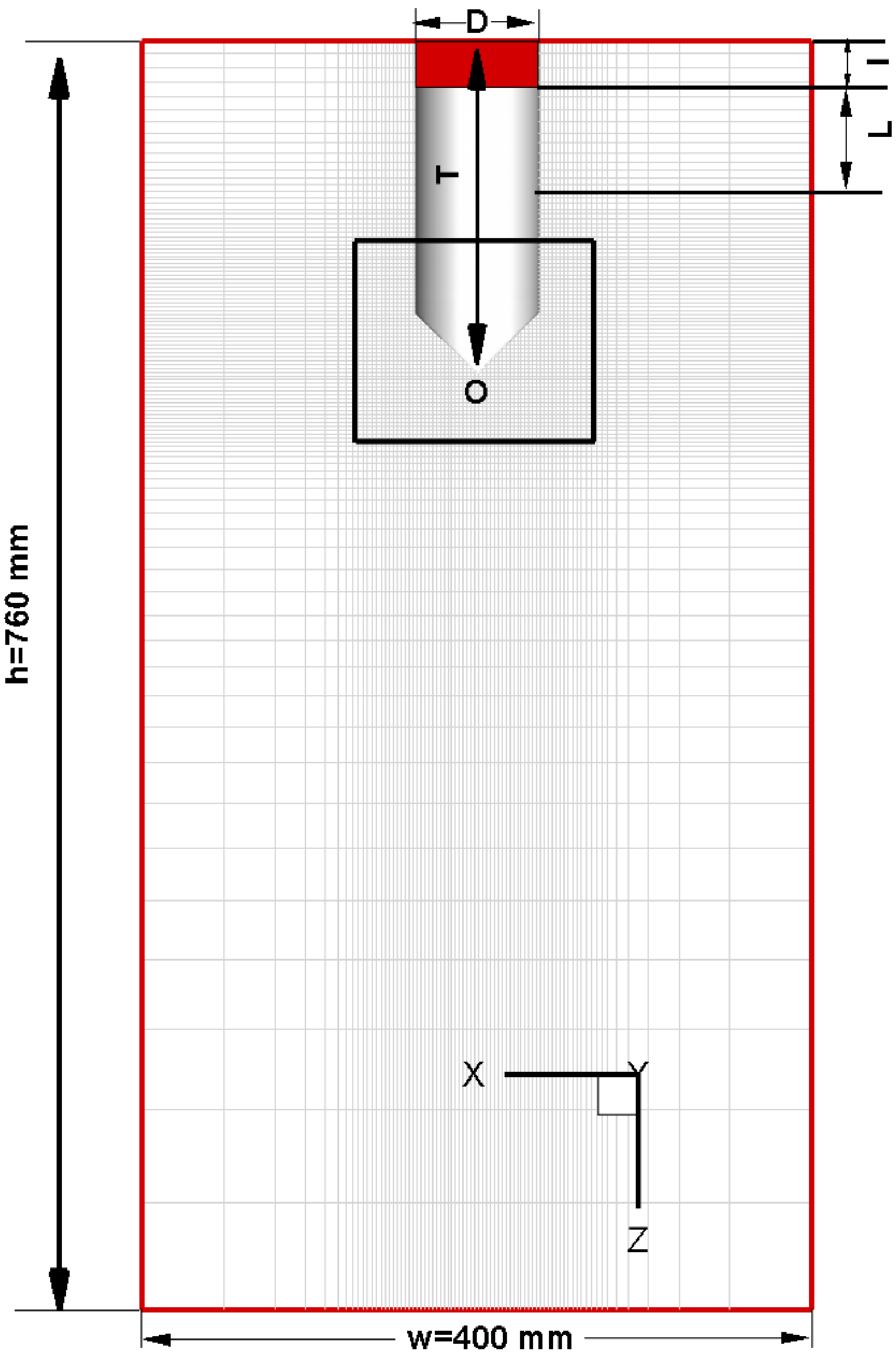

**Figure 1 : The diagram illustrates the computational setup with the A-notched nozzle placed inside the tank. The particular A-notched nozzle case is shown as one example of all cases. The detail of each computational case is summarized in Table ?. The computational grid (Grid 3) of size 301 x 301 x 413, only one of every four grid points is shown. The grid is uniform close to the nozzle exit and stretched toward the top, the bottom of the tank as well as side walls. The rectangular box at the nozzle exit indicated by thick solid line is the region of interest (ROI). The grid sensitivity analysis focuses in that region.**

The simulated cases consist of notched nozzles as shown in Figure 1. Following procedure of (New et al., 2005), each notched nozzle is derived from a circular pipe by two cutting planes with an inclined angle C to the nozzle axis symmetrically across the centerline. Thus the notched nozzle exit comprises of two half-ellipses formed at an angle $\alpha$ =2C to each other. Controlling geometrical parameters are illustrated in a cross-section through the nozzle axis as shown in Figure 1 including the inclined angle, nozzle length and its exit type. Notched nozzles are denoted as V-notched (Case 1), A-notched (Case 2) and W-notched (Case 3). The V-notched and A-notched nozzles are created intentionally so that they have the same maxium and minimum axial distances at peaks and troughs, respectively. Using this special design, it is possible to examine the effect of the nozzle exit's shape on the flow dynamics. Here both V-notched and A-notched nozzles have the same number (two) of peaks and troughs. The difference between V-notched nozzle and A-notched nozzle is the concave shape of V-notched nozzle and the convex shape of the A-notched nozzle. Therefore it is possible to infer the role of the nozzle exit's curvature on the flow features. To further examine the effect of troughs, a saw-tooth nozzle is created by merging the geometries of both V-notched and A-notched nozzles into a single surface (W-notched nozzle). Thus the W-notched nozzle has four troughs with both smooth and sharp peaks as shown in Figure 3. The common origin O in this case is chosen on the nozzle's centerline at the axial distance i.e $x = 0, y = 0, z = 0$ at the center of the tank top. The computational details for each simulated case are shown in Table 1.

For computational purposes, the nozzles and piston in the current study are represented as triangulated surface. The piston surface in these simulations is flat and has the diameter equal to the inner diameter of the nozzle $D_{piston} = 72.8\ mm$. We use 8032 triangles for representing the piston surface. Depending on the case, the nozzle surface is triangulated with different number of elements as shown in Table 1.

We use the cylindrical coordinate $(r,\psi,z)$ as in Le.2011 to indicate the location of one point on the nozzle surface as shown in Figure 2. The angle $\psi$ marks circumferential location of the point. With $\psi=0$ indicates the furthest location of the nozzle exit along the main nozzle axis (z) and other locations are marked accordingly in counter-clockwise direction. With this convention, the notched nozzles have two peaks at $\psi=0$ *and* $\psi=\pi$ and two troughs at $\psi= \pi/2$, $\psi=3\pi/2$. The difference between V-notched and A-notched nozzle is therefore the sharpness of peaks and troughs. The V-notched nozzle has smooth peaks and sharp troughs but the A-notched nozzle has sharp peak and smooth troughs. With this definition of $\psi$ for the notched nozzle cases, we follow the convention of (New et al., 2005) to define two important planes which cut through the nozzle centerline. First, the plane *peak-to-peak* is defined as the plane through two vertical lines $\psi = 0$ and $\psi = \pi$ (or plane $Y=0$). Second, the plane *trough-to-trough* goes through the line $\psi = \pi/2$ and $\psi = 3\pi/2$ (or plane $X=0$).

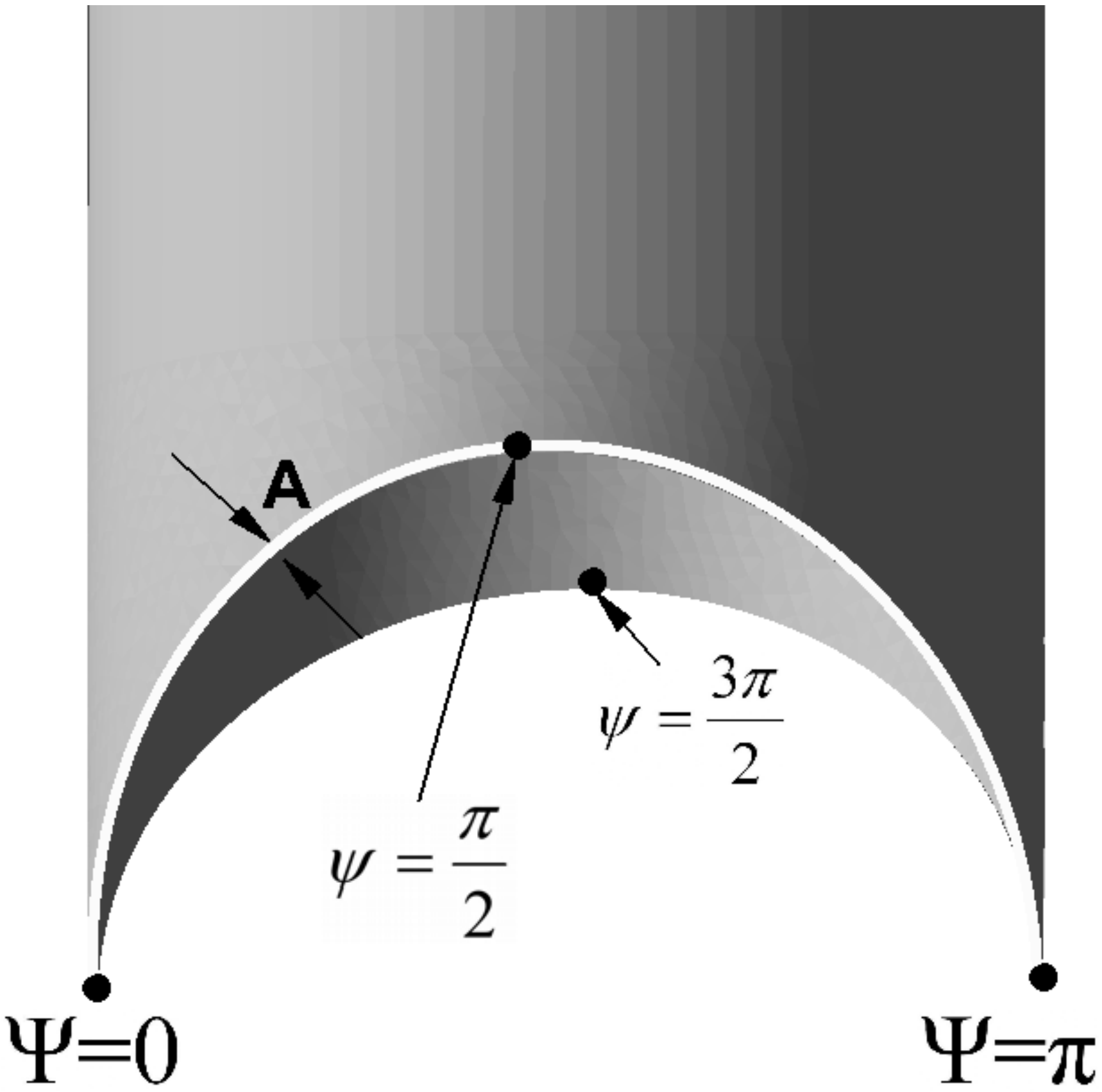

**Figure 2 The close-up view for the exit of A-notched nozzle. The longest lip of nozzle is marked with ψ=0. The plane *peak-to-peak* is defined as the plane cut through locations of ψ=0 and ψ=π. The *trough-to-trough* plane is the plane of ψ= π/2 and ψ=3π/2. The thickness of the nozzle is denoted as A=1.78 mm.**

We prescribed the motion of the piston as translational motion along Z direction only. The piston starts from location *z =+0.25D* and is displaced a distance (stroke length) L along the Z direction as shown in Table 1. In all cases, the stopping location of the piston is located 1D upstream of the common origin O. The piston velocity profile *U(t)* is reconstructed from the experimental data set as shown in Figure 4.

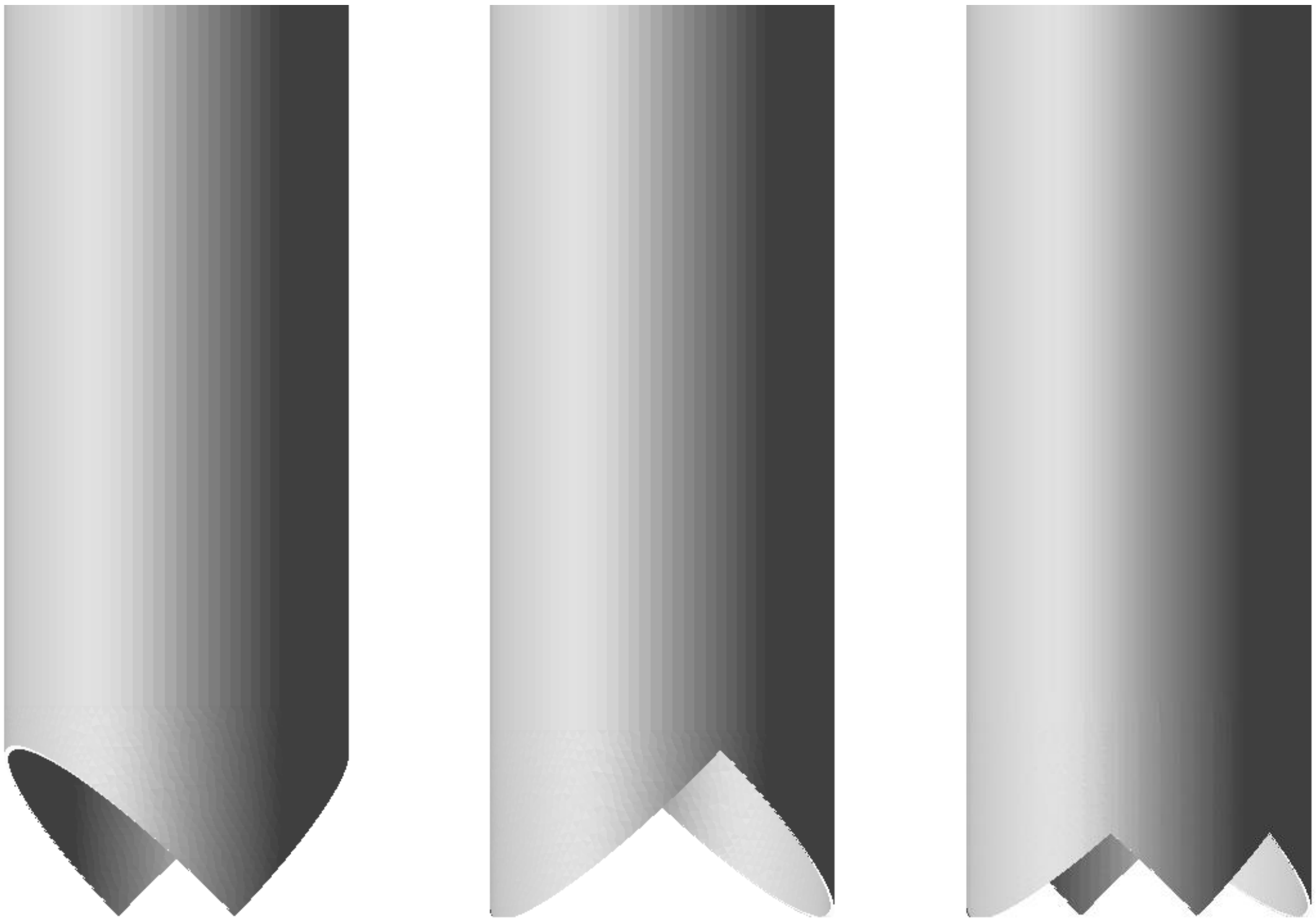

(a) Case 1 (b) Case 2 (c) Case 3

**Figure 3 Notched nozzles in the simulations: a) Case 1: A-notched nozzle b) Case 2: V-notched nozzle c) Case 3: W-notched nozzle. Note that the surface of Case 3 (W notched nozzle) is the merging of the surface for A and V notched nozzles.**

If piston starts its motion from t=0 to the time t=T, the time-averaged velocity of the piston $\bar{U}$ can be calculated as:

$$\bar{U} = \int_0^T \frac{1}{T} U(t)dt$$

The stroke length $L = \int_0^T U(t)dt$ is calculated from the piston velocity profile by integrating the area under the curve and gives *L/D = 1.01*. The maximum velocity of the piston is $U_{max}$ = 80.75 mm/s while the time-averaged velocity is $\bar{U}$ =73.6 mm/s. The characteristic velocity is chosen as the nearly constant piston velocity $U_0$ = 77 mm/s. Thus nominal Reynolds number $Re_D$ based on nozzle inner diameter and $U_0$ is $Re_D = \frac{U_0 D}{\upsilon} = 5600$. We use the "slug model" of (Glezer, 1988) to characterize the initial momentum thickness in different simulated cases. Although this model is a simplified approximation of the boundary layer it provides a reasonable approximation of initial circulation injected to the primary ring core. The characteristic Reynolds number based on "slug" model $Re_\Gamma = \frac{\Gamma_0}{\upsilon} = \frac{U_0^2 T}{2\upsilon} = 2900$.

The time is non-dimensionalized as $t^* = \frac{(t-t_0)U_0}{D}$ with $t_0$ is the initial time when the piston starts to accelerate. Since the stroke length is L = 1D the piston motion starts from $t^*$ =0 and stops at $t^*$ =1. The non-dimensional time step used in this study otherwise noted is 0.01 (associated with physical time step of 10 ms).

No slip boundary conditions are applied on the side walls and the surface of the cylinder. Since the piston surface area is comparably small to the tank surface the displaced volume of fluid from the nozzle is insignificant compared to the total tank volume. Thus the free surface does not does not vary significantly at the tank top (the free surface). We therefore apply the Neumann boundary condition at the domain exit (the tank top).

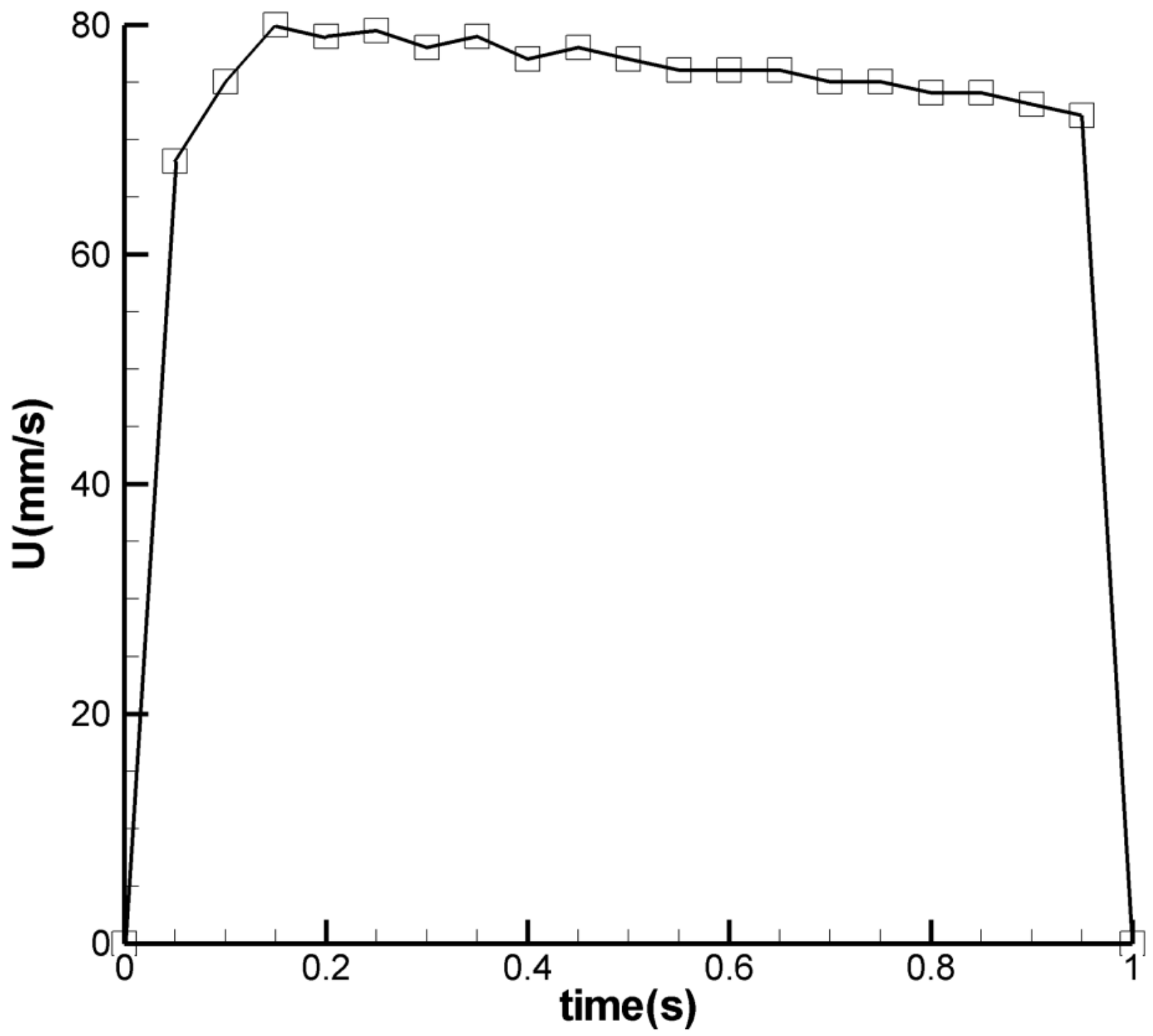

**Figure 4: The piston velocity profile is taken from experimental data of Webster.1998. It is used for all simulation cases in current study.**

## 4. Results

### 4.1 Flow entrainment

The entrainment of ambient fluid into the starting jet's core is shown in Figure 5. From t*=0 to t*=1.0, the piston displaces the fluid inside the interior part of the nozzle with the piston's velocity as seen in Figure 4. During this period, the fluid exits from the nozzle continuously. As seen in Figure fig:Exit_Profile, the exit velocity profiles at the shortest axial distance (i.e cross-section at troughs) *z/D = +2.25* are shown for two different planes: i) peak-to-peak and ii) trough-to-trough. Since the piston surface is flat, the resulting velocity profile has a blunt profile (i.e nearly uniform) inside the nozzle's interior. However, the velocity profile is altered when the flow starts to exit the nozzle. In both V-notched and A-notched nozzles, the velocity profile at *t*=0.5* shows that the velocity profile in the *peak-to-peak* plane is nearly uniform as shown in

Figure 5. However, there exists significant entrainment of ambient fluid into the jet's core from the troughs ($\psi = \pi/2$ and $\psi = 3\pi/2$). This entrainment induces the larger velocity near the nozzle's edge and lower velocity at the jet's center. Comparing the velocity profile in the *trough-to-trough* plane of V-notched and A-notched shows that the entrainment is stronger in the A-notched nozzle, which has in larger axial velocity $W/U_0 = 1.6$ near the edge than one in the V-notched nozzle ($W/U_0 = 1.4$). Comparing the velocity profile of the current case with planar elliptic nozzles(O'Farrell and Dabiri, 2014)  shows that the notched nozzles modify more severely the profile due to entrainment. Therefore the non-planar exit has the significant effect on the velocity profile in general.

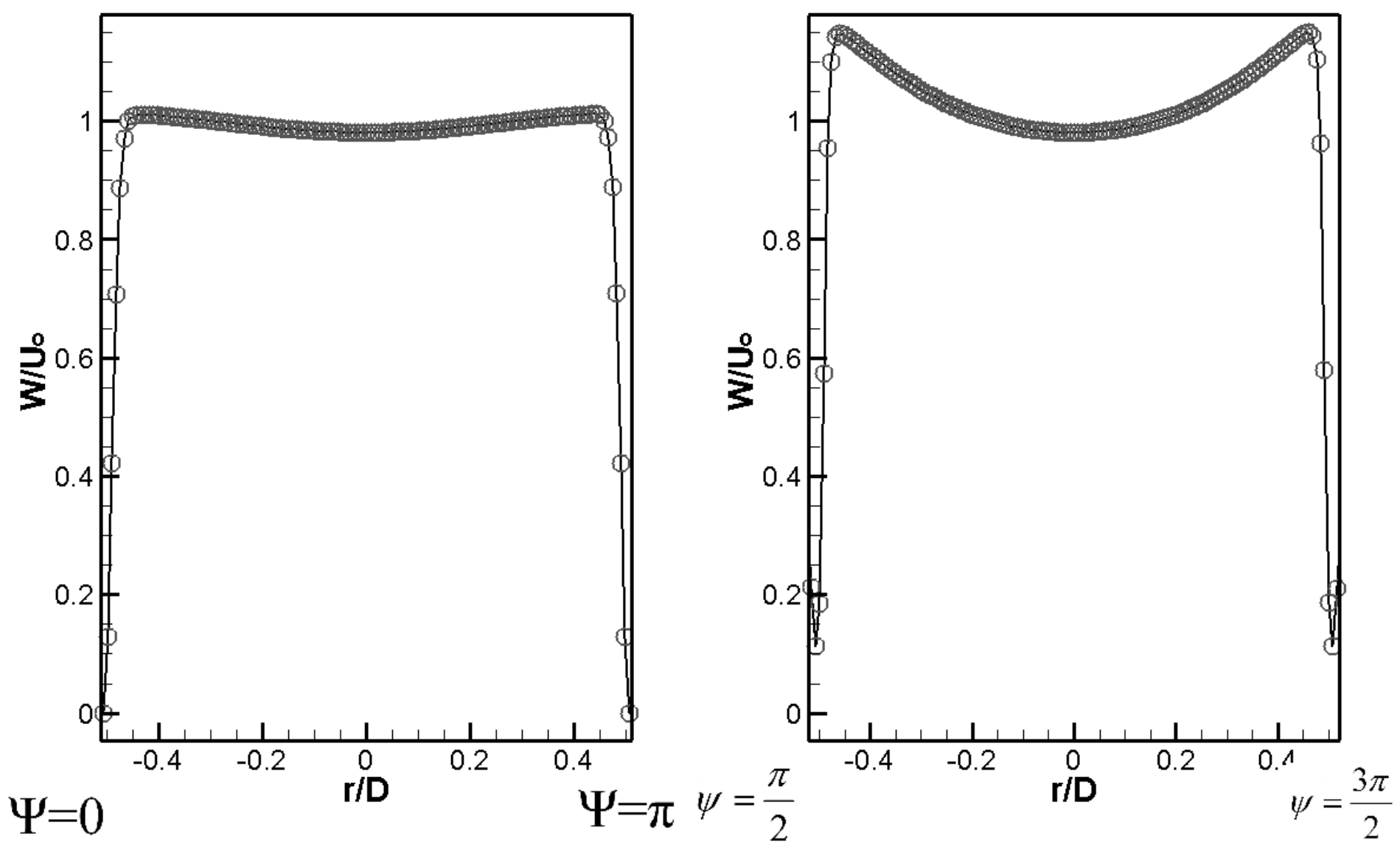

(a) A-notched nozzle

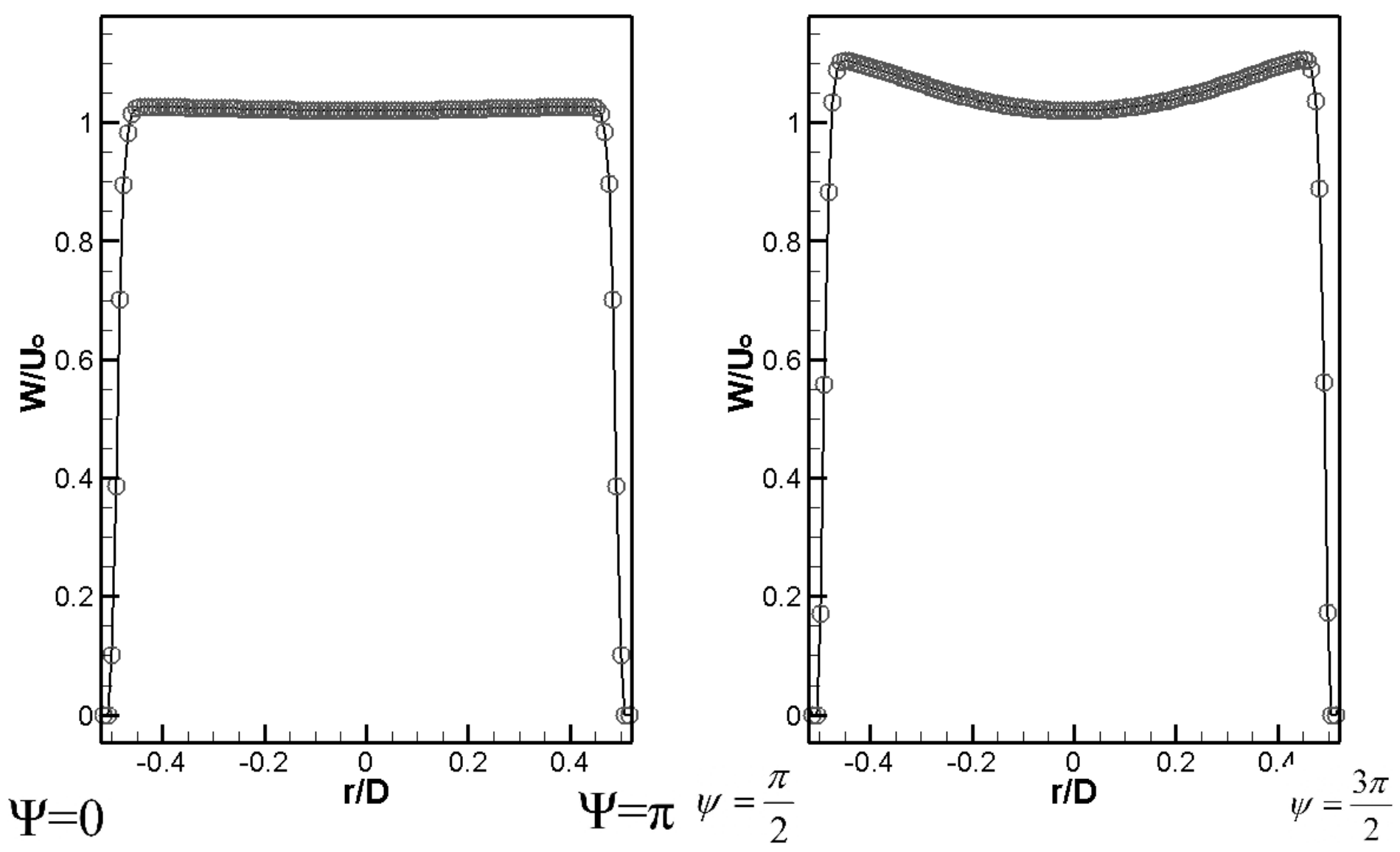

(b) V-notched nozzle

**Figure 5: Exit velocity profile on the *peak-to-peak* (left) plane and *trough-to-trough* (right) plane the shortest axial cross-section (i.e at troughs) for (a) A-notched nozzle and (b) V-notched nozzle at t*=0.5.**

## 4.2 Vorticity dynamics

The vortex ring formation from nozzles (Le et al., 2011) includes three main structures: i) the main vortex ring (*R1*); ii) the stopping ring (*R2*) and the piston ring (*R3*). As discussed in details in (Le et al., 2011) for the inclined nozzle cases, the *R1* is created by the rolling-up of the shear layer at the nozzle tip due to the swift acceleration of the piston. The *R2* is created by the entrainment of ambient fluid into the nozzle's interior due to the sudden stop of the piston. *R3* is the interaction of piston and the boundary layer inside the nozzle. For inclined nozzles, the dynamics in the near-field region of the nozzle is the result of interaction between *R1* and *R2*. In this section, we examine the formation sequence in notched nozzles.

*A-notched nozzle*

The vorticity dynamics of A-notched nozzle is different from that of a circular vortex ring. As the piston accelerates from t*=0 to t*=0.5, a shear layer is formed on the interior surface of the nozzle. The shear layer rolls up at the nozzle exit to form the first ring (*R1*) as shown in Figure 6. Since the nozzle shape is A-notched nozzle, the axial location varies from point to point on the circumference. Thus there is a variation of vortex core size around the circumference from the plane *peak-to-peak* to *trough-to-trough* plane. At t*=0.5 as shown in Figure 6a, vorticity entrains significantly large on the *trough-to-trough* plane into a strong vortex core at $\frac{z}{D}$=+0.1 whereas the rolling-up of shear layer has just started at $\frac{z}{D}$= 0.5 on the *peak-to-peak* plane. This large variation of vorticity entrainment indicates that the vortex's core is distorted and bent upward. We denote this bent vortex ring as *R1*.

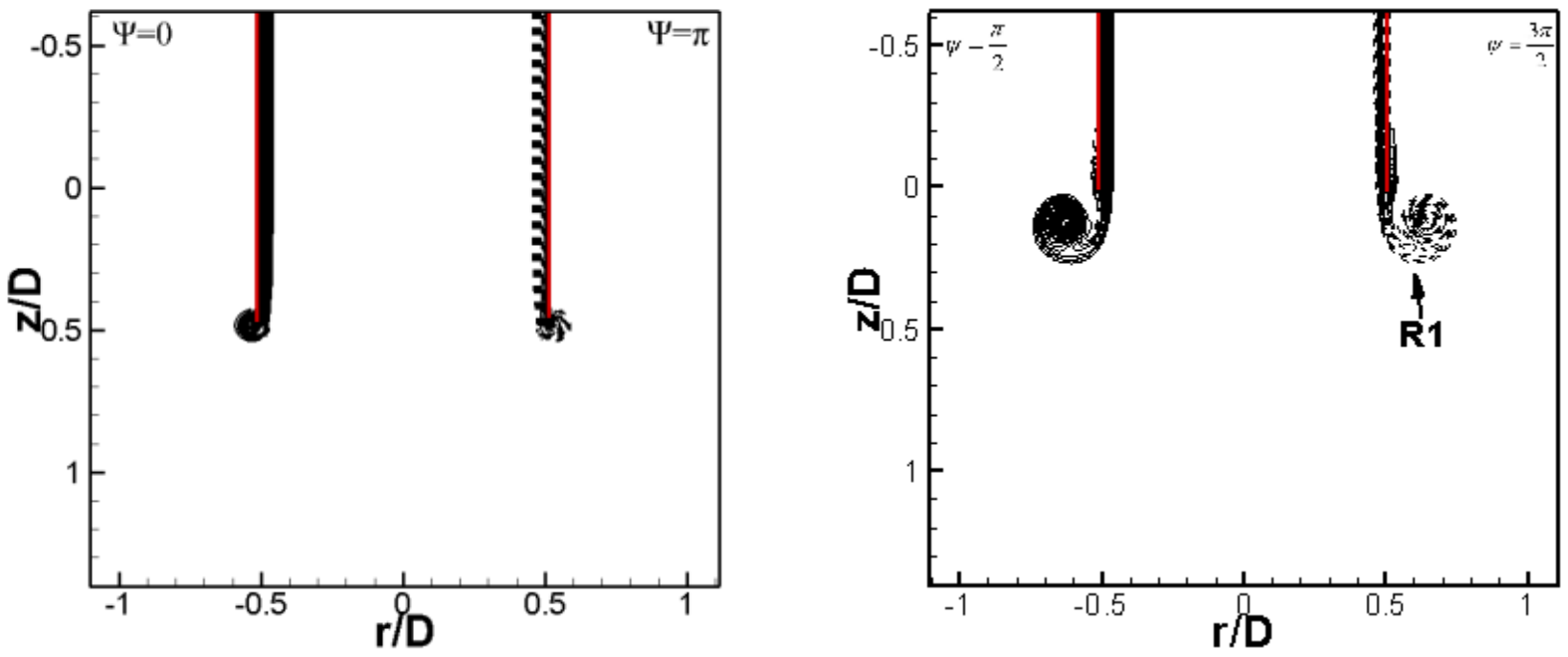

(a) t * = 0.5

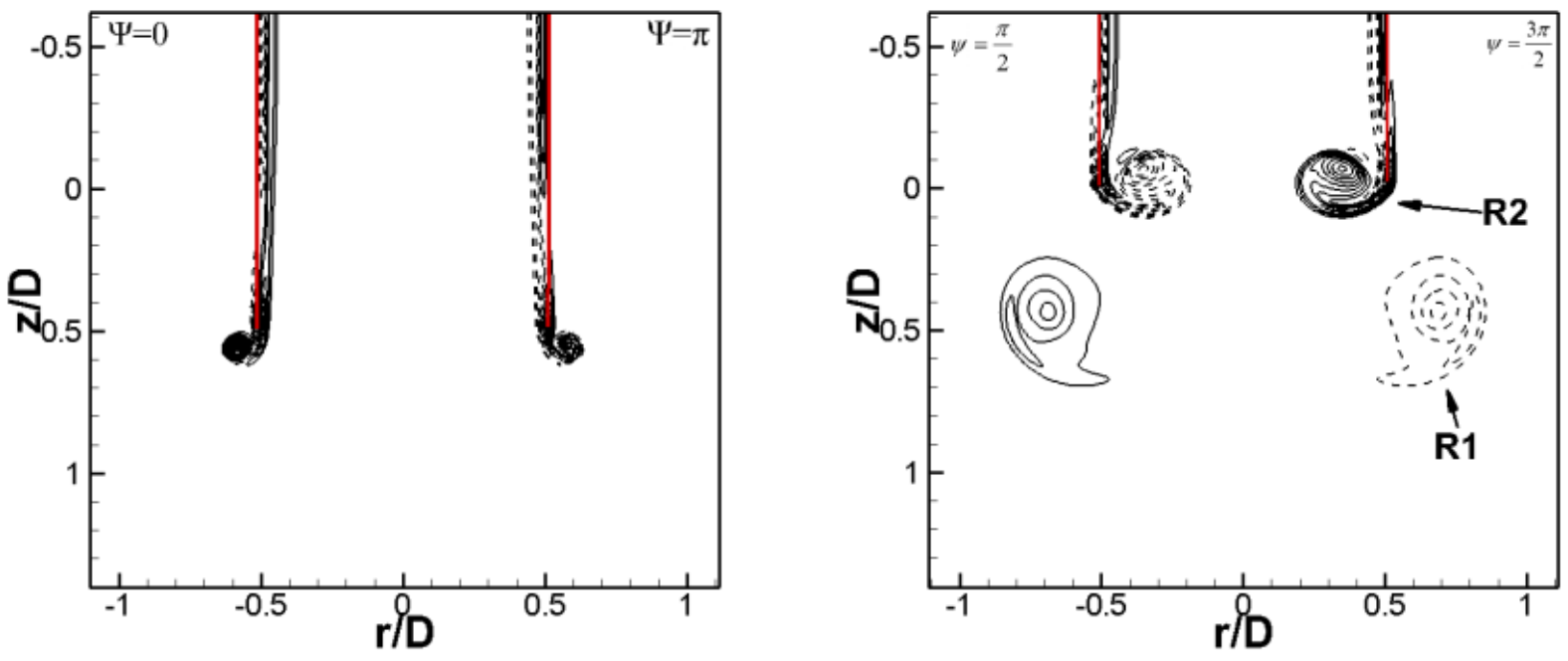

(b) t * = 1.56

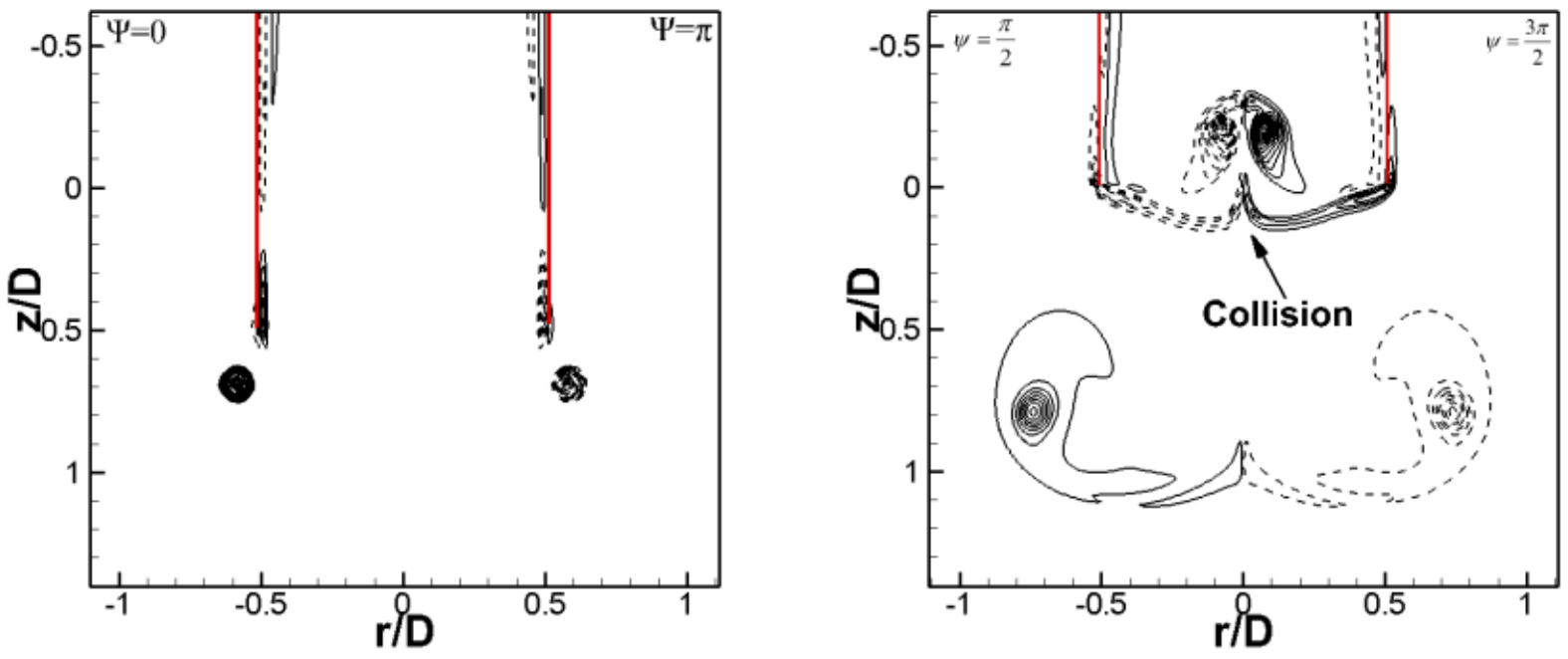

(c) t * = 2.62

**Figure 6: The vorticity fields at the exit of A-notched nozzle at different instances depict the formation of the primary ring as well as the collision of trailing rings. Left column is displayed in *peak-to-peak* plane which is orthogonal to the *trough-to-trough* plane (right column). The first contour level is 2.5 with the increment is $2.5DU_0$. The dash lines indicate negative values.**

The effect of the asymmetric exit is further exacerbated after the piston stops at $t^*=1$. The main vortex ring *R1* grows into a large vortex core as shown in the *trough-to-trough* plane but remains relatively small size on the *peak-to-peak* plane as shown in Figure 6b. Comparing Figure 6b and Figure 6a shows that *R1* flattens its shape since its vortex core now resides nearly on the same plane $\frac{z}{D} = 0.5$ on both *peak-to-peak* and *trough-to-trough* planes. In addition, the entrainment of ambient fluid induces the second vortex structures (*R2*) to form as seen in $t^*=1.56$. This vortex structures (*R2*) are formed symmetrically at both *trough* positions. Note that there is no apparent connection between the vortex cores of *R1* and *R2* shown in Figure 6b. The entrainment process here is similar to the stopping ring's dynamics of inclined nozzles (Le et al., 2011). The role of trough here is equivalent to the shortest lip of the inclined nozzle. Two cores of *R2* propagate toward each other heading to the centerline.

At $t^*=2.62$ in Figure 6c, *R1* completely separates from the nozzle's exit. Observation in Figure 6 shows that vortex core on the *trough-to-trough* plane $\frac{z}{D} = +0.8$ moves faster than that on the *peak-to-peak* plane. Therefore *R1* is bent downward. Two vortex cores *R2* collide at the centerline as shown clearly on the *trough-to-trough* plane but there is no vorticity exchange visible on the *peak-to-peak* plane.

*V-notched nozzle*

Similar to A-notched nozzle, a distinct vortex ring (*R1*) is formed at the exit of V-notched nozzle as shown in Figure 7 as the piston is displaced from $t^*=0$ to $t^*=0.5$. At $t^*=0.5$ the shear layer on the *trough-to-trough* plane rolls up earlier and the respective vortex cores grow larger in size near the location $z/D=0.25$. At the same time, the smaller vortex cores on the *peak-to-peak* plane stay close to the location $z/D=0.5$. The whole vortex structure is thus bent downward around the circumference and continues to propagate far from the nozzle as the piston displaces.

At $t^*=1.56$ as shown in Figure 7, the *R1* has already separated from the nozzle exit. The *R1* is now near the axial location $z/D=0.65$ on the *peak-to-peak* plane while it resides near $z/D=0.6$ on the *trough-to-trough* plane. *R1* thus reorient its direction to be orthogonal to the main nozzle axis. The entrainment of ambient fluid at both trough locations ($\psi= \pi/2$ and $\psi =3\pi/2$) creates two trailing vortex cores (*R2*) propagating toward each other. Note that *R2* rotates in opposite direction to *R1*. As observed on the *trough-to-trough* plane, there exists a footprint of additional structures that connects *R1* and *R2* that does not exist in A-notched nozzle. This connecting structure is the result of large sweeping from the trough position toward the centerline.

The Figure 7c shows a striking feature of this flow at $t^*=2.62$. The vortex cores of *R2* collide at the nozzle centerline. This collision also shows up both on the plane *peak-to-peak* and on the *trough-to-trough* plane. The *R1* at this moment locates near the location $z/D=0.9$ in both *peak-to-peak* and *trough-to-trough* planes. Exchange of vorticity between the cores of *R2* induces the additional structure showing up on the *peak-to-peak* plane.

Comparing the vorticity dynamics of inclined nozzles (Le et al., 2011; Troolin and Longmire, 2010; Webster and Longmire, 1998) with those of A-notched and V-notched nozzle in Figure 6 and Figure 7 shows a distinctive difference. In inclined nozzles, there is only one minimum and maximum of vorticity at the longest lip (peak) and the shortest lip (trough), respectively. In the current study, there exist two minima and extrema at two peaks and troughs. Therefore the structure of the flow is inherently more complex.

A close examination of *R1* on Figure 6 and Figure 7 shows that the exit geometry does have impact on the propagation speed of the vortex ring. The results show that vortex ring from V-notched nozzle exhibits slightly higher propagation speed. Since the structure of *R1* is rather complex as discussed in the following section, this comparison is rather qualitative than quantitative.

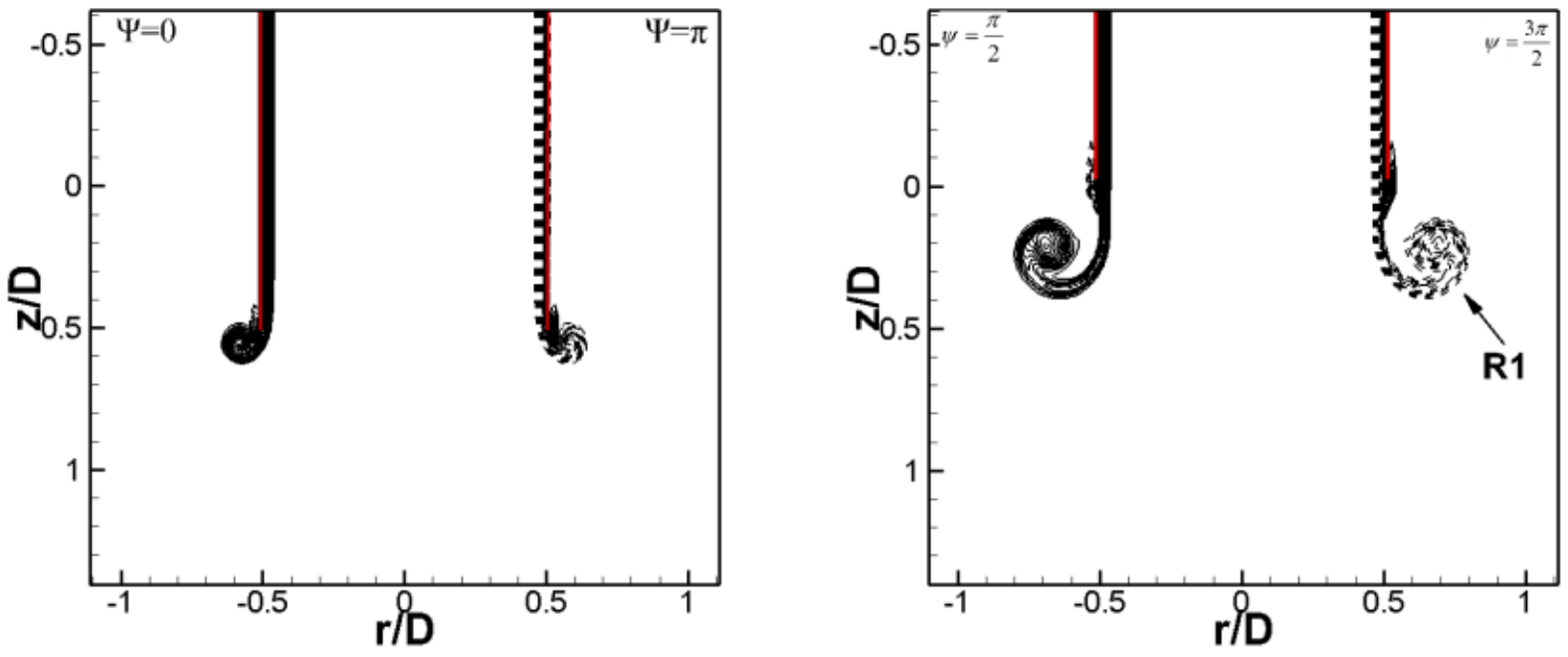

(a) t* =0.5

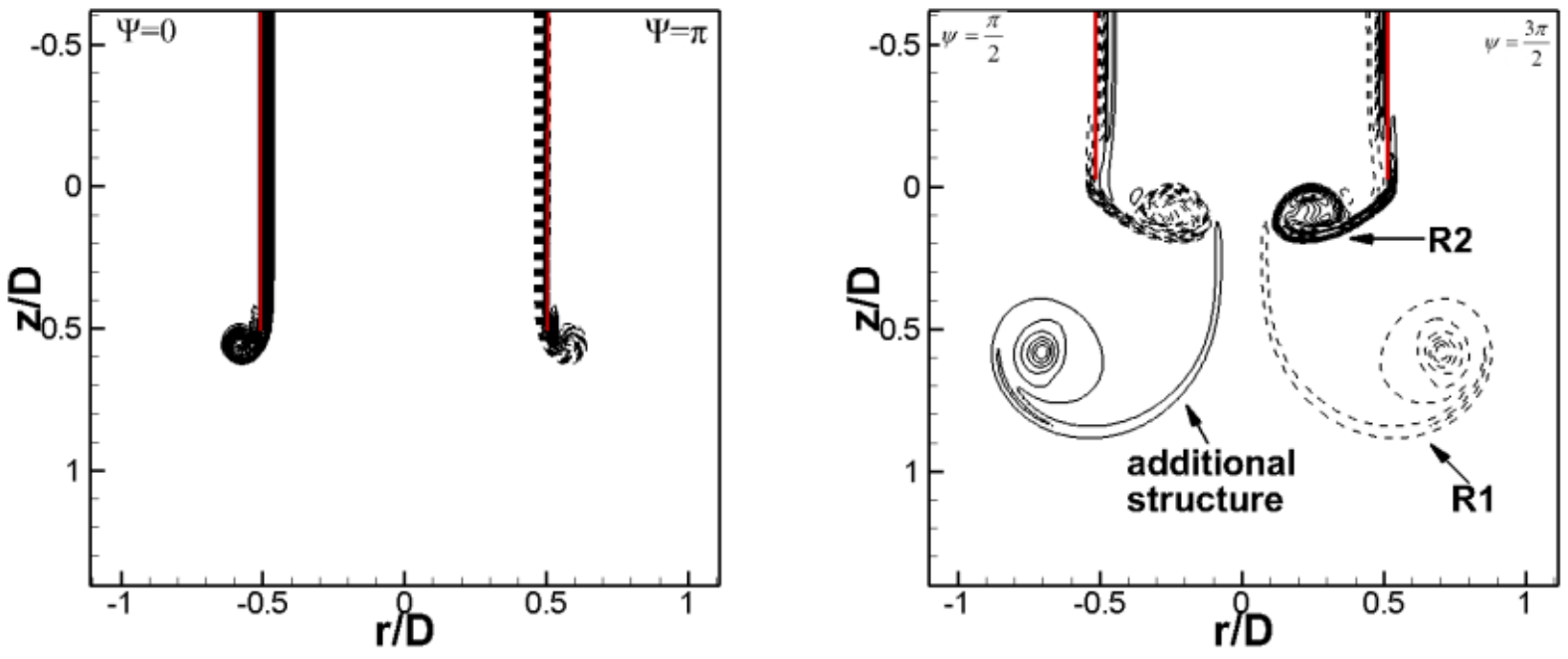

(b) t*=1.56

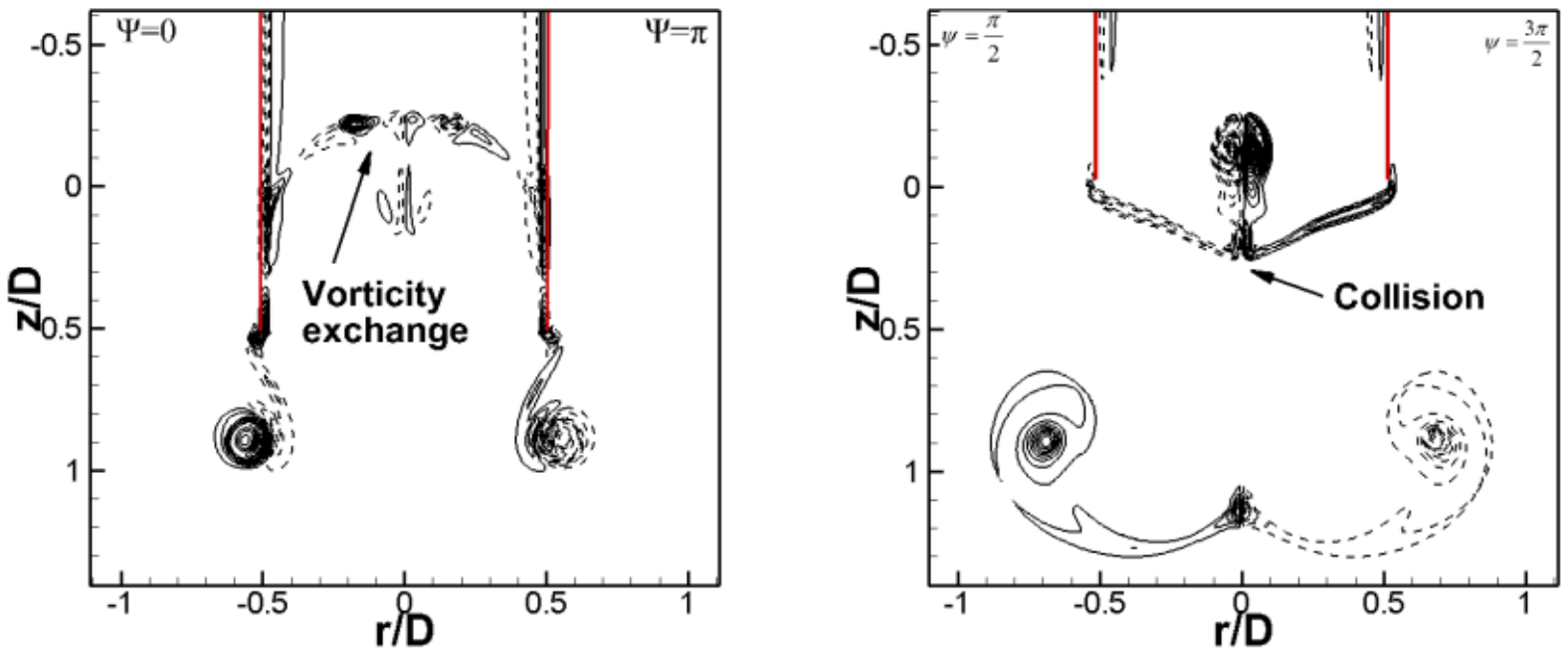

(c) t*=2.62

**Figure 7: The vorticity fields at the exit of V-notched nozzle at different instances depict the formation of the primary ring as well as the collision of trailing rings. Left column is displayed in *peak-to-peak* plane which is orthogonal to the *trough-to-trough* plane (right column). The first contour level is 2.5 with the increment is $2.5DU_0$. The dash lines indicate negative values.**

## 4.3 Three-dimensional structure

In this section, we reconstruct the three-dimensional structure of the vortex rings from A-notched and V-notched nozzle. Since these nozzles have the same number of peaks and troughs, it is possible to highlight the effect of nozzle's shape on the vortex formation process. Following (Le et al., 2011) we visualize the structure of the ring using iso-surface of vorticity magnitude colorized with helicity density $h = u.\omega$. In the region of the non-zero values of h, there exist significant changes in the topological structure of the vortex tube (Moffatt, 2001).

*A-notched nozzle*

In A-notched nozzle, there exist a distinct leading vortex ring (*R1*) is formed at the initial time *t*=0 to t*=1* as the piston displaces as shown in Figure 8. In this case the convex shape of the nozzle exit creates a bent vortex ring following the nozzle's exit shape (i.e A-shape). Here the vortex cores at peaks are shown to be significantly smaller than ones in the troughs. During this process, a thin vortex sheet is formed on the nozzle's interior surface and connects with the leading ring. During its propagation, the leading vortex ring deforms from the original A-shape to flatten its shape to be orthogonal to the main nozzle's axis at later stage as discussed earlier in Figure 6.

After the piston stops, the leading vortex ring *R1* continues to propagate along the nozzle's axis in a self-induced fashion as seen at *t*=1.56* in Figure 8b. There exists a significant variation of the vortex core size along the circumference of *R1*. In addition, Figure 8 shows a striking feature of this flow. As the piston stops at *t*=1*, the entrainment of ambient fluid occurs at both trough locations ($\psi = \pi/2$ and $\psi = 3\pi/2$). In addition to the leading ring, this entrainment (see also Figure 6) induces the pair of trailing vortex tubes to form. We denoted them as *R2a-I* and *R2a-II*. This structure has opposite vorticity sign to the main ring (*R1*) as discussed in Figure 6. This vortex tube *R2a* starts rolling up at trough and consequently reorganizes the remnant vortex sheet to create two pairs of vortex tubes which has the *λ-shape*. Since the strong entrainment directs toward the nozzle's axis, the *R2a* is stretched while it is linked to the leading ring via the remnant vortex sheet. The vortex sheet is symmetrically split at $\psi = \pi/2$ and $\psi = 3\pi/2$ as seen in the plane view of Figure 8b. The splitting of the vortex sheet in adjunction to the strong rotation at trough leads to the self-reorganization (folding) of the sheet into distinct tubes. These *R2a* pairs of vortex tubes finally complete their formation and wrap around the leading ring at $t^* = 2.0$ as seen in Figure 8c at four locations $\psi = \pi/4$, $\psi = 3\pi/4$, $\psi = 5\pi/4$ and $\psi = 7\pi/4$. The wrapping location ("*leg*") moves from the four original locations toward the troughs ($\psi = \pi/2$ and $\psi = 3\pi/2$) throughout the propagation of the leading ring. This wrapping process also deforms and stretches the vortex core of the leading ring.

Due to the geometrical symmetry, two vortex tubes *R2a* propagate toward each other at $t^*= 2.00$. Two vortex tubes *R2a* approach each other on the *peak-to-peak* plane at $t^*=2.62$ as shown in Figure 9a. The lower portion of *R2a* ("*leg*") is further stretched and displaces the R1's core further outward. The lower portion ("*leg*") of *R2a* grows significantly large and disrupt the *R1*'s core at $\psi = \pi/2$ and $\psi = 3\pi/2$. At these locations, the tubes of *R2a* tend to touch each other. At t*=3.0 the two *R2a* collide and exchange vorticity as shown in Figure 9b. The vortex tubes of *R2a* impinge on the cylinder's wall and at the same time interact with the opposite *R2a* portion. This process leads to the highly stretched upper portion of *R2a* as nearly a U-shape tube. Finally the upper portion of *R2a-I* finally pinch-offs from the lower portion and reconnect to the opposite *R2a-II* at $\psi=0$ and $\psi=\pi$. The tubes of lower *R2a-I and R2a-II* reconnect to create a hollow cross-shape as seen in Figure 9c at $t^*=4.0$. The impingement of *R2a-I* and *R2a-II* on the wall also excite smaller tubes at peaks. Though severely distorted at ($\psi = 0$ and $\psi = \pi$) by this small tubes, the core of *R1* maintains its coherency during this collision.

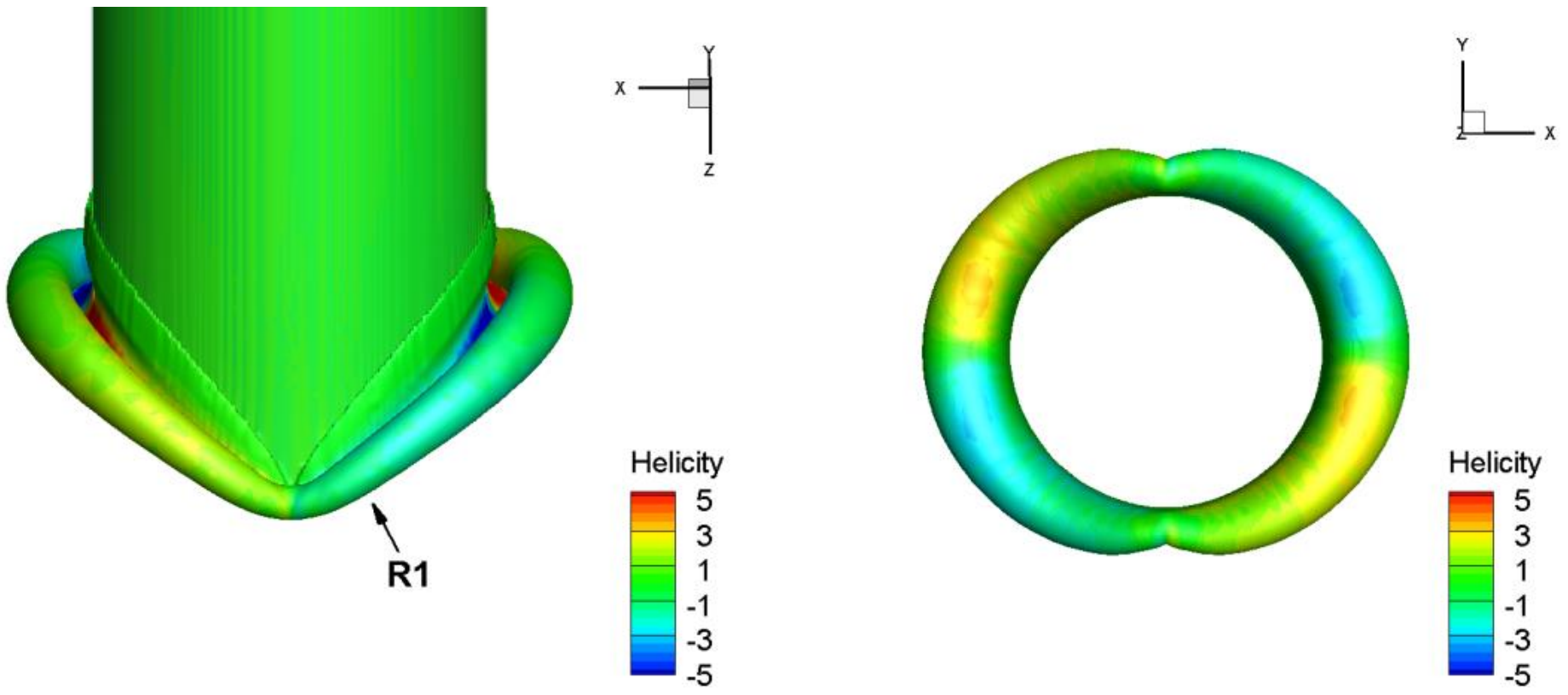

(a) t* = 0.5

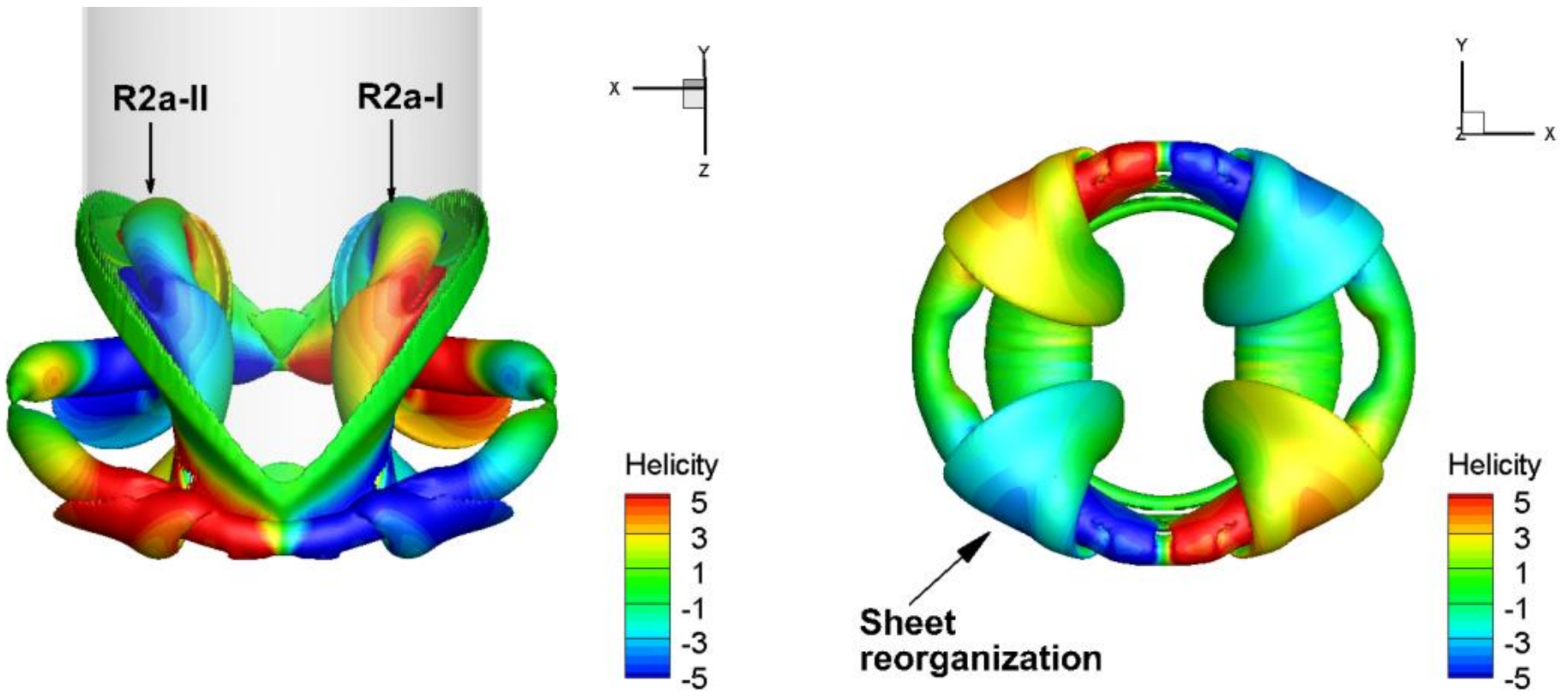

(b) t*=1.56

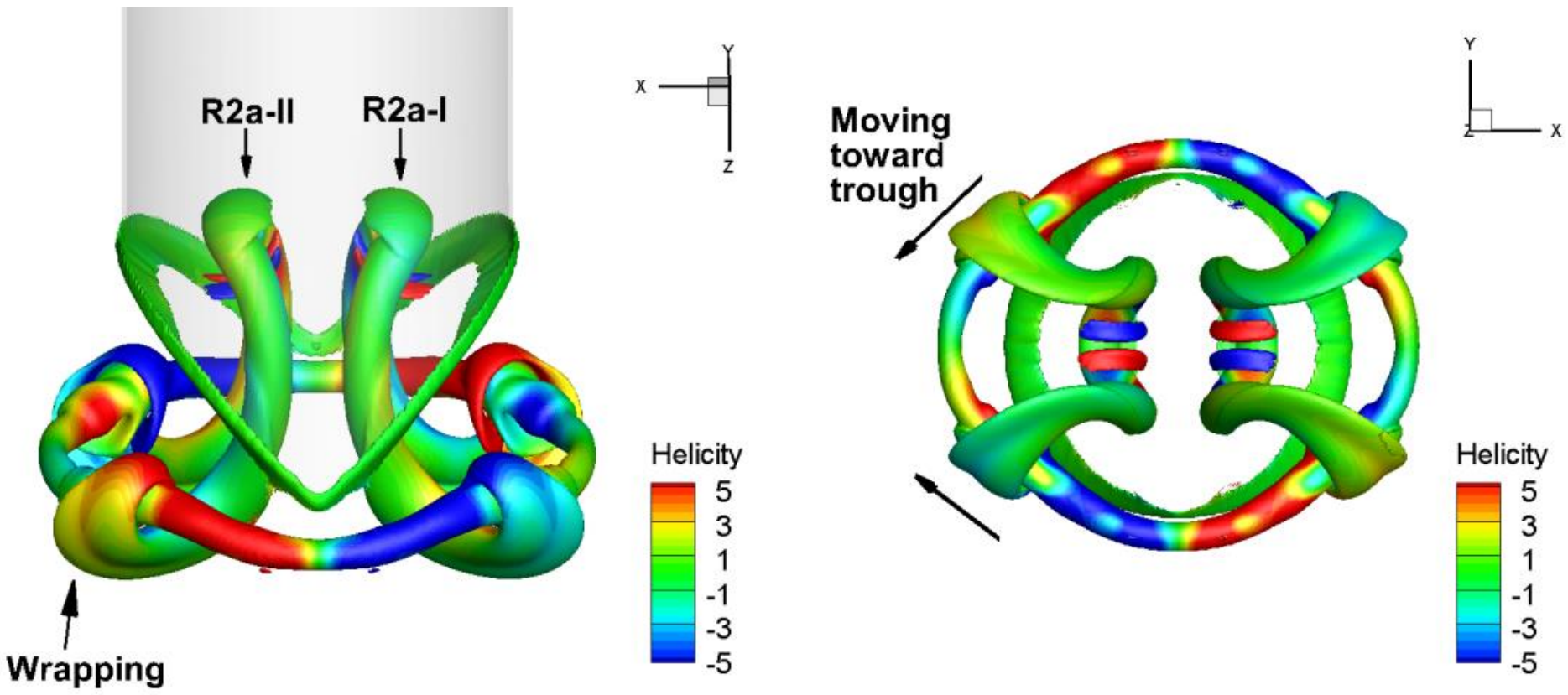

(c) t*=2.00

**Figure 8: The evolution of vortex ring interaction is visualized by vorticity magnitude colorized by helicity density for A-notched nozzle. Each column displays the side view and bottom view at the same time instance. From left to right, the column display the time: t*=0.5, t*=1.56 and t*=2.00. The primary ring (R1) is formed following the A-shape of the exit. Two distinct trailing rings (R2a-I and R2a-II) are created at trough locations after the piston stops.**

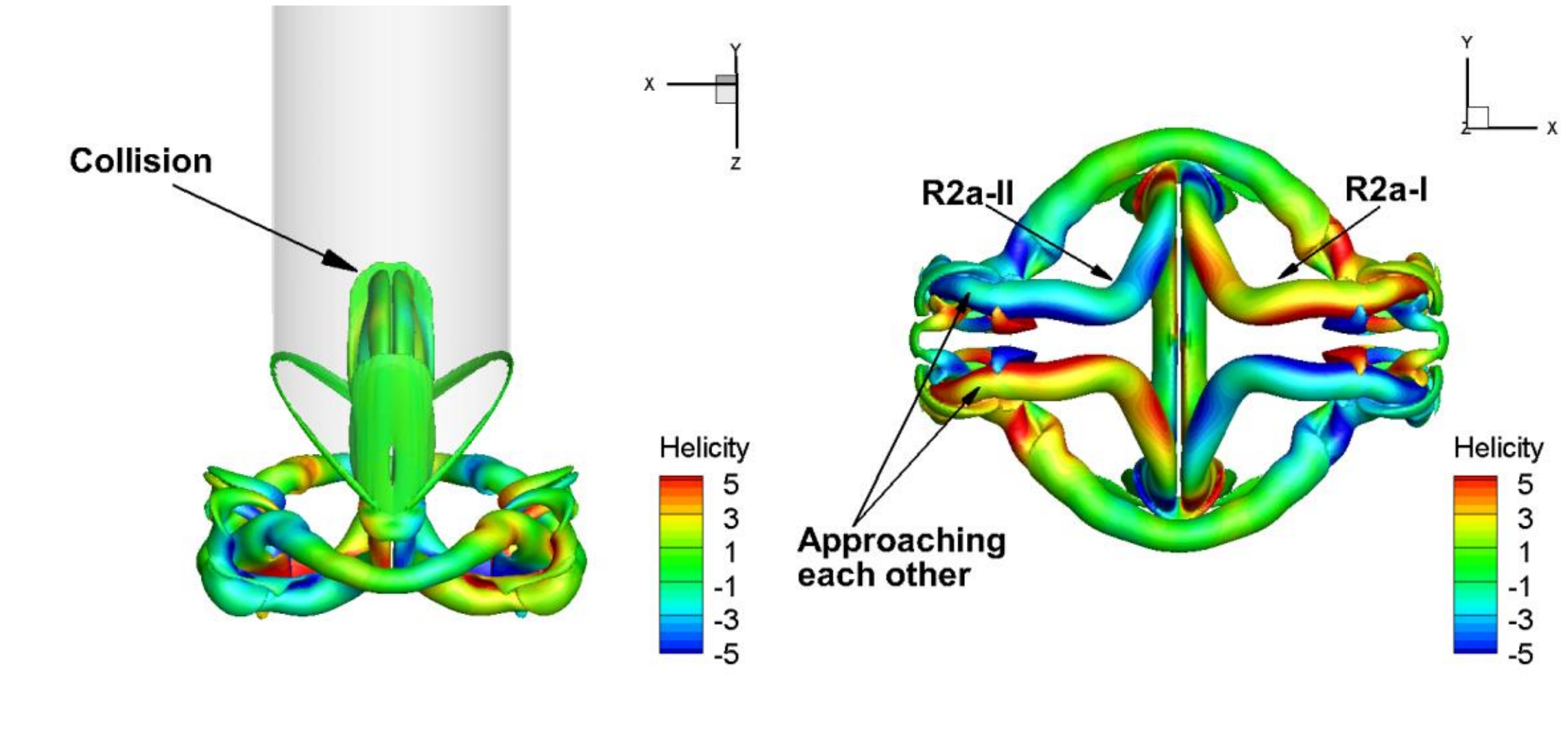

(a) t*=3.0

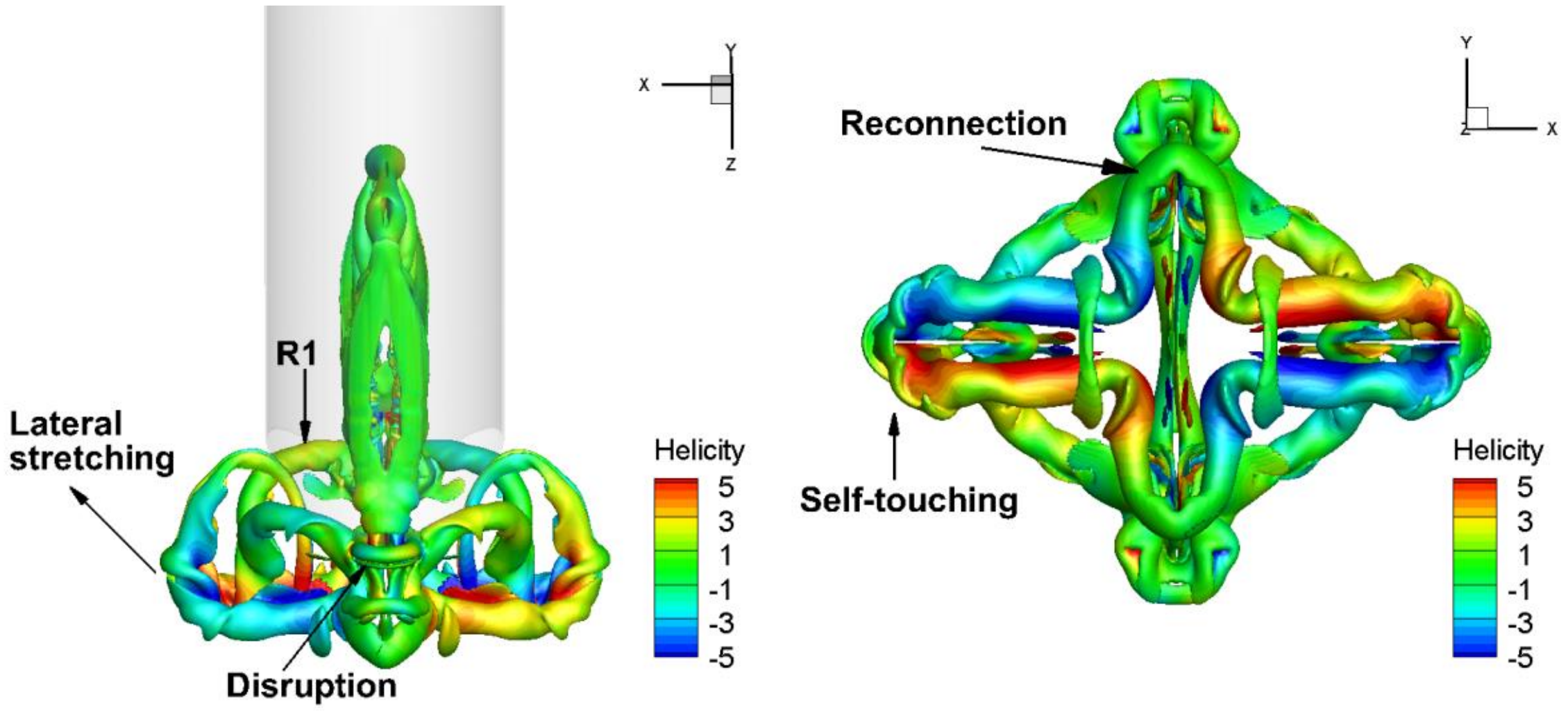

(b) t*= 4.0

**Figure 9: The collision of two secondary rings in A-notched nozzle are visualized by iso-surface of vorticity magnitude colorized by helicity density. The spanwise vortices are formed at peaks while streamwise vortices at troughs are remaining parts of secondary rings**

*The V-notched nozzle*

The three dimensional structure of the vortex cores from V-notched nozzle is visualized as seen in

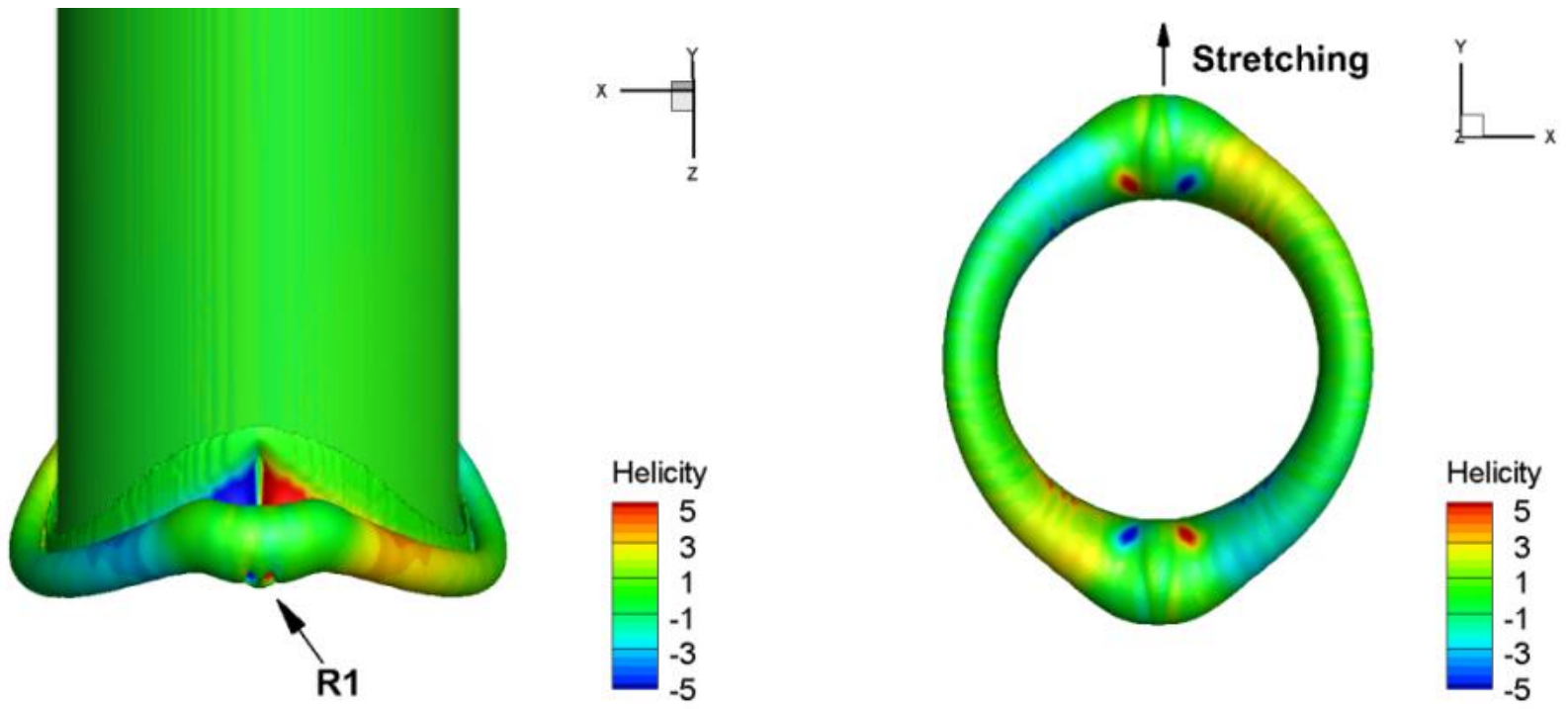

(a) t*=0.5

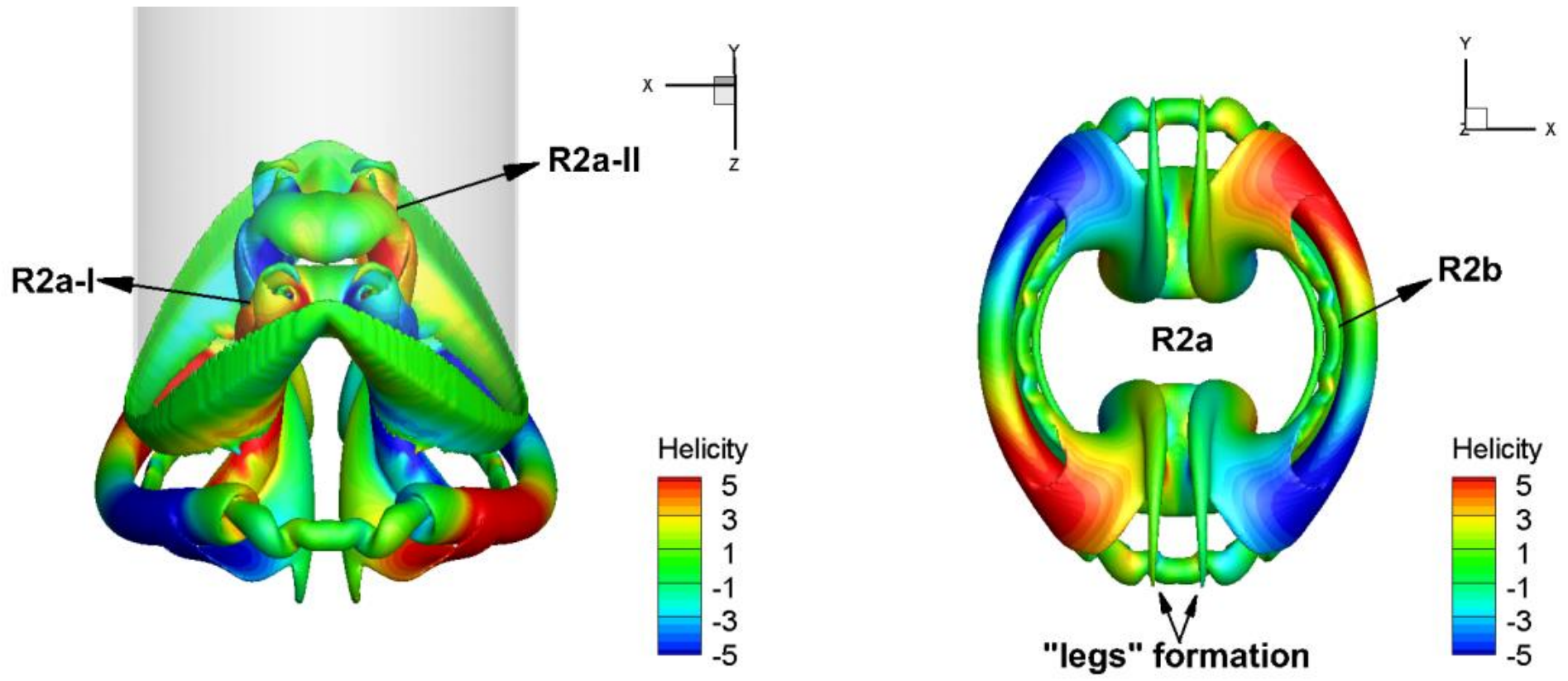

(b) t* = 1.56

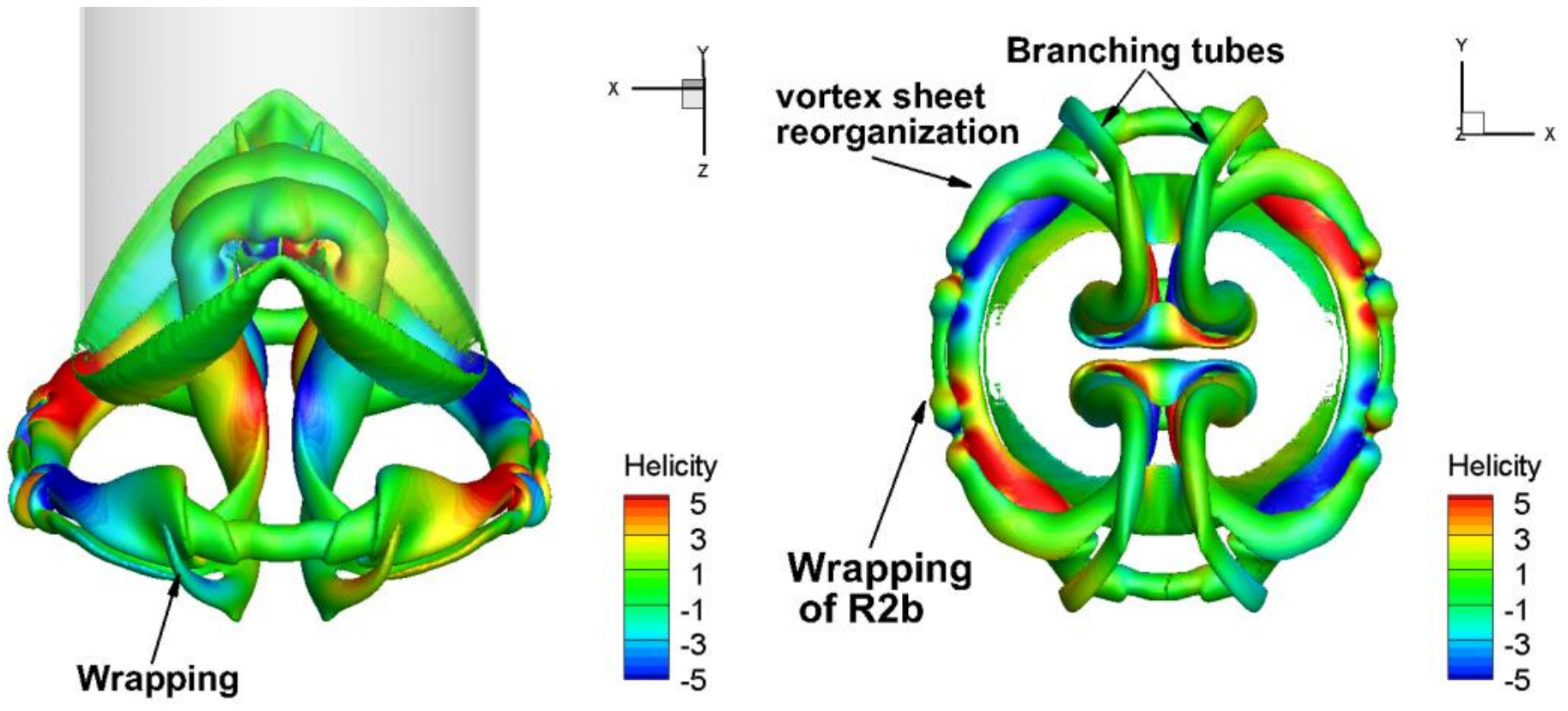

(c) t* = 2.0

Figure 10. Initially from $t^*=0$ to $t^*=0.5$, the shear layer rolls up at the nozzle exit edge as shown

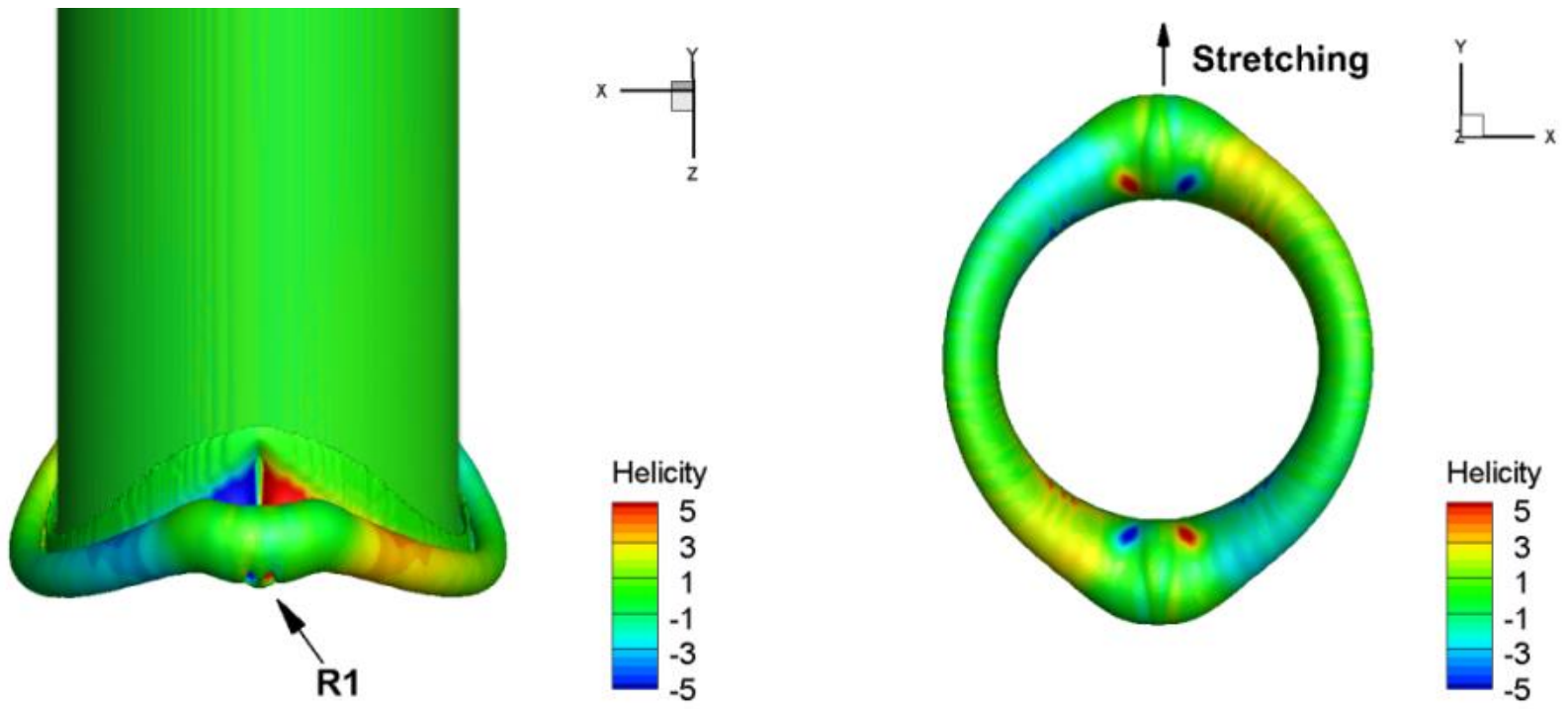

in

(d) t*=0.5

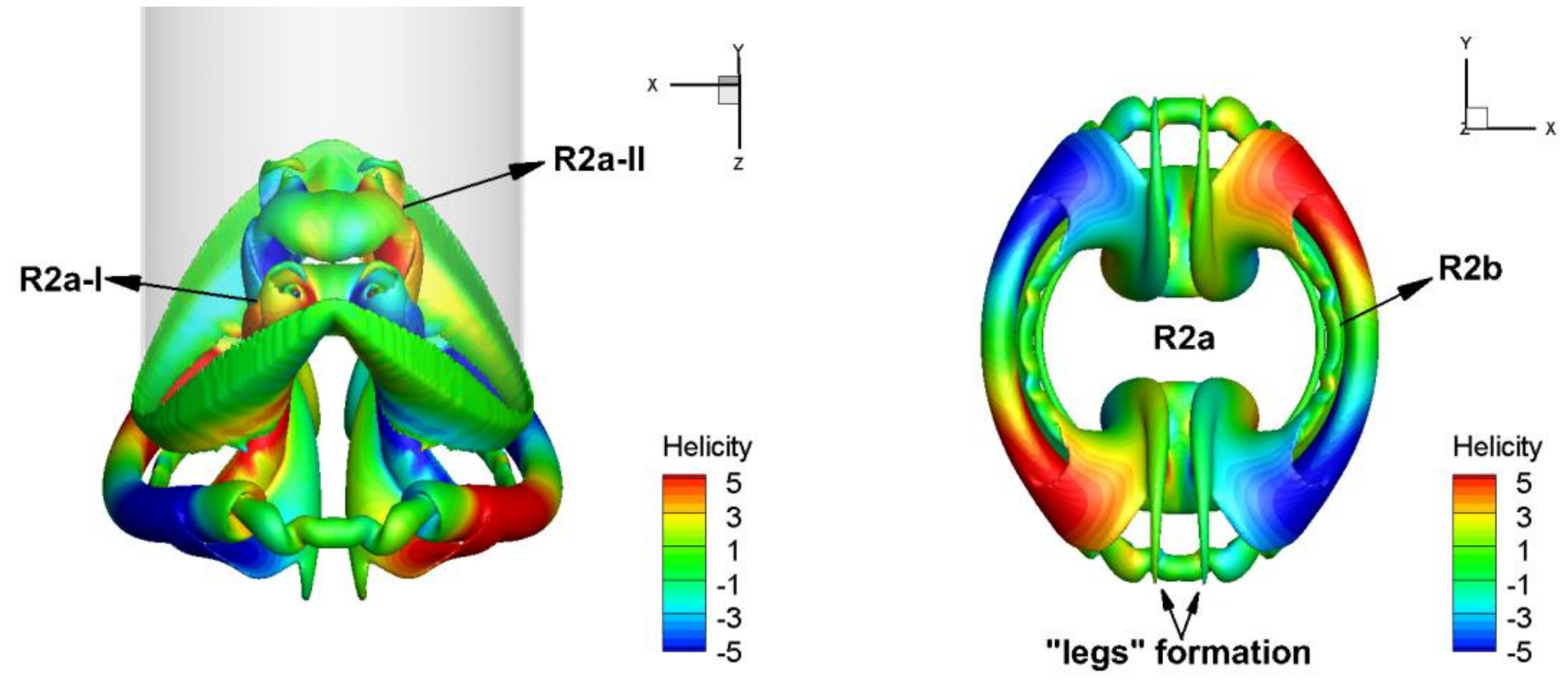

(e) t* = 1.56

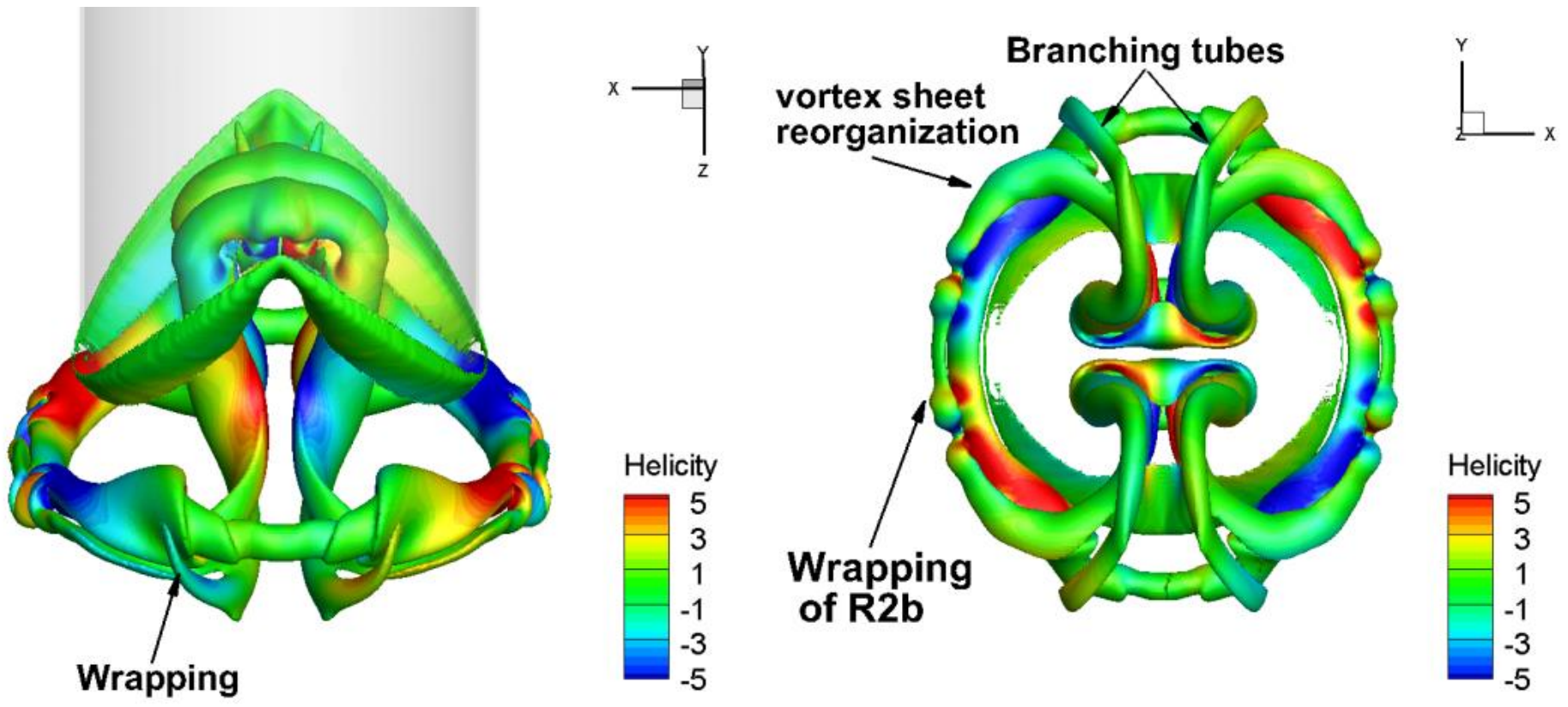

(f) t* = 2.0

Figure 10a. There exists a primary vortex core (*R1*) following the shape of V-shape exit closely. Similar to A-notched nozzle, the core's size in this case also continuously varies along the circumference. As indicated in Figure 7, the vortex core reach its maximum size at trough locations ($\psi = \pi/2$ and $\psi = 3\pi/2$) and its minimum at peak locations ($\psi = 0$ and $\psi = \pi$). At ($\psi = \pi/2$ and $\psi = 3\pi/2$), the ring corners are rounded off and extended with larger sizes in comparison with ones at peaks locations ($\psi = 0$ and $\psi = \pi$). The rounding-off at troughs leads to the lateral stretching as seen in the plane view of

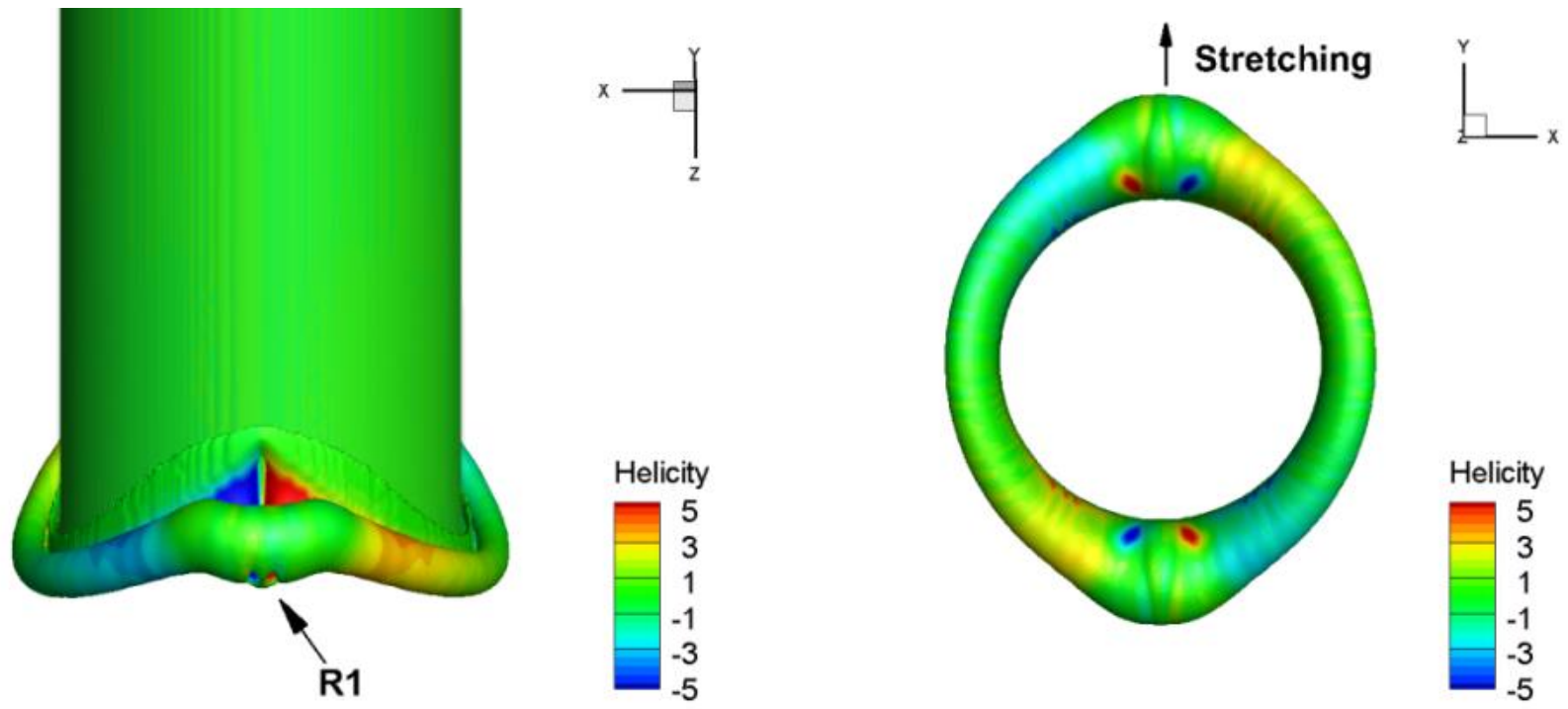

(g) t*=0.5

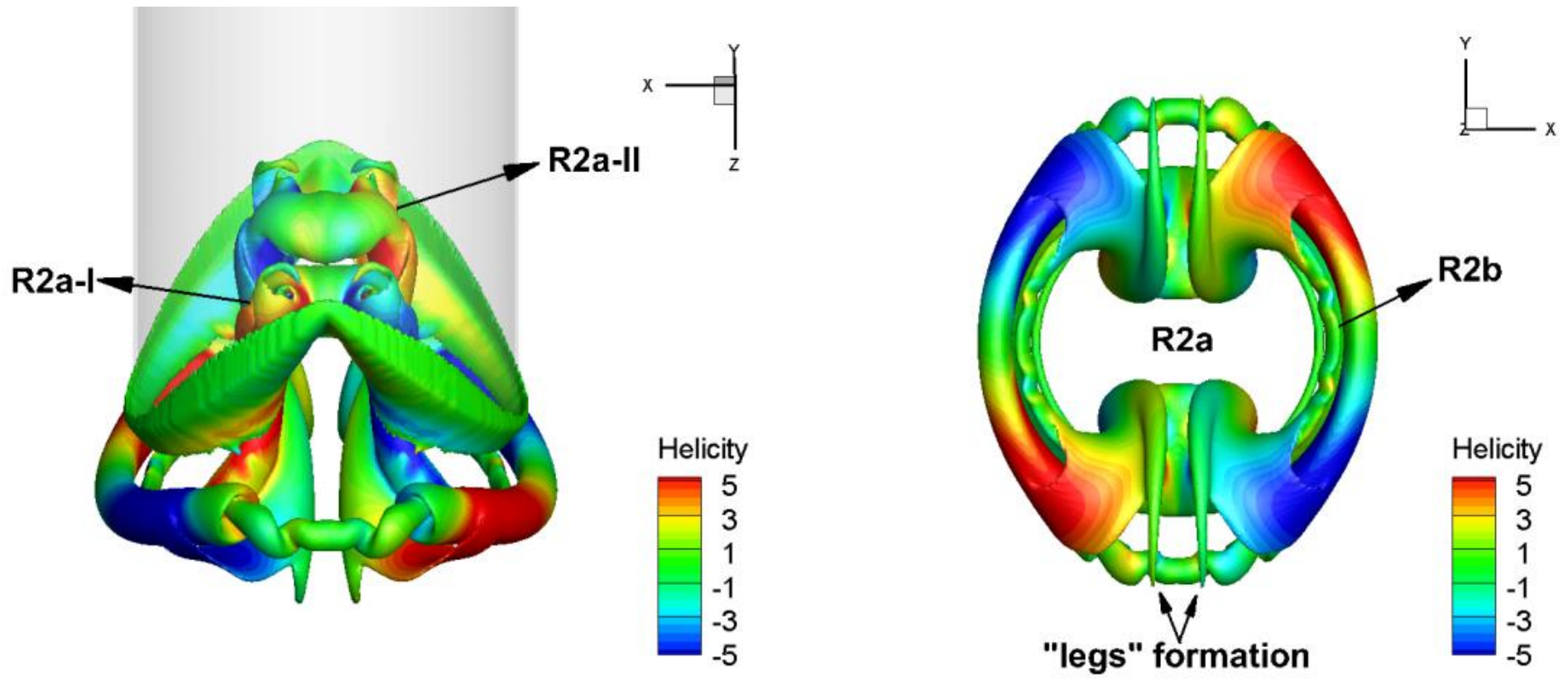

(h) t* = 1.56

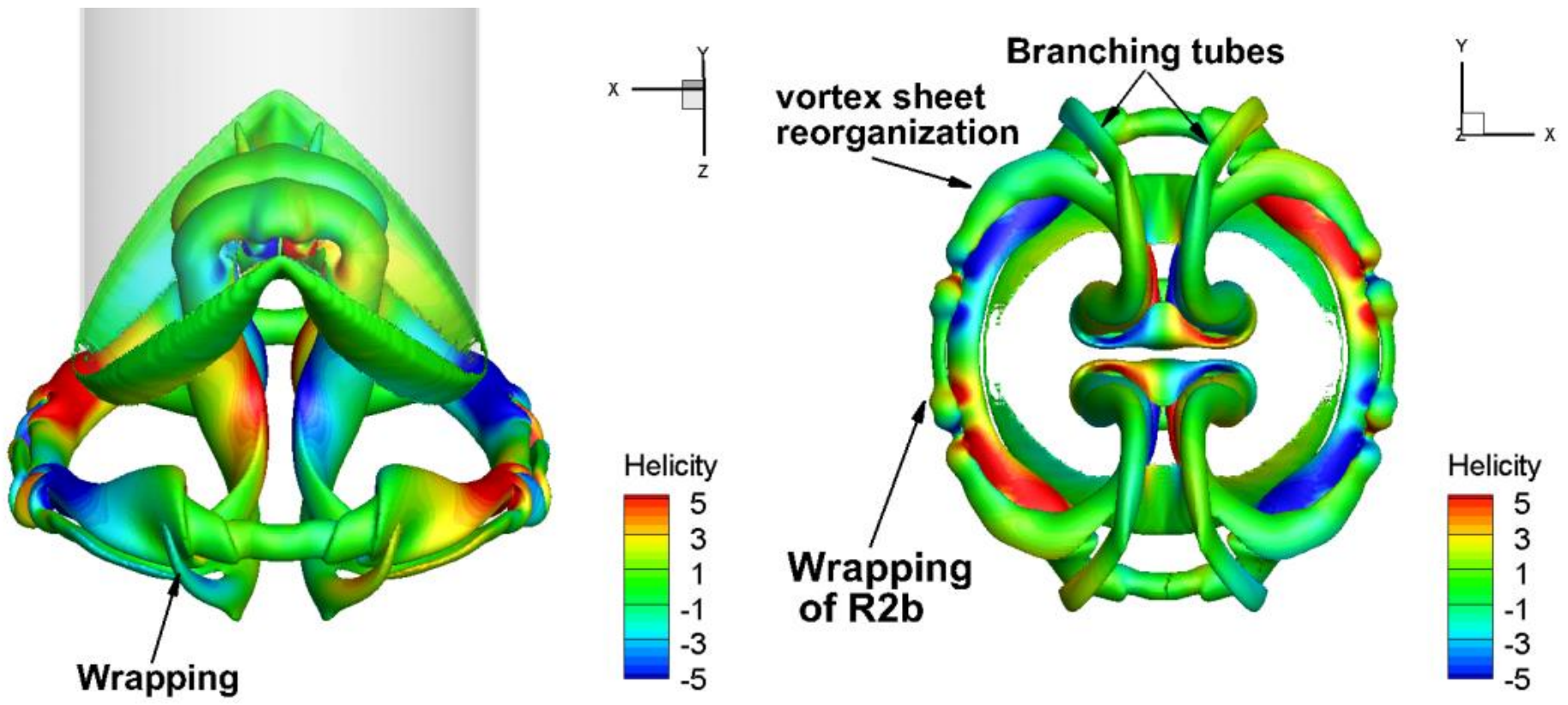

(i) t* = 2.0

Figure 10a.

At *t*=1.56*, *R1* detaches fully from the exit's edge as shown in

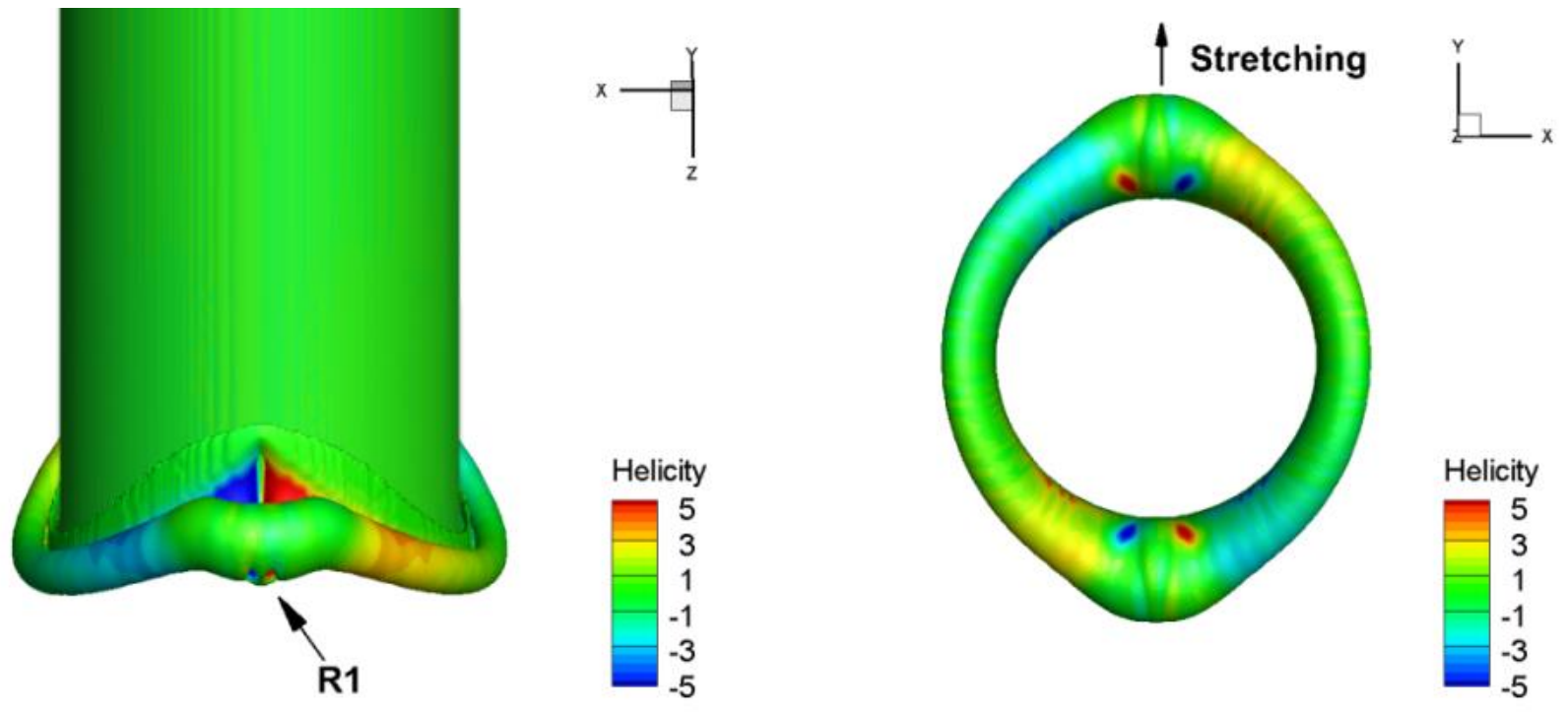

(j) t*=0.5

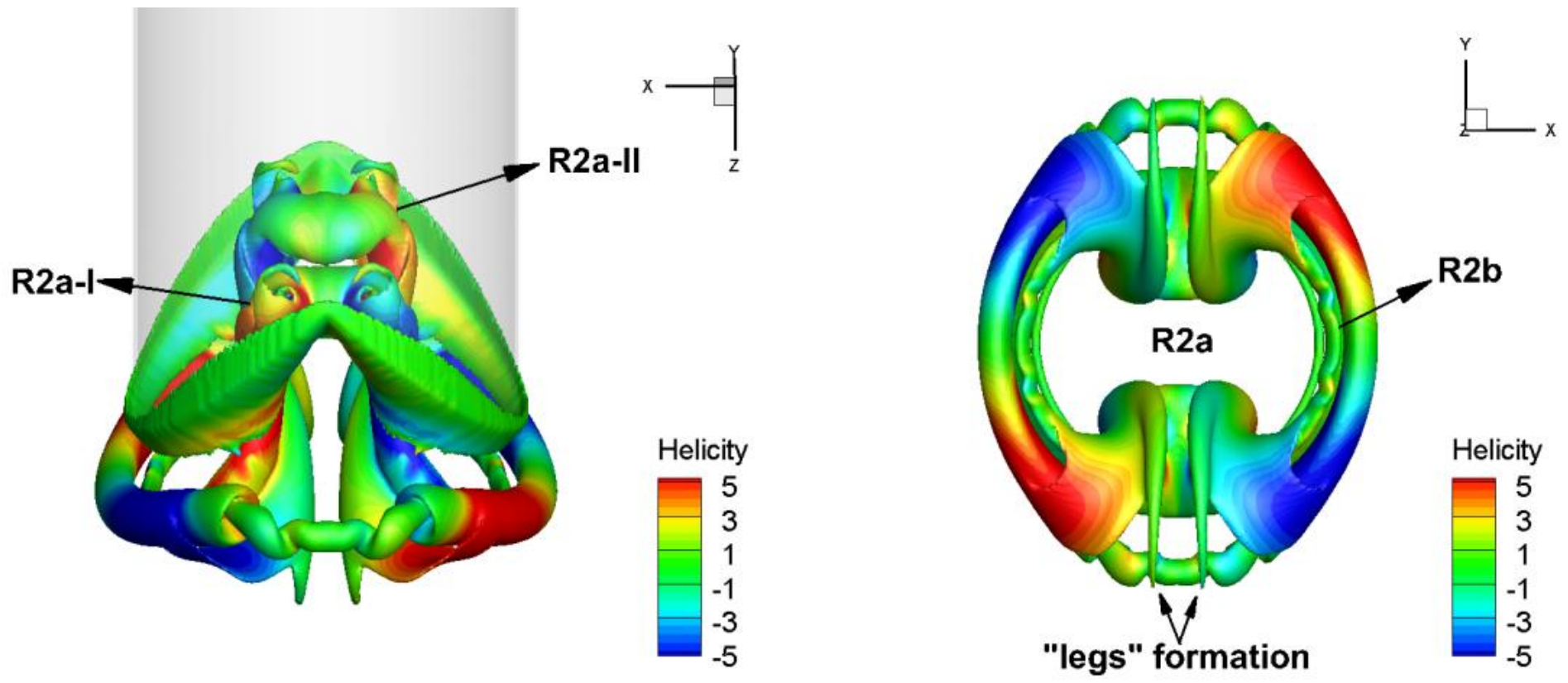

(k) t* = 1.56

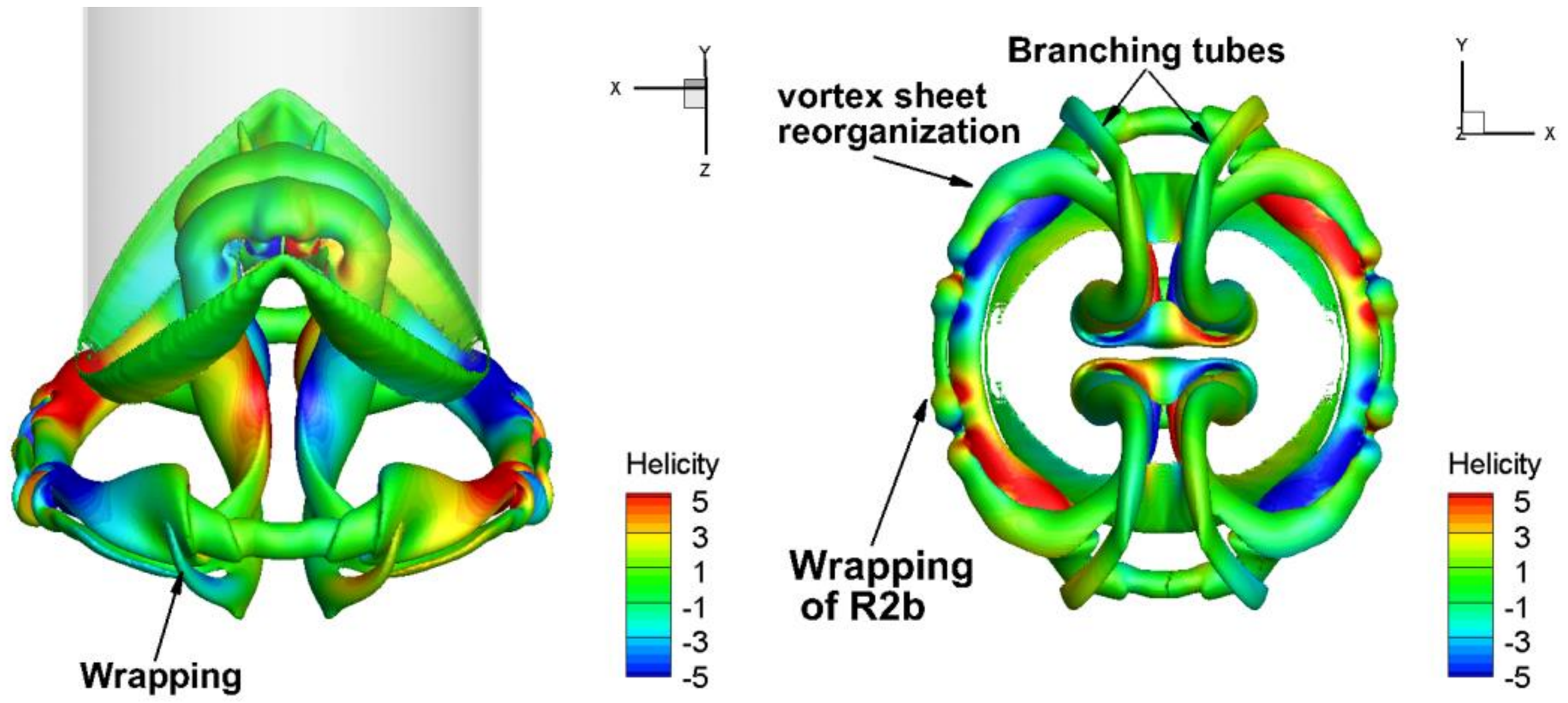

(l) t* = 2.0

Figure 10b. Since the piston has already stopped at *t*=1*, the entrainment of ambient fluid now creates the pair of trailing vortex tubes *R2a* at troughs similar to the A-notched nozzle case. However, the smooth curvature of the nozzle exit at peaks now adds further complexities to the flow structure. The original vortex sheet, which supplies to the leading ring, splits into several portions after the piston stops. Similar to A-notched nozzle, there exists a pair of *λ-shape R2a* connecting to the leading ring *R1* via a pair of tubes ("*leg*") at troughs ($\psi = \pi/2$ and $\psi = 3\pi/2$). In addition, the remnant of the above-mentioned vortex sheet reorganizes into small arches denoted as *R2b* as seen in

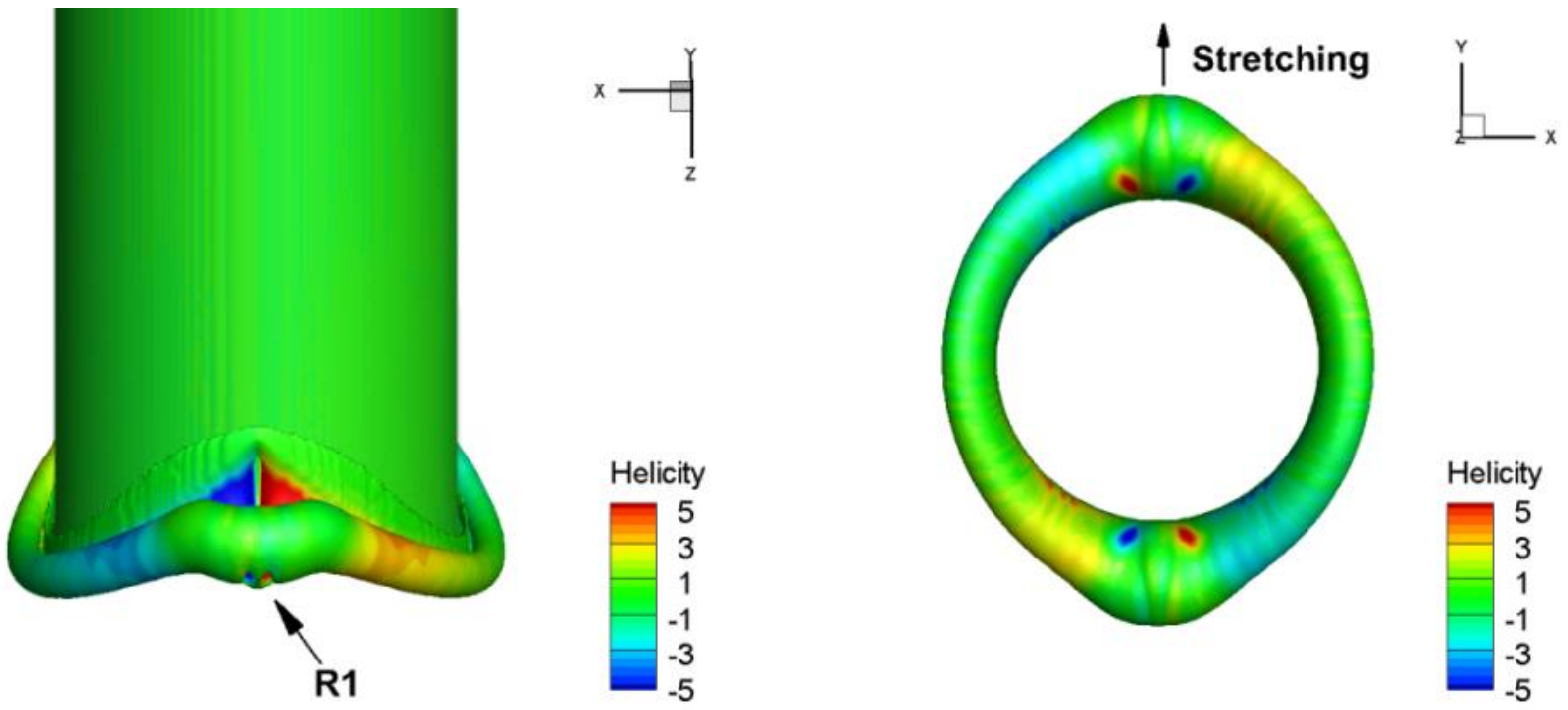

(m)t*=0.5

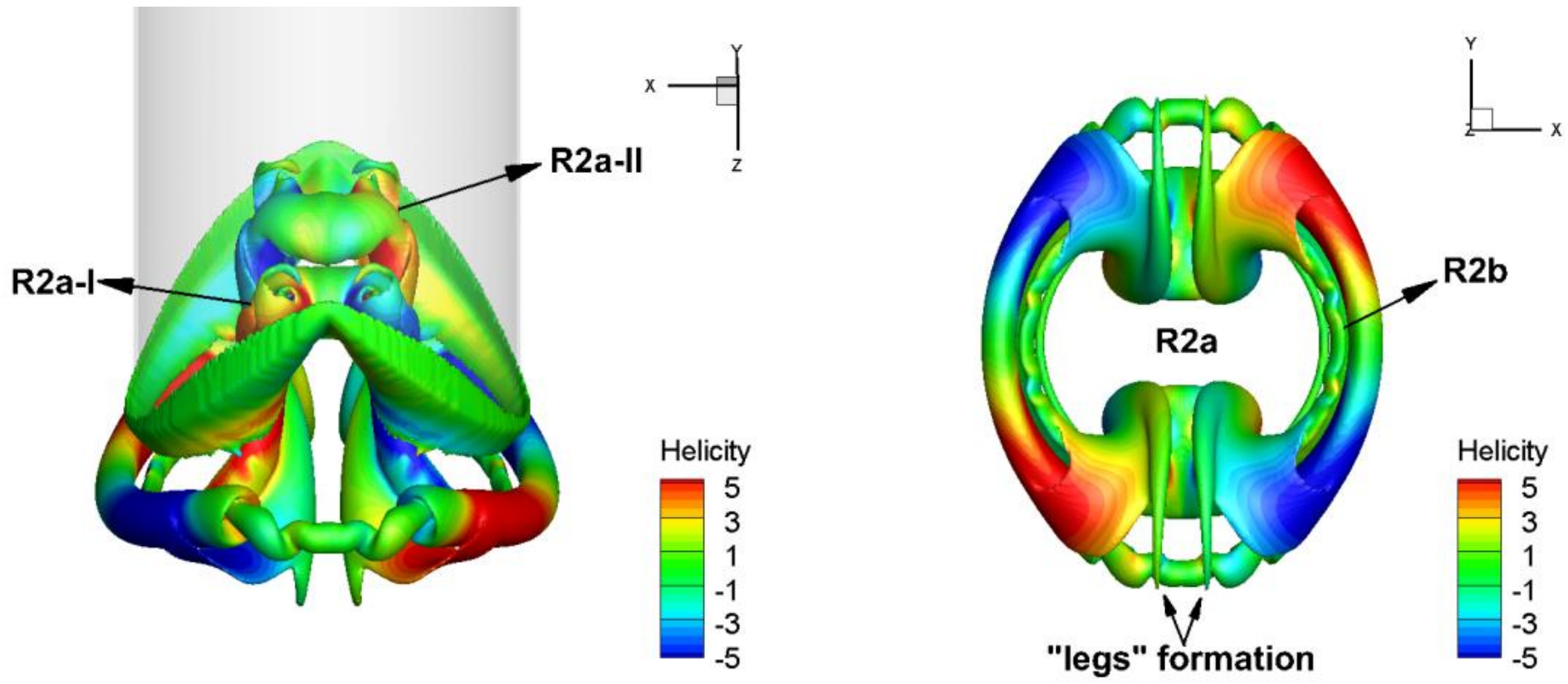

(n) t* = 1.56

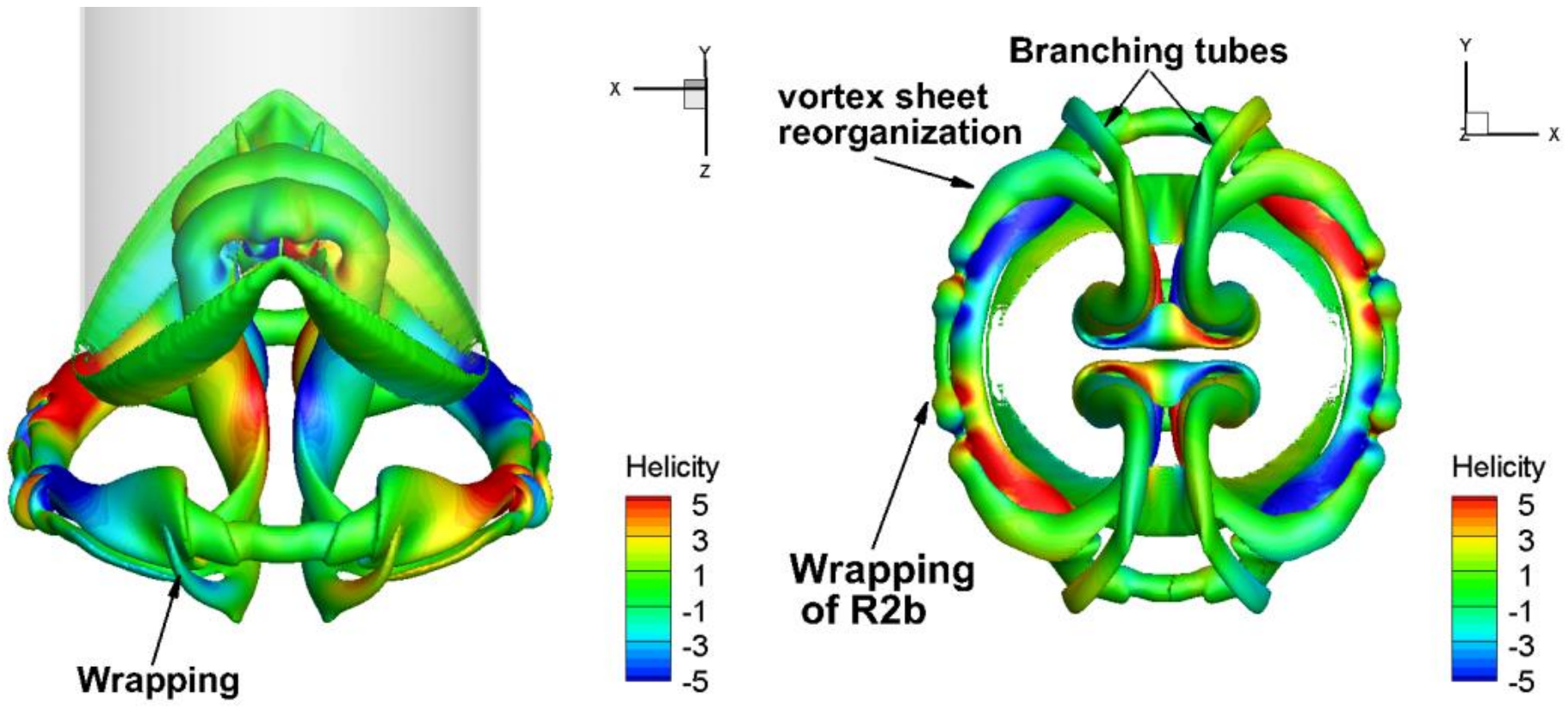

(o) t* = 2.0

Figure 10b. These *R2b* structures are formed symmetrically across the *trough-to-trough* plane and parallel to the curvature of the main ring *R1*. As *R2b* splits from the vortex sheet, it starts to exhibit wavy instability. As it fully splits from the sheet, *R2b* approaches the main ring *R1* and intertwines with the *R1* during its growth as observed in the plane view of

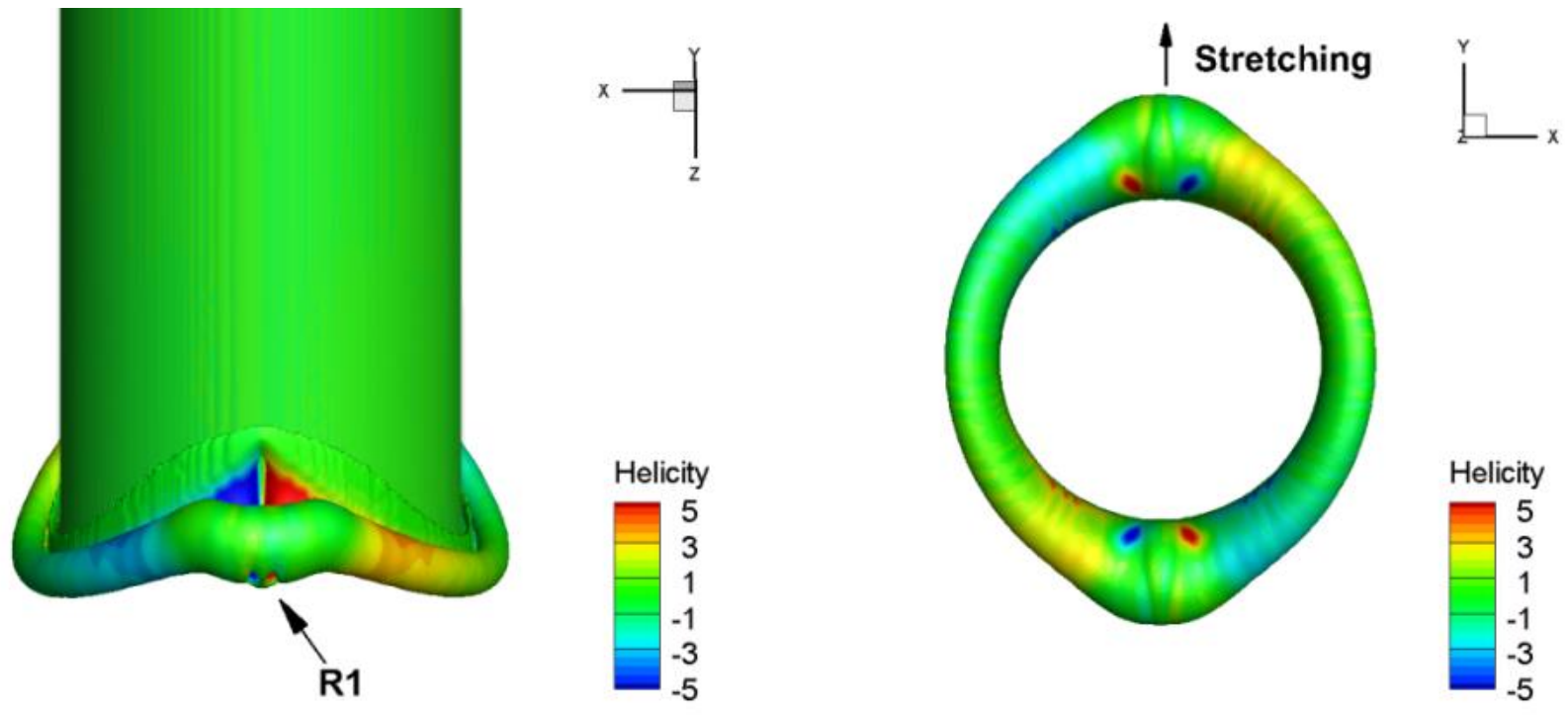

(p) t*=0.5

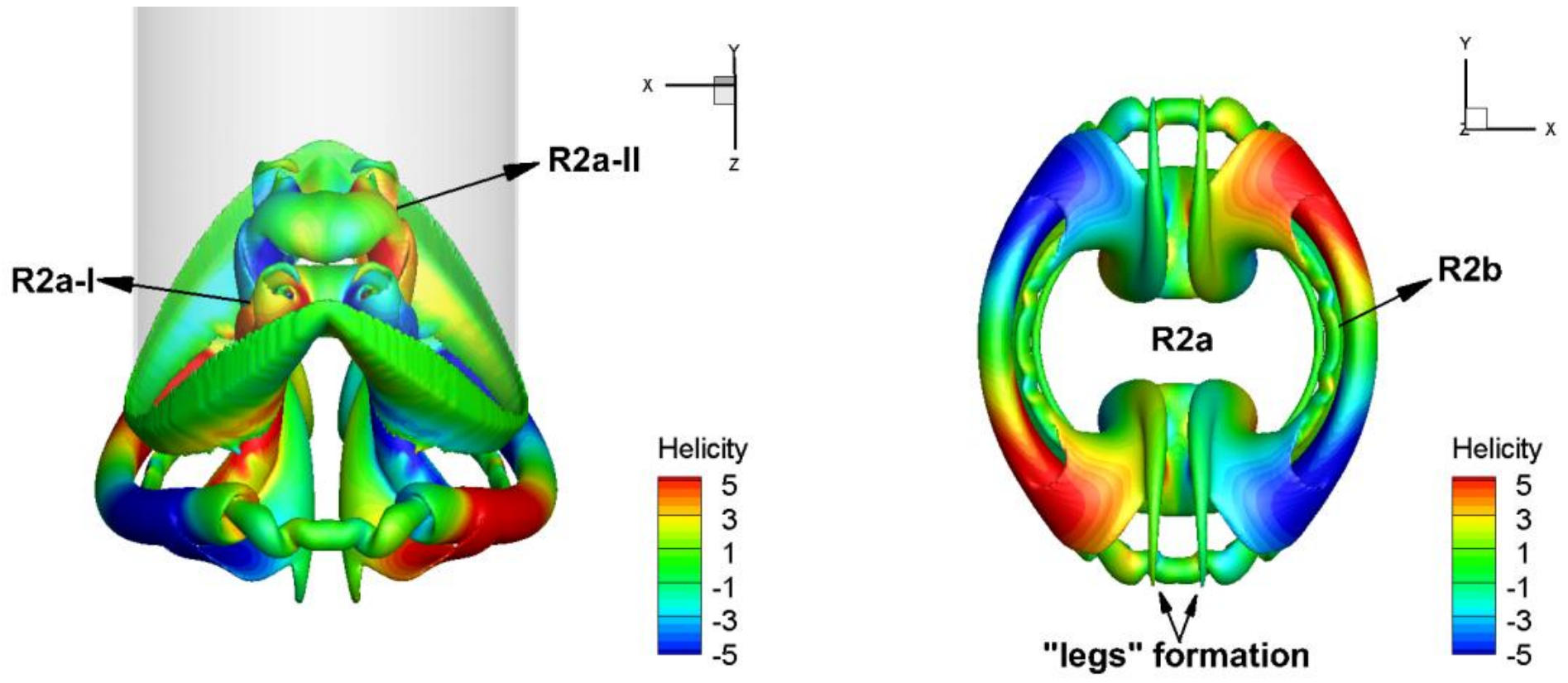

(q) t* = 1.56

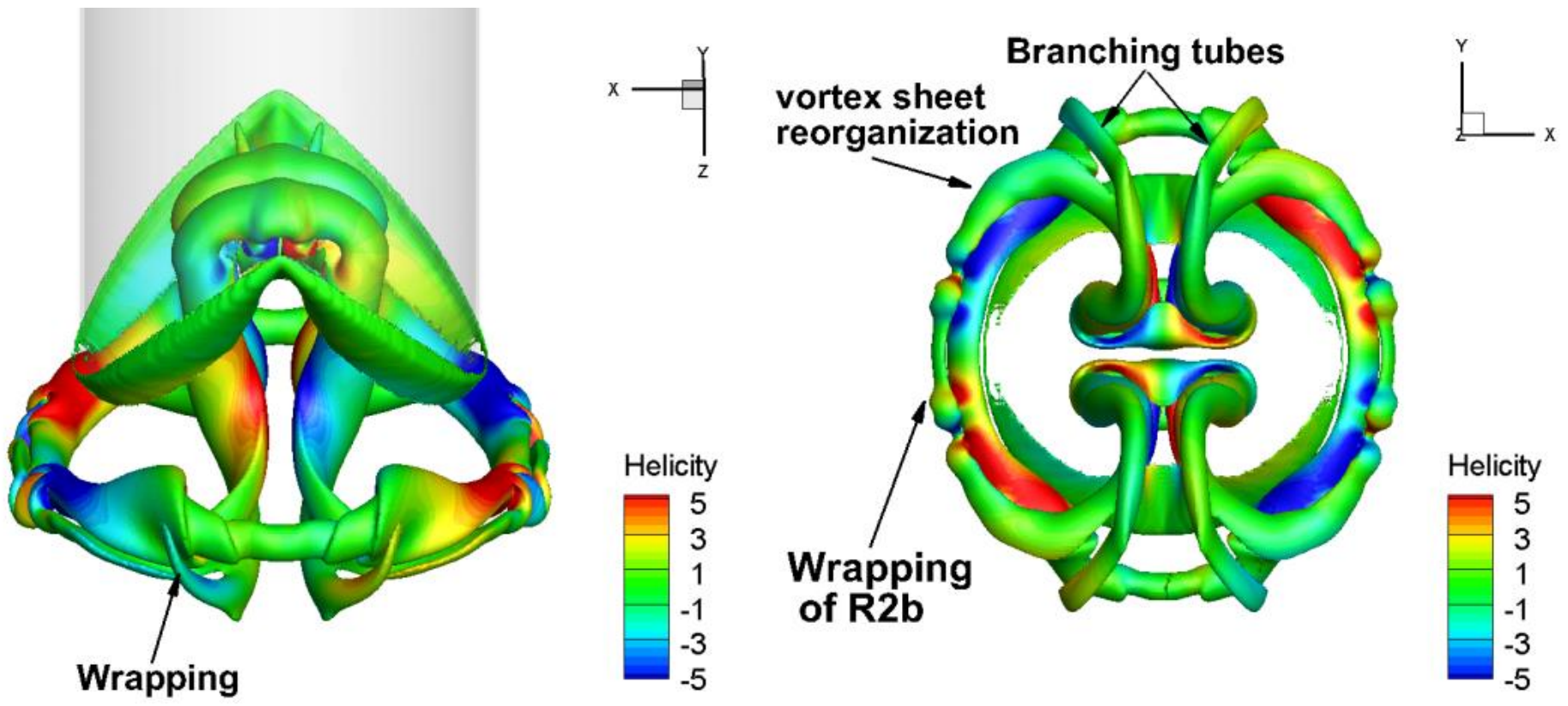

(r) t* = 2.0

Figure 10b.

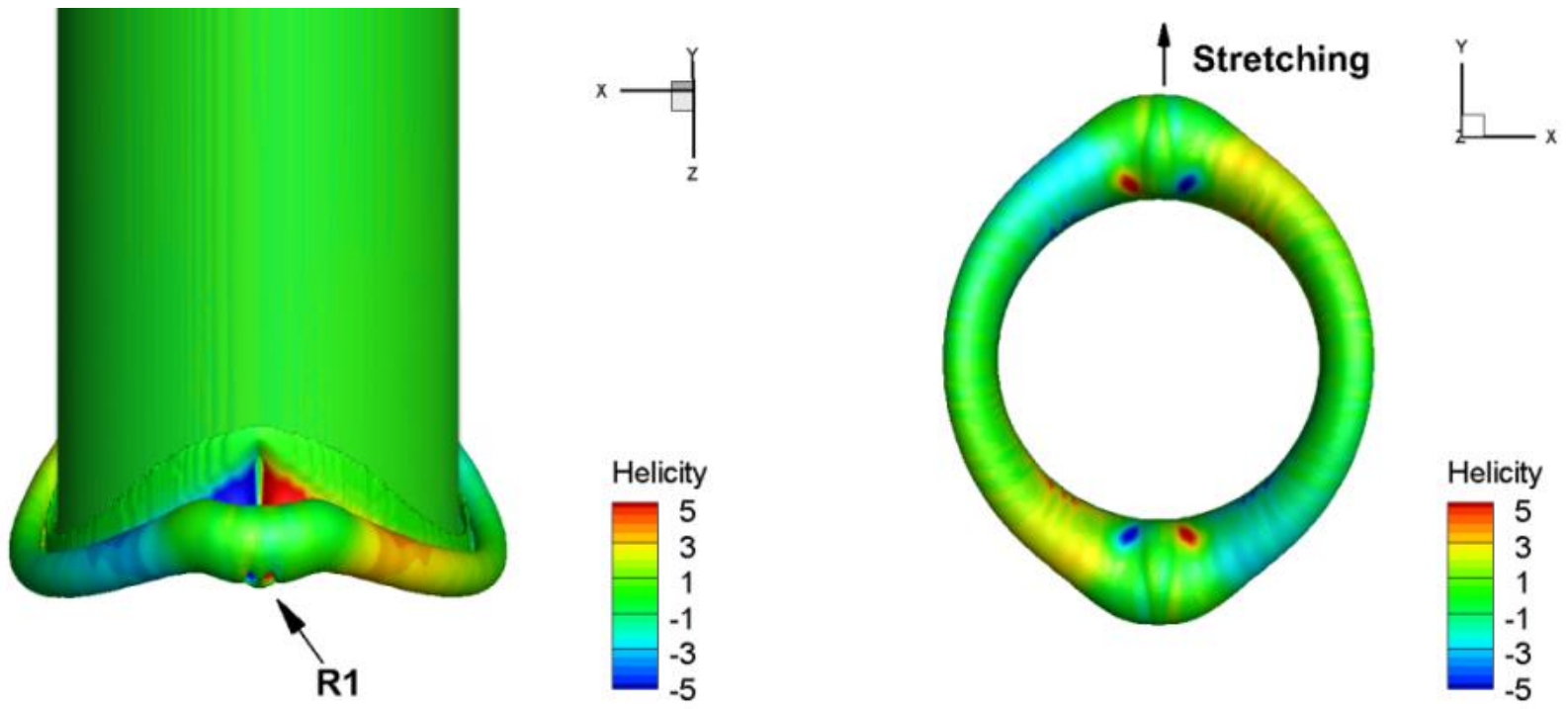

At *t*=2.0* as shown in

(s) t*=0.5

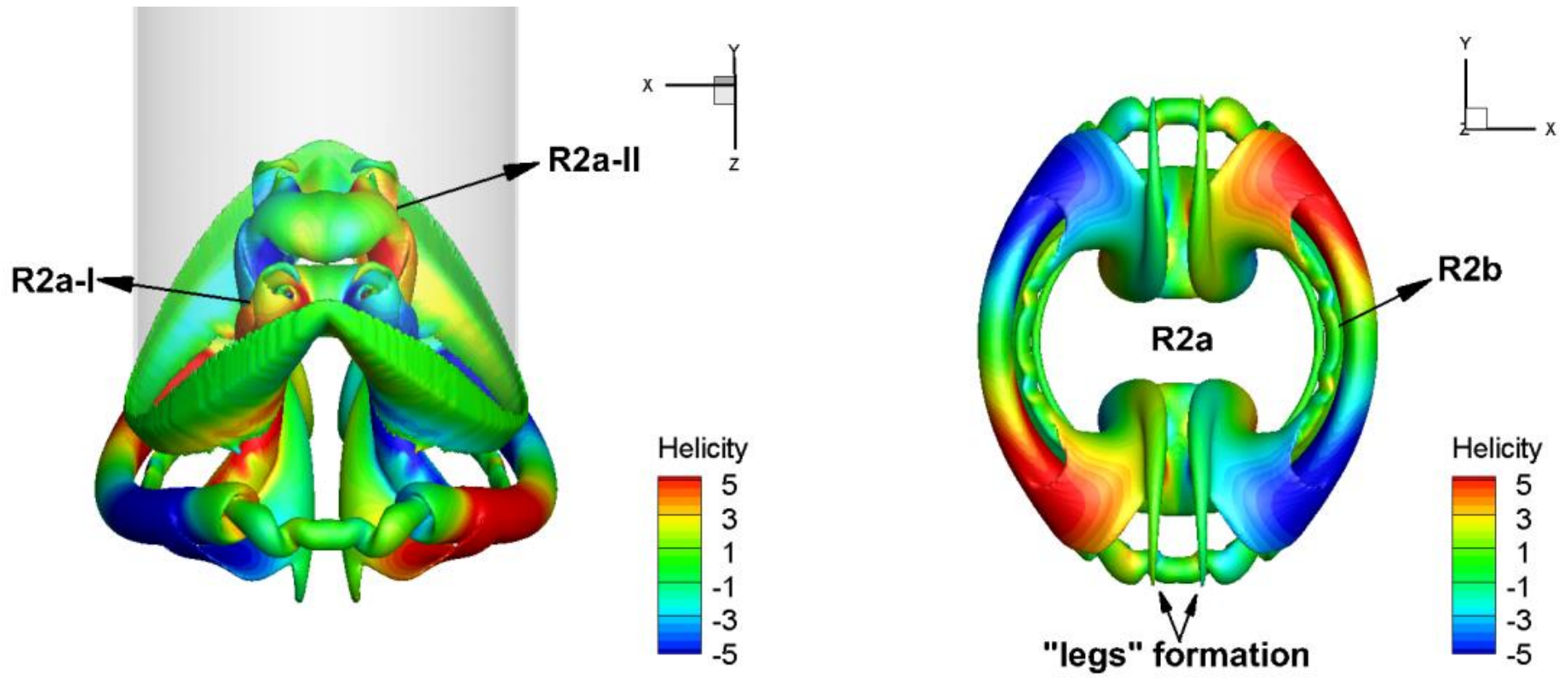

(t) t* = 1.56

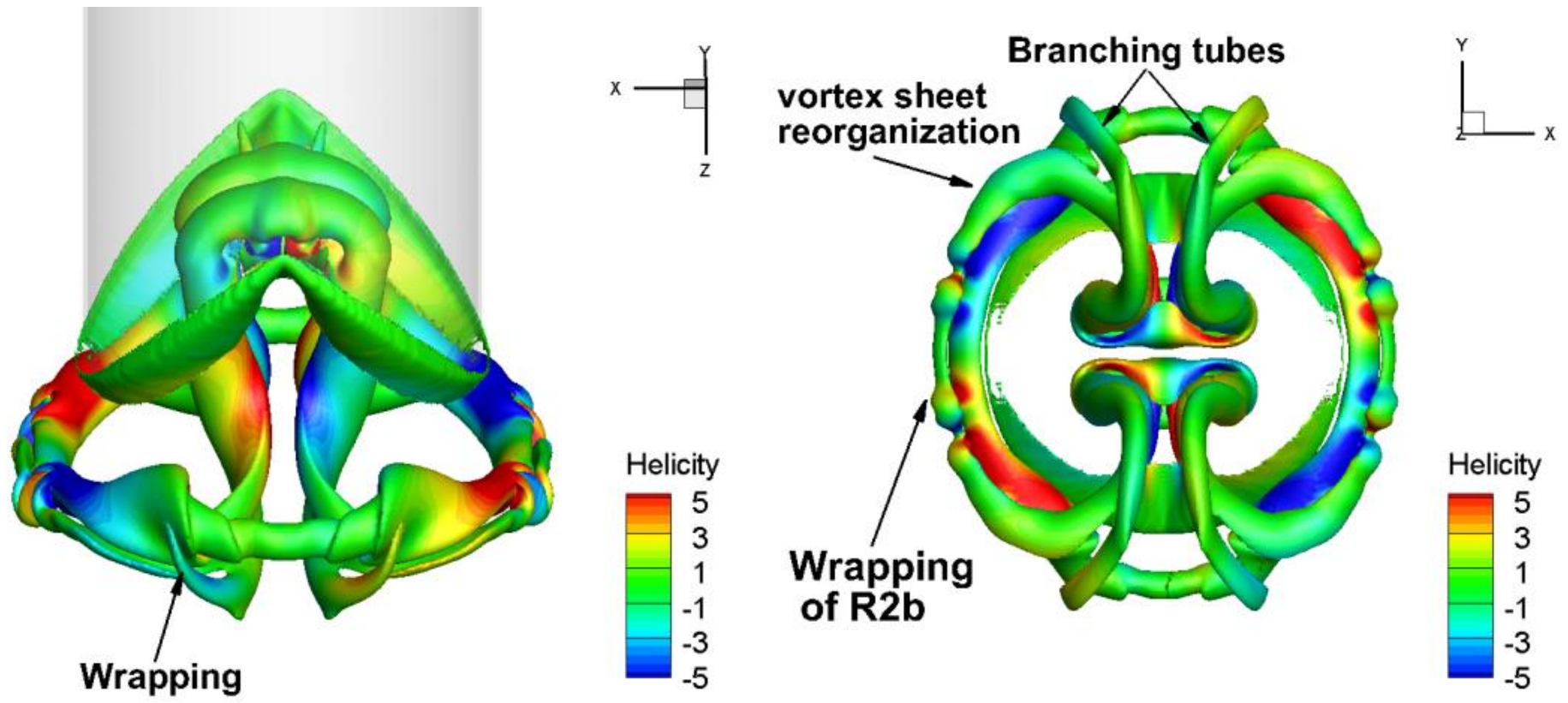

(u) t* = 2.0

Figure 10c, the *R2b* structures are fully formed and twist around the main core *R1*. As intertwining to the main ring *R1*, the wavy instability of *R2b* cores further progresses and distorts its own tubes. *R2b* wraps around the *R1* and excite high level of helicity density at points of attachments. The vortex tubes *R2a* now approach closer each other and thus the upper portion of *R2a* is highly stretched on the *peak-to-peak* plane. During the approaching process, the symmetry of the vortex tube configuration is preserved across the *peak-to-peak* plane.

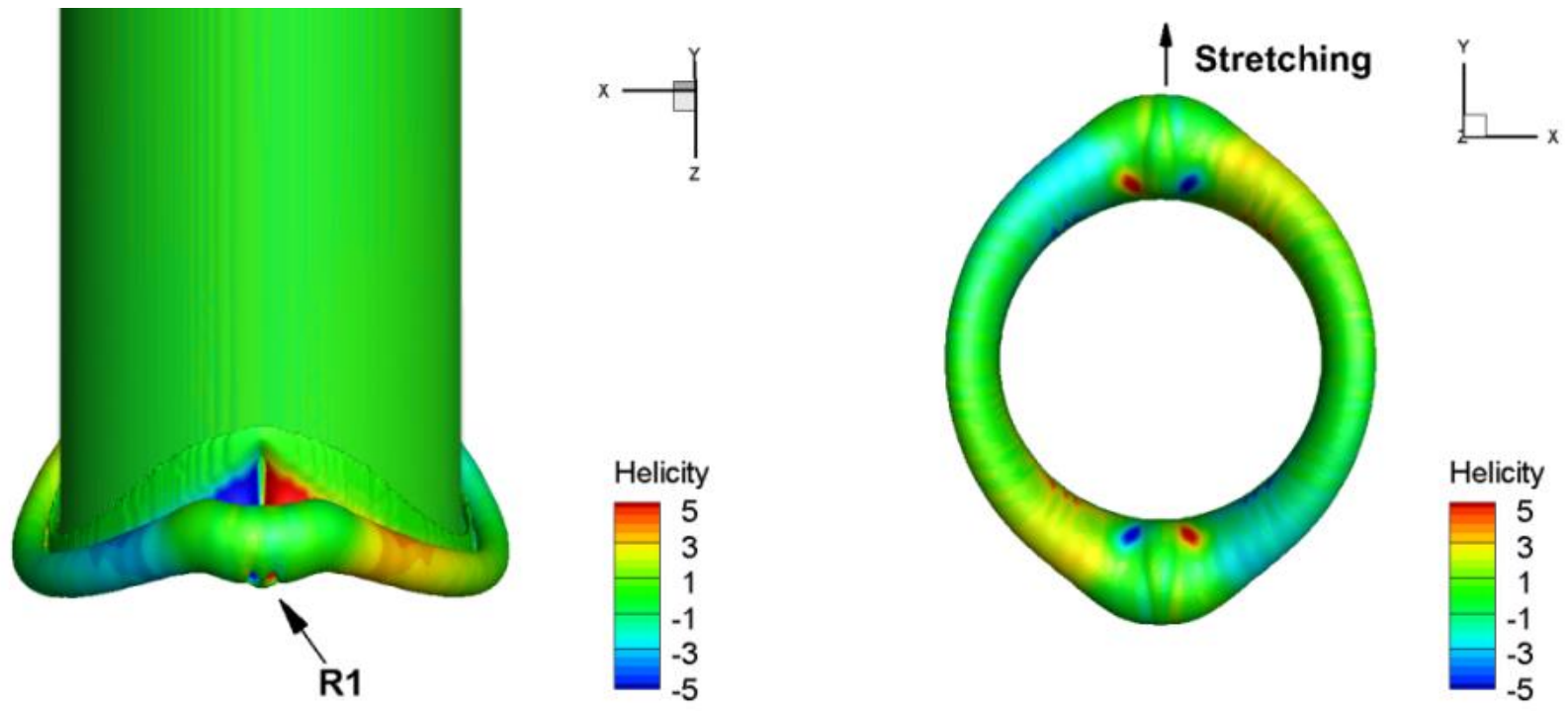

(v) t*=0.5

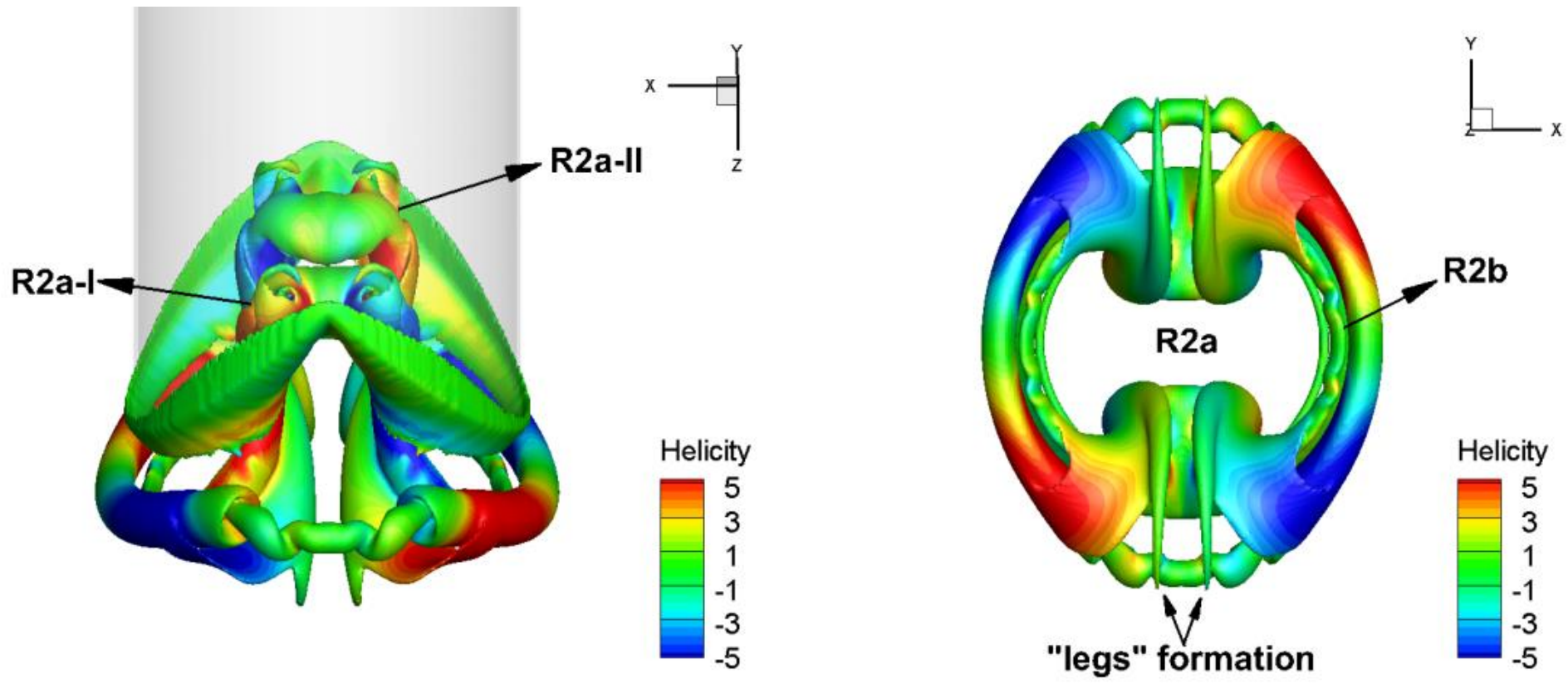

(w) t* = 1.56

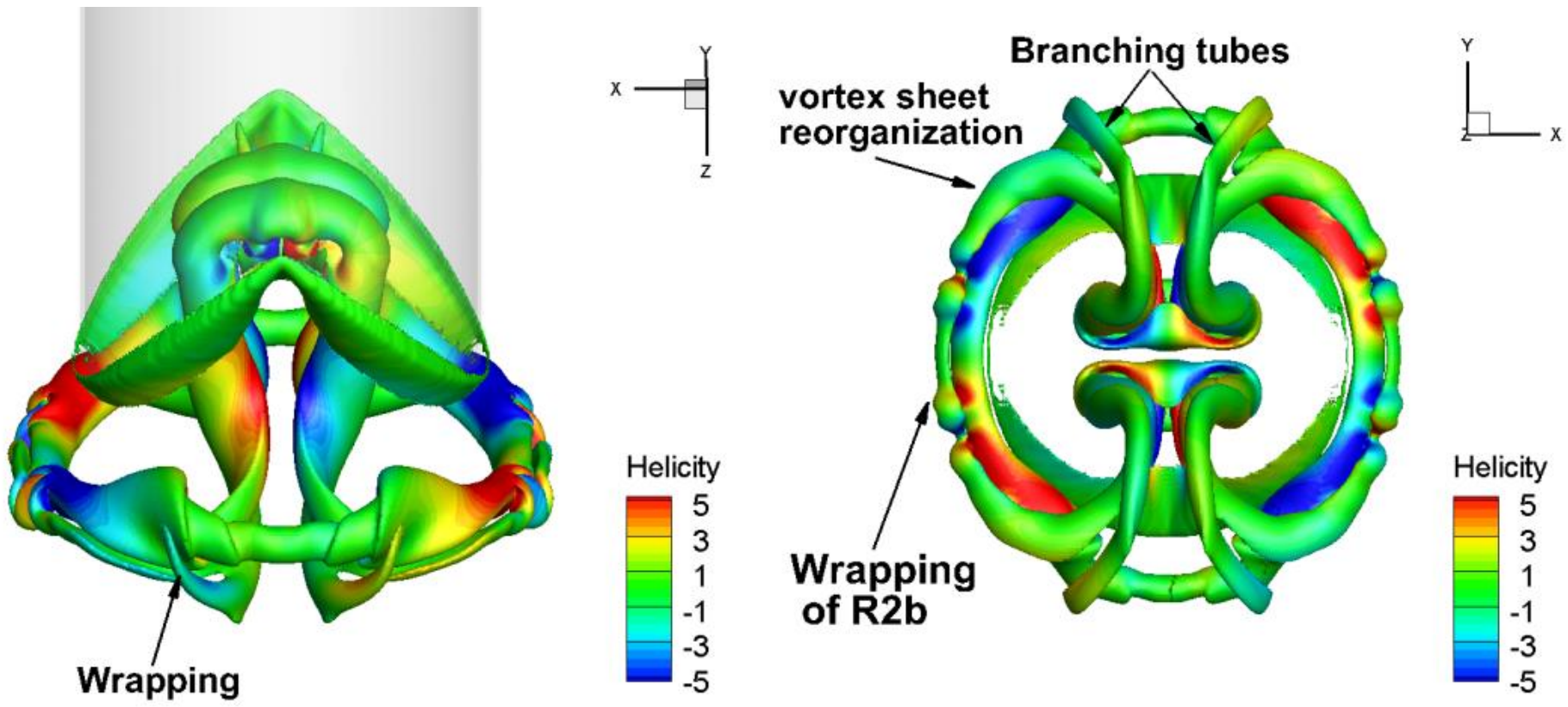

(x) t* = 2.0

**Figure 10: The evolution of vortex ring interaction is visualized by vorticity magnitude colorized by helicity density for V-notched nozzle. Each column displays the side view and bottom view at the same time instance. From left to right, the column display the time: t*=0.5, t*=1.56 and t*=2.00. The primary ring (R1) is formed following the V-shape of the exit. Two distinct trailing rings (R2a-I and R2a-II) are created at trough locations after the piston stops.**

At *t*=2.62*, the pair of structures *R2a* finally collide on the symmetry plane as seen in Figure 11. The vortex tubes are first stretched on the *peak-to-peak* plane. This stretching induces the vortex tubes *R2a* to be closer to the leading ring *R1*. As *R2b* twists and fully merges with *R1*. The remnant of *R2b* vortex cores can be observed on the plane view of Figure 11a. The annihilation of cores *R2b* induces the strong twisting of *R2a* and remnant of *R2b* as shown in Figure 11b. The *"leg"* of *R2a* keeps wrapping around the remaining core of *R1*. As the tube *R2a* starts to approach the nozzle's surface and interacts with the solid wall.

At *t*=3.0*, the tubes of *R2a-I* and *R2a-II* interact to each other as their cores approach each other near the peaks ($\psi = 0$ and $\psi = \pi$). A vortex-wall interaction occurs just outside the nozzle exit at $\psi = 0$ and $\psi = \pi$ where the advancing *R2a-I* and *R2a-II* gets interrupted by the peaks. Two lower portions of *R2a-I* and *R2a-II* finally collapse.  The collapse of *R2a* at $\psi = 0$ and $\psi = \pi$ is seen to induce new, arch-like secondary vortex tubes (denoted as *T-I* and *T-II* in Figure 11 and also seen at later times in Figure 12) that wraps around the original vortex tubes *R2a*. The formation of T is similar to what has been observed during the collapse of Lissajous-elliptic (LE) vortex rings studied by (Fernandez et al., 1995).

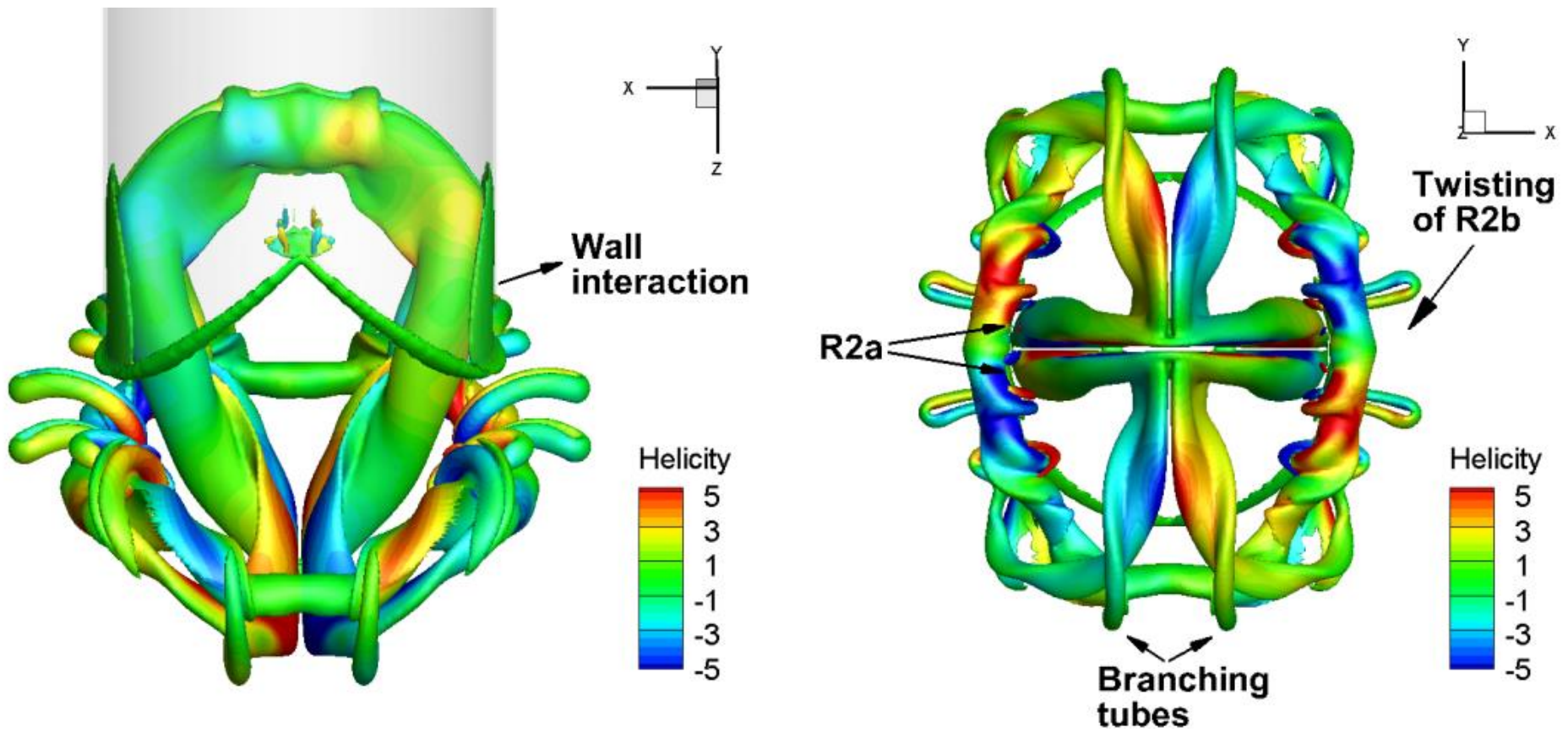

(a) t* = 2.62

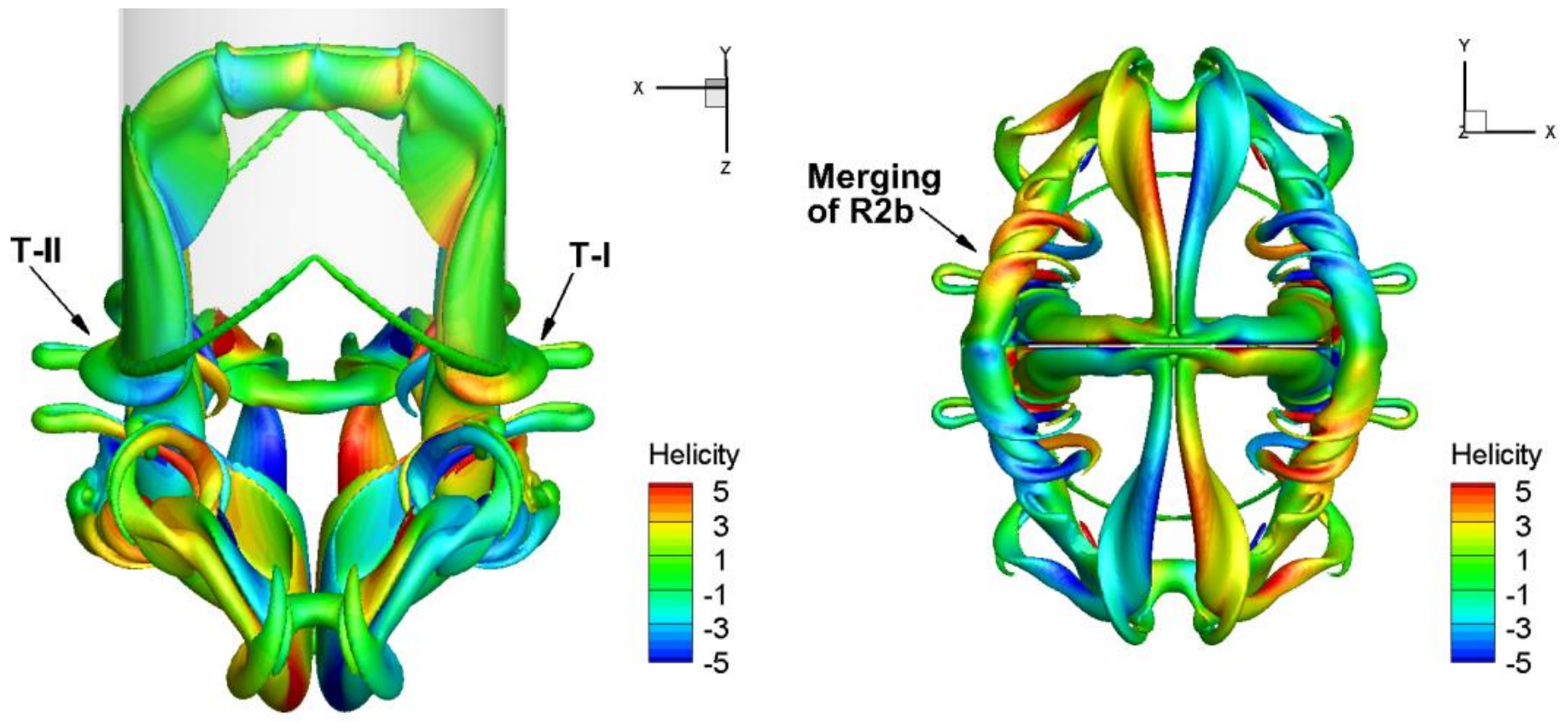

(b) t*=3.0

**Figure 11: The collision of two secondary rings in V-notched nozzle are visualized by iso-surface of vorticity magnitude colorized by helicity density. The spanwise vortices are formed at peaks while streamwise vortices at troughs are remaining parts of secondary rings**

At *t*=4.0*, vortex reconnection occurs as shown in Figure 12. The *R2a* cores twist around each other and exchange vorticity. The *R2a-I* and *R2a-II* are stretched and merged into the larger core. The strong twisting finally induces the pinch-off of the merged cores into two separated parts. The upper portions of *R2a-I* and *R2a-II* are separated and rejoined into a separated structure, which still connects to the cylinder. Similar process has been reported in several previous works (Aref and Zawadzki, 1991; Ashurst, 1987; Chatelain, 2003; Kida et al., 1991) on vortex ring collision. However, here the pinch-off process is largely under effect of twisting rather than stretching along the axial direction. Note that two branches *R2a-I* and *R2a-II* are two anti-parallel vortex tubes and bridges do not exist during the reconnection process. Two structures *T-I* and *T-II* are seen to grow larger and propagate further downstream. The remnant tubes of *R2* reconnect to each other while maintaining their links with the descendent branching tubes of *R2*.

The evolution at later time of the vertical structures involves the diffusion of the tubes due to viscosity.

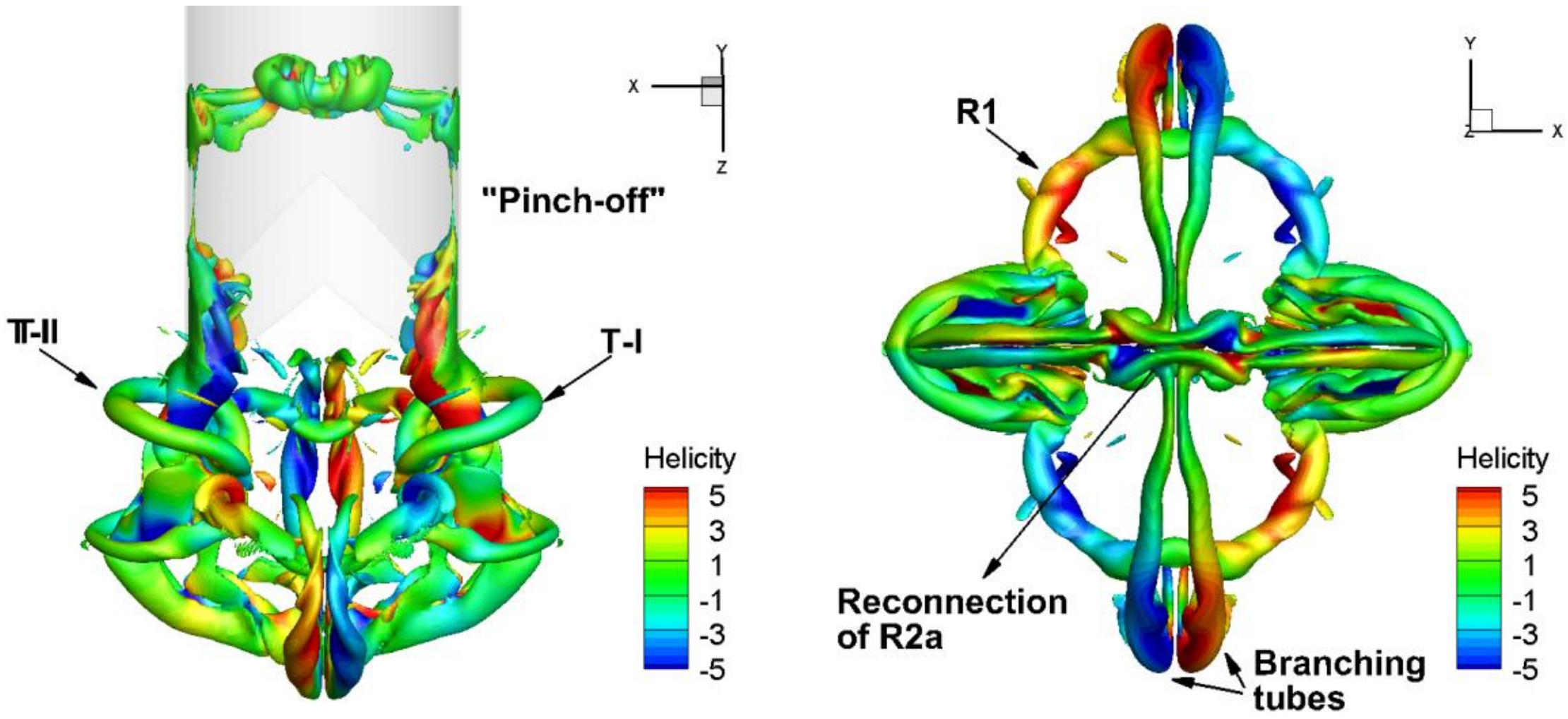

**Figure 12: The collision of two secondary rings in V-notched nozzle are visualized by iso-surface of vorticity magnitude colorized by helicity density. The spanwise vortices are formed at peaks while streamwise vortices at troughs are remaining parts of secondary rings at t*=4.0**

*The W-notched nozzle*

The vortex formation process of *W-notched* nozzle has the same sequence as the *A-notched* and *V-notched* nozzles. The first step of the process includes the formation of the leading ring. In this case, there are four notches and thus the entrainment at the troughs stretch the leading ring (*R1*) outward as shown in Figure 14a at $t^*=0.5$. This stretching perturbs the formation of kinks around the circumference as predicted by (LONGMIRE et al., 1992a). These four kinks induce the lateral deformation of *R1* as it is formed.

As the piston stops in Figure 14b at $t^*=1.56$, *R1* is now slightly bent but deforms into square shape. As expected, the stopping rings *R2* are formed following the nozzle's shape as the fluid entrains into the nozzle's interior. There are four duplications of *R2a* at four troughs which are denoted as *R2a-I*, *R2a-II*, *R2a-III* and *R2a-IV* in Figure 14b. These four *R2a* are symmetrically located across the *peak-to-peak* and *trough-to-trough* planes. Note that the nozzle's curvature is continuous and discontinuous at two smooth peaks (at $\psi=0$ and $\psi=\pi$) and sharp peaks (at $\psi=\pi/2$ and $\psi=3\pi/2$), respectively. Therefore, the *R2a-III* is connected to *R2a-IV* via a continuous vortex tube as shown in Figure 14b. Here we denote such tubes as *R2b*. On the contrary, the vortex tube that connects *R2a-II* and *R2a-III* is disrupted abruptly at the sharp peaks ($\psi=3\pi/2$) at $t^*=1.56$ leading to a thin connection between the sharp peaks and the leading ring via the smaller tubes (*R2d*). The tube *R2b* also connects with *R1* via an independent tube (*R2c*). At $t^*=2.62$, *R2b* further evolve and approach the leading ring *R1* and start to intertwine with its main core as shown Figure 14c. *R2b* and *R2c* continue to wrap around *R1* as it propagates downward. Vortex reconnection is observed as the upper portions of *R2a* collide. The propagation speed of *R1* is observed approximately equal to one in V-notched and A-notched nozzles as shown in Figure 13.

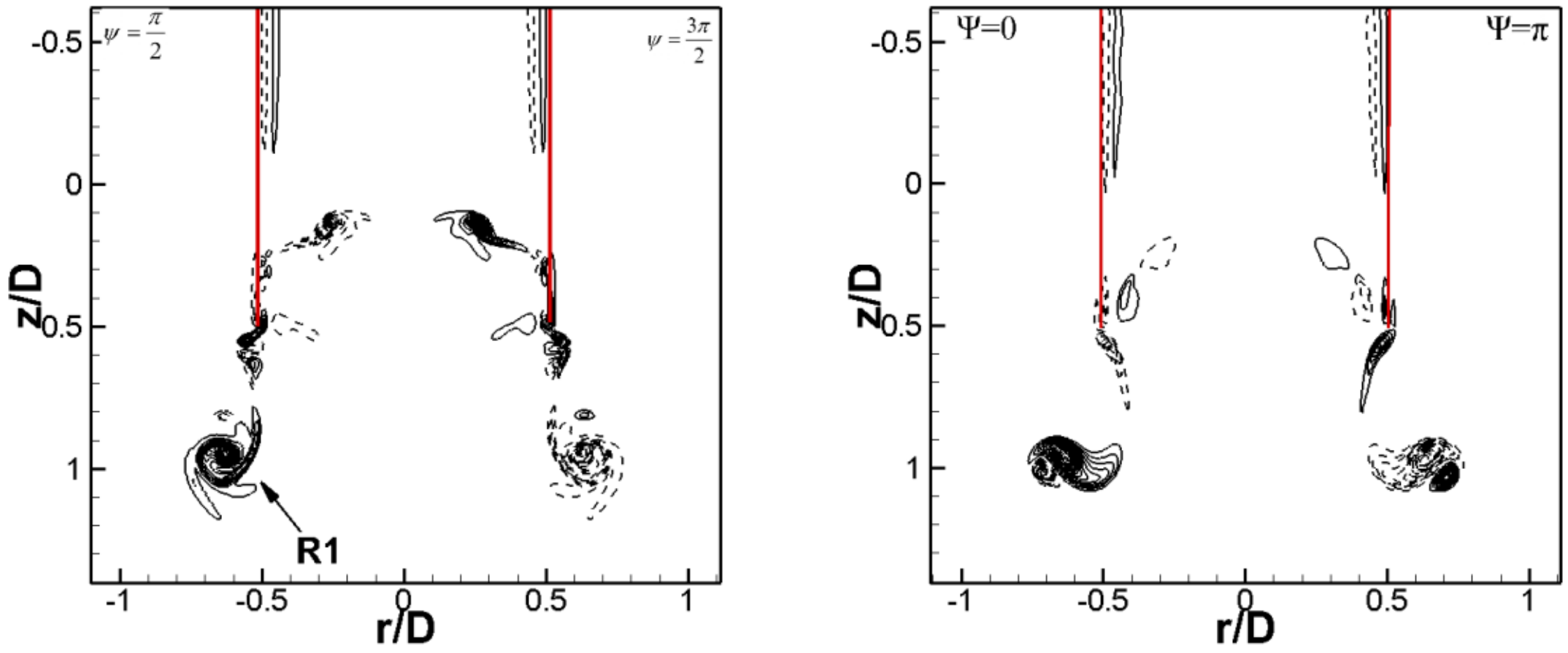

**Figure 13 The vorticity fields at the exit of W-notched nozzle at $t^*$=2.62 depicting the formation of the primary ring. Left column is displayed in *sharp peak-to-sharp peak* plane which is orthogonal to the *round peak-to-round peak* plane (right column). The first contour level is 2.5 with the increment is $2.5DU_0$. The dash lines indicate negative values.**

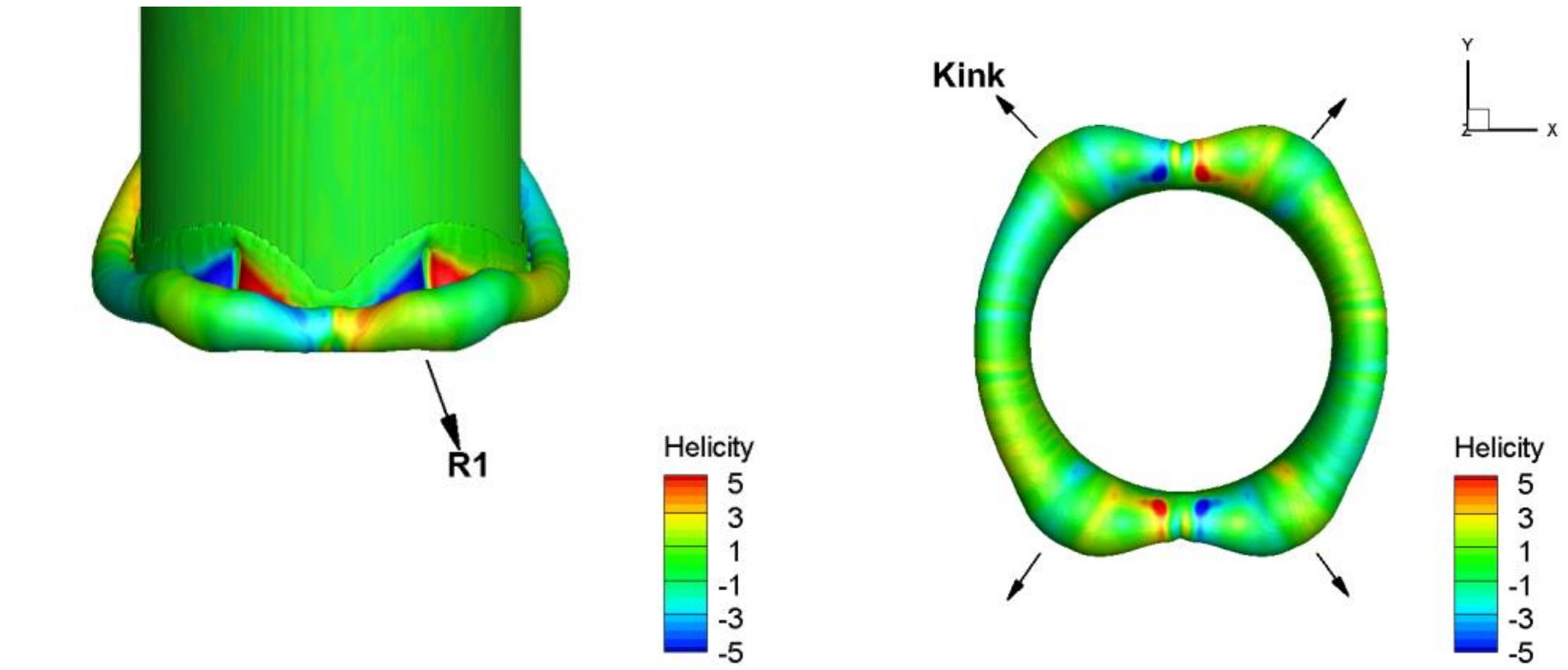

(a) t*=0.5

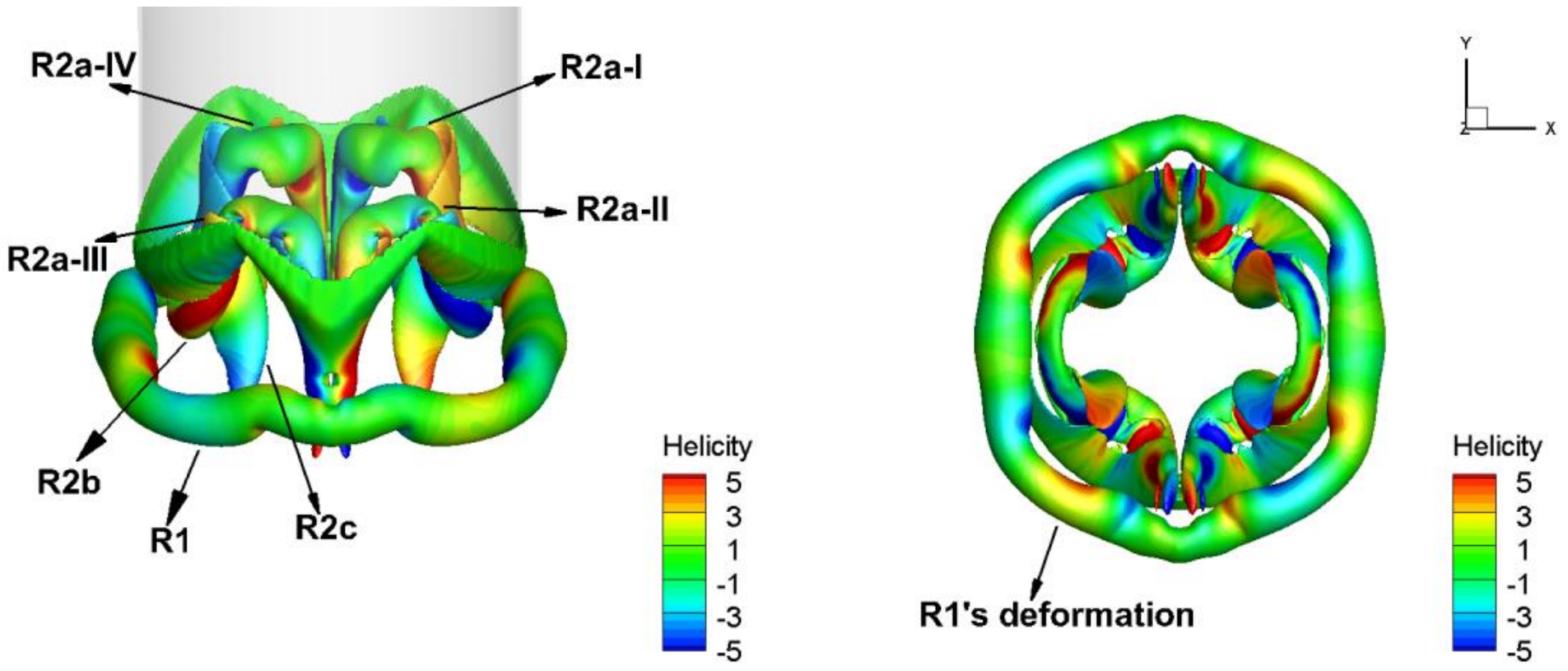

(b) t* = 1.56

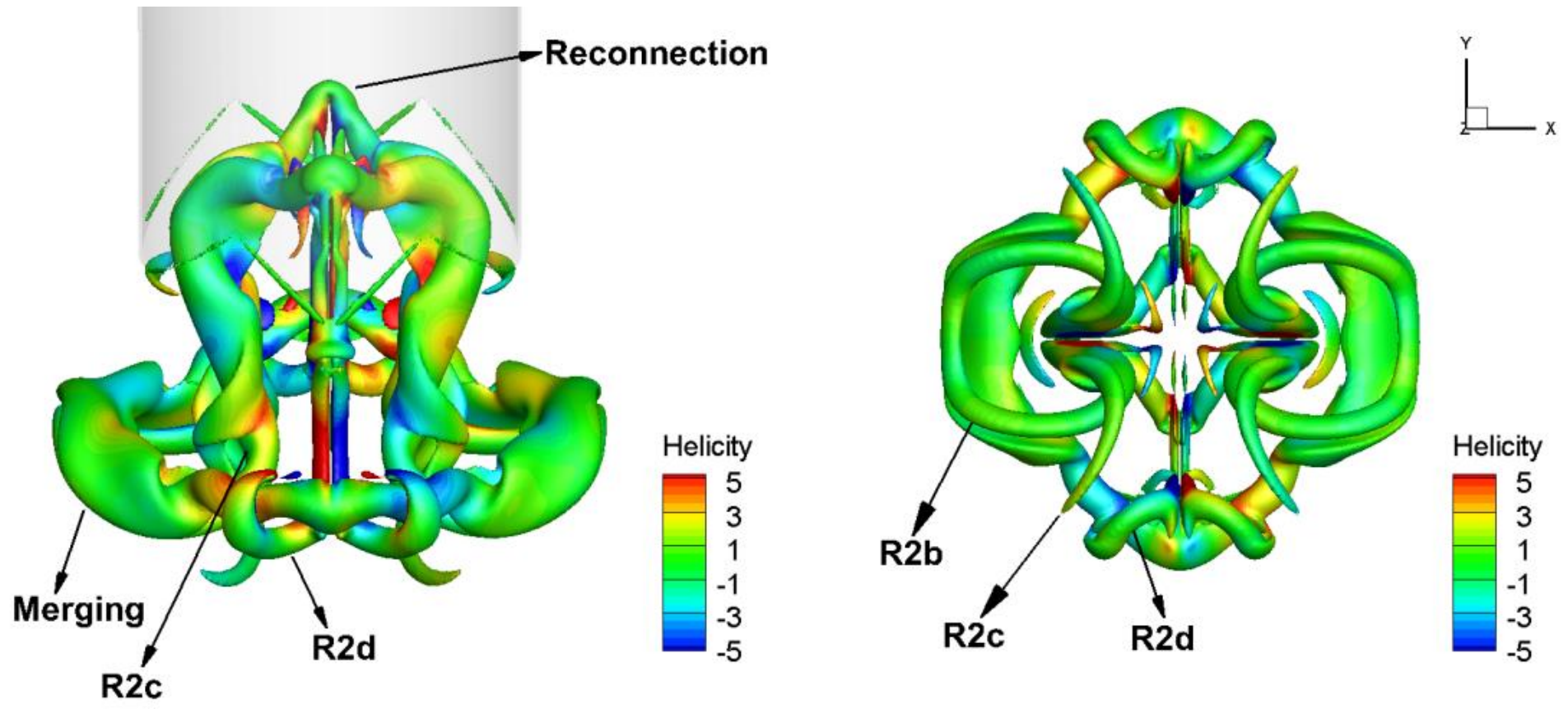

(c) t*=2.62

**Figure 14: The evolution of vortex ring interaction is visualized by vorticity magnitude colorized by helicity density for W-notched nozzle. Each column displays the side view and bottom view at the same time instance. From left to right, the column display the time: t*=0.5, t*=1.56 and t*=2.62.**

## 4.4 Kinematics of flow structure

The vortex structure of an asymmetrical vortex ring deviates far from that of a circular vortex ring. For a perfect circular ring, the ring has a simple toroidal shape with a closed vortex tube. For elliptic vortex rings, the vortex tube first has the bent elliptical shape (Dhanak and Bernardinis, 1981). However, the vortex tube tends to approach and touch itself when the ring exhibits oscillation. This self-touching might lead to the partial bifurcation or full bifurcation of the elliptic ring where it is separated into two independently smaller rings. In partial bifurcation, the vortex lines reorganize themselves into a chain of two inter-connected vortex loop (ADHIKARI et al., 2009). During that process, additional branches of vortex line (cross-linked vortices) might be generated and destroyed in time (Oshima et al., 1988). For an inclined vortex ring, experimental (Troolin and Longmire, 2010; Webster and Longmire, 1998) and computational data (Le et al., 2011) reported that there exist an additional vortex tube formed besides the main vortex ring. This additional tube is created under the entrainment of ambient fluid into the nozzle after the piston stops (Le et al., 2011). The evolution of the additional tube and its interaction with the main ring induces the large deformation of the main ring as well as formation of smaller secondary structures.

The vortex tube topology of A-notched and V-notched nozzles is shown in Figure 16. The main vortex ring is represented by the vortex tube following the shape of the nozzle's tip at the initial time. The pair of secondary vortex tube branch *R2* is shown near $\psi=\pi/2$ and $\psi=3\pi/2$. This secondary structure connects to the main ring *R1* symmetrically to the *peak-to-peak* and *trough-to-trough* planes.

For inclined nozzles, experiments (Lim, 1998; Troolin and Longmire, 2010) and computations (Le et al., 2011) also identified the presence of circumferential flow within the core of the main vortex ring shedding from the nozzle edge *R1*. As previously discussed in (Le et al., 2011), the circumferential flow inside the main vortex ring from the inclined nozzles is caused by the asymmetrical geometry configuration. There exists a distribution of pressure along the nozzle's circumference due to the different of entrainment rate from the ambient fluid into the nozzle interior. The key feature of the kinematics is a pair of diametrally opposite spiral saddle foci which drives the flow from the unstable manifold toward the stable manifold.

We observe the existence of circumferential flow inside the core of the main ring *R1* for both A-notched and V-notched nozzles. The primary ring here is created under the effect of the nozzle exit asymmetry which deviates from the perfect circular counterpart. This asymmetry leads to the creation of the circumferential flow inside the core of the primary ring. Since the symmetry is respected, the circumferential flow inside the core of *R1* is split into 4 separated parts as shown in Figure 16. The circumferential flow directions for both types of notched nozzles are:

$$\psi = 0 \rightarrow \psi = \frac{\pi}{2}$$

$$\psi = \pi \rightarrow \psi = \pi/2$$

$$\psi = \pi \rightarrow \psi = 3\pi/2$$

$$\psi = 0 \rightarrow \psi = \frac{3\pi}{2}$$

Therefore, there exist two pairs of saddle foci in notched nozzles as SF-0, SF-π, SF-π/2 and SF-3π/2. Observations show that the flow from unstable manifold (SF-0, SF-π) is split into two branches directed toward the stable manifold (SF-π/2 and SF-3π/2). In Figure 8, it is seen that the direction of circumferential flow drives the wrapping dynamics of R2a. Initially, R2a is formed as the remnant of the supplying vortex sheet. However, the circumferential flow of R1 helically reorganizes its contact region with R2a and pushes it toward $\psi=\pi/2$ and $\psi=3\pi/2$. This driving activity finally results in the self-touching and reconnection of the lower part of R2a as shown in Figure 9 and Figure 12.

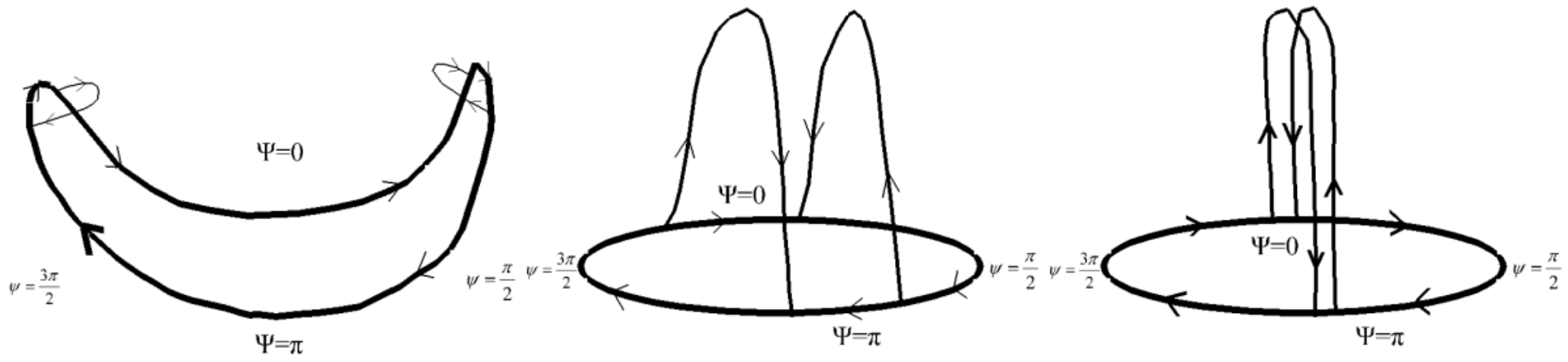

**Figure 15: The vortex structure of the asymmetrical ring from notched nozzle. The lines indicate the vortex line and the arrow indicates the vorticity vector direction.**

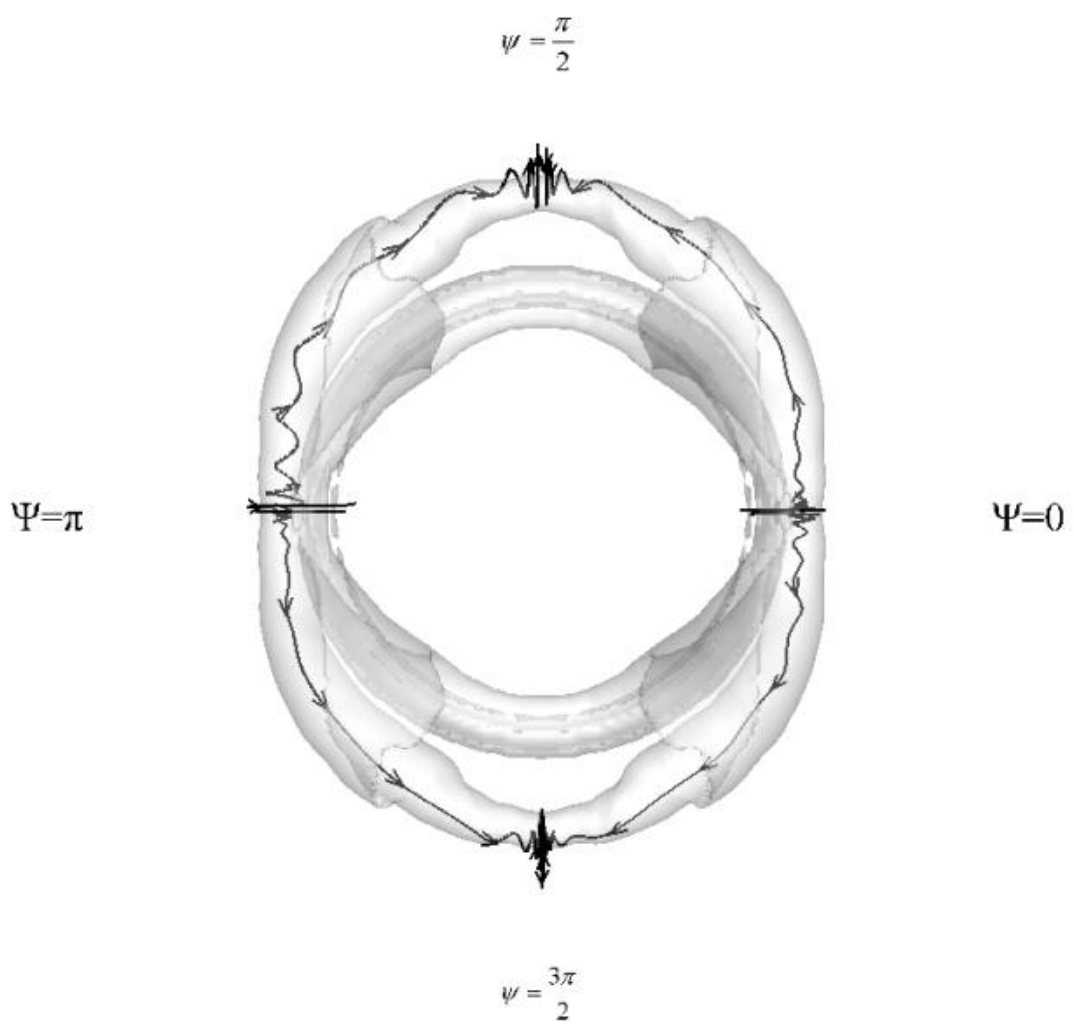

(a) A-notched nozzle

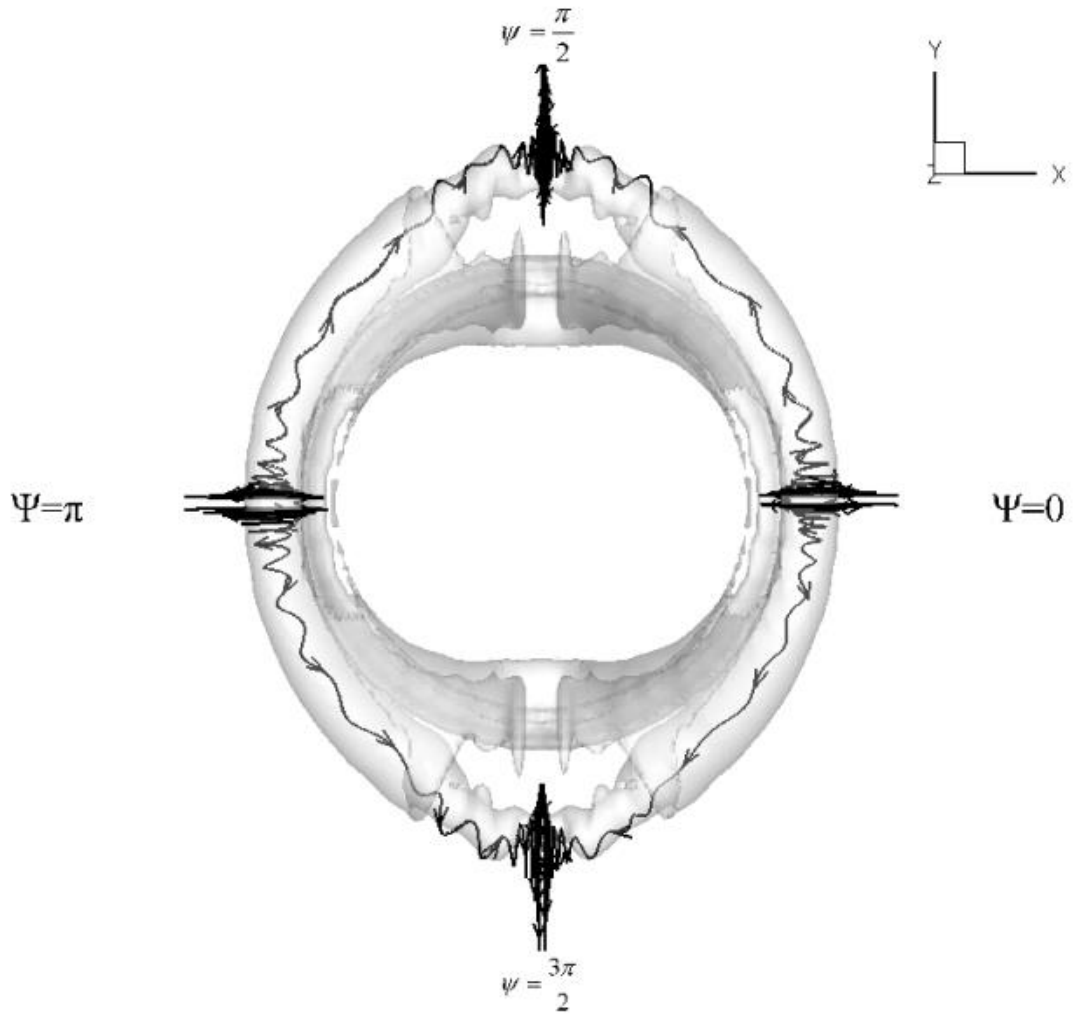

(b) V-notched nozzle

**Figure 16: The direction of the circumferential flow inside the core of R1 at t*=1.1. The vortex core is visualized by vorticity magnitude $|U_0/D|=12$. The flow direction is indicated using instantaneous streamline.**

## 5. Discussions

Asymmetrical vortex ring is formed in many natural and engineering processes including biological (Gharib et al., 2006; Le et al., November 2010; Le et al., 2012) and engineering (Troolin and Longmire, 2010) flows from complex nozzles. In such flows, the shape of the nozzle's exit deviates far from the perfect circle. From engineering viewpoint, these nozzles are classified broadly into two categories: i) planar nozzle and ii) non-planar nozzle (in-determinate origin nozzle) based on the general nozzle's exit shape. The first class of nozzles includes non-circular nozzles in which the nozzle's exit plane is orthogonal to the main nozzle's axis (i.e elliptic, triangular, rectangular or square nozzles). The second class of nozzles includes nozzles with arbitrary exit shapes. In this class, the exit's circumference is fully three-dimensional and thus does not reside on a particular two-dimensional plane.

Owing to the complexity of generating condition, asymmetrical vortex ring dynamics has only been reported mostly for the planar nozzles. The most well-known case is the formation of elliptic vortex ring (Dhanak and Bernardinis, 1981) from elliptic nozzles. It has been shown that the dynamics of this ring exhibits complex phenomena including: i) axis-switching; ii) cross-links and iii) bifurcation. The most significant dynamic is the axis-switching phenomenon where the elliptic ring oscillates periodically. During its propagation the major and minor axis of the ring lengthens and shortens harmonically leading to the "switching phenomenon" at low stroke length ($L/D < 2$) (ADKIHIRI, 2009). (Domenichini, 2011) reported the first systematic investigation of the nozzle exit's shape on vortex formation process for slender orifice. He found that the flow has two dimensional features at low aspect ratio of the orifice. A distinct vortex ring is formed following the shape of the orifice from the shear layer. However, the two-dimensionality quickly breakdowns as the aspect ratio increase. There exists a significant variation of vortex core size around the circumference of the orifice. This variation leads to the three-dimensionality of the flow which is characterized by the large deformation of the leading ring and the formation of secondary vortex tubes. (O'Farrell and Dabiri, 2014) have recently examined the formation of vortex rings from elliptic and oval nozzles. They found that the leading vortex ring is bent and curved as it exhibits the "axis-switching" phenomenon. They also reported the formation of additional structure (crescent vortex) at large stroke ratio in addition to the leading vortex ring.

Axis-switching is a common phenomenon in elliptic rings with low aspect ratio (Dhanak and Bernardinis, 1981; Oshima et al., 1988). Our results in Figure 8,

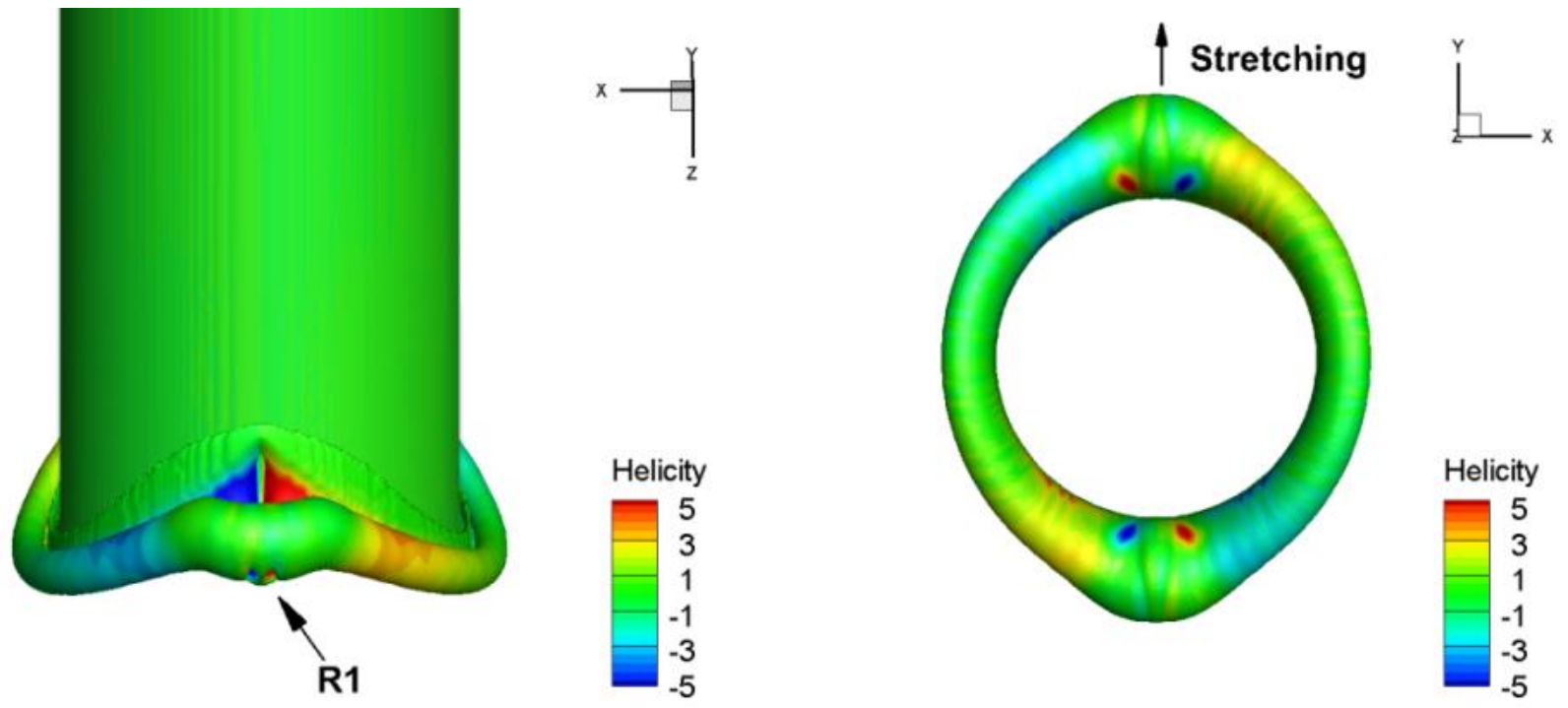

(y) t*=0.5

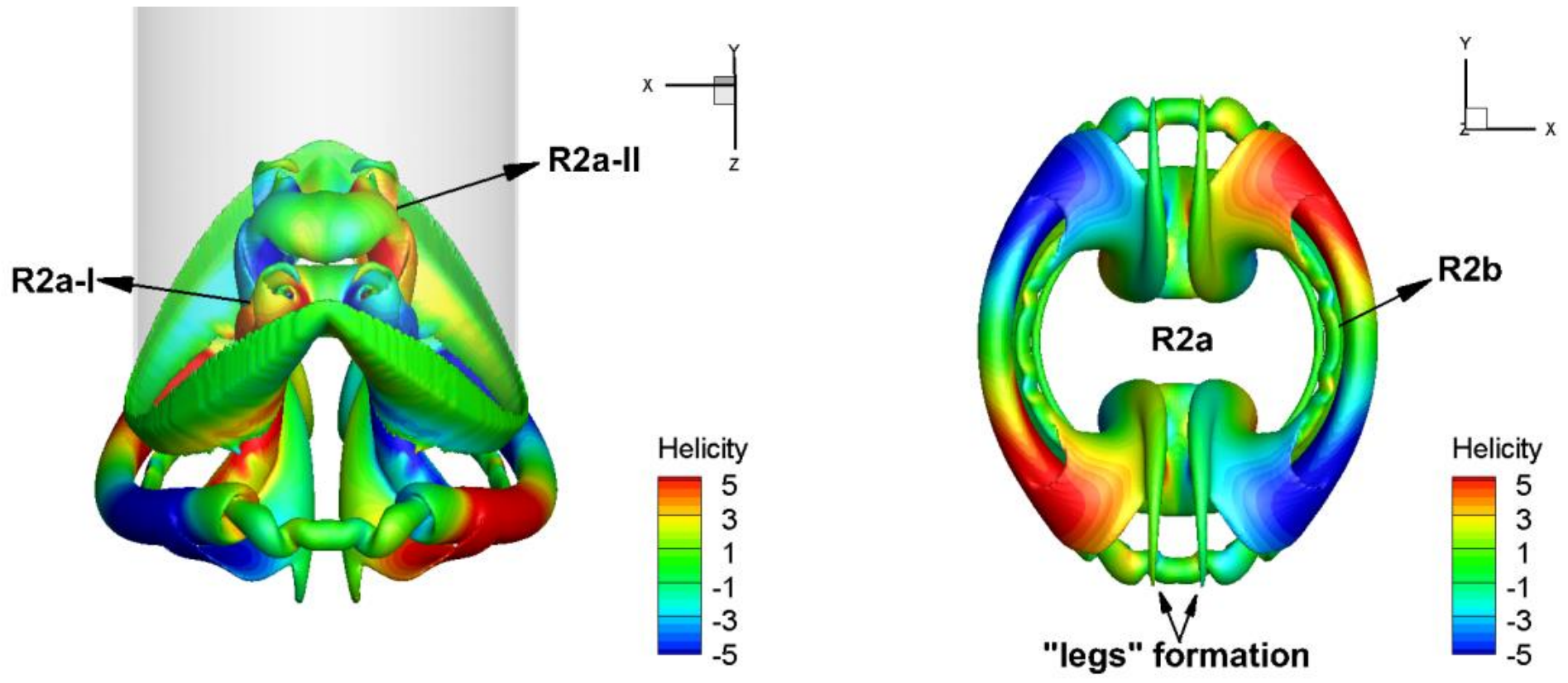

(z) t* = 1.56

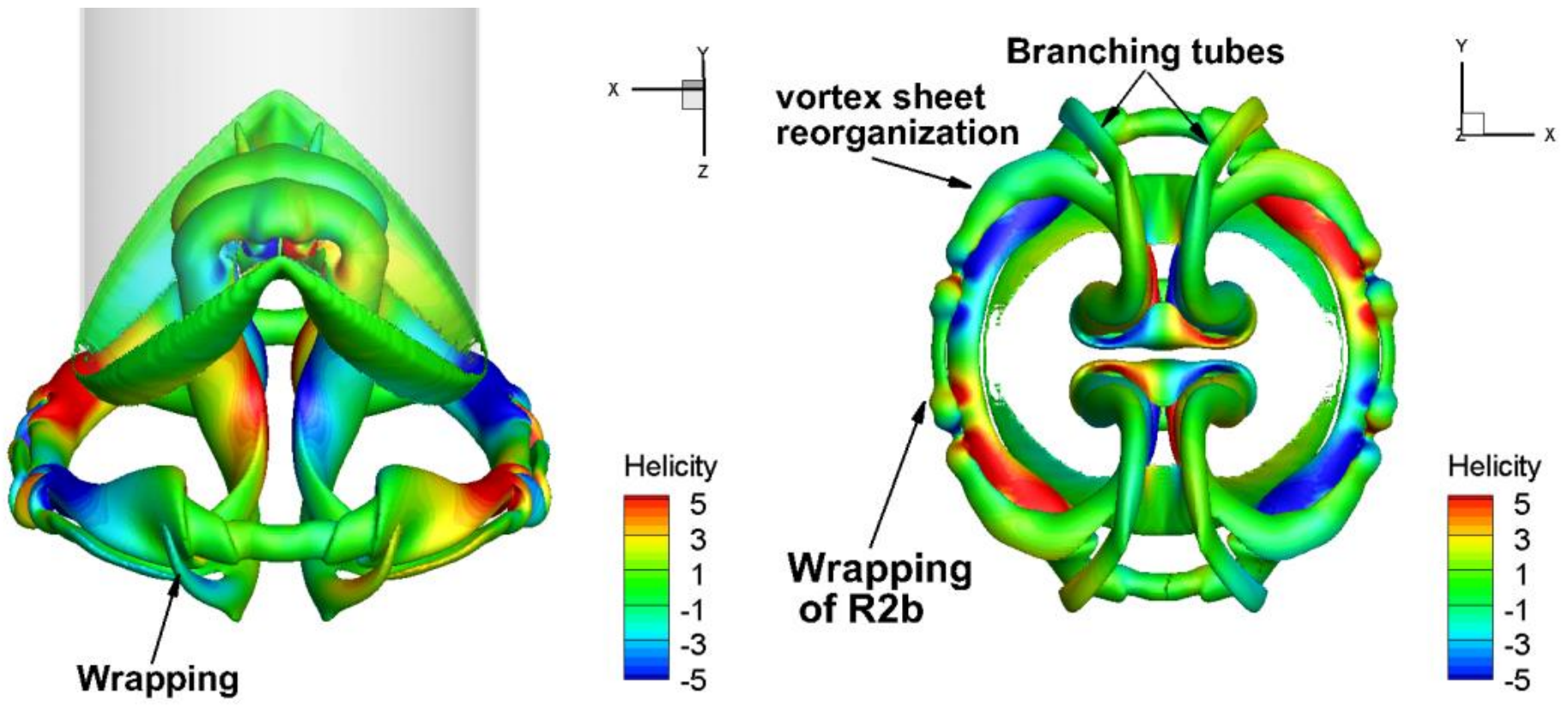

(aa) t* = 2.0

Figure 10 and Figure 14 show that axis-switching phenomenon is not observed in notched nozzles even at our stroke length *L/D=1*. The leading ring only exhibits the flattening (reorientation) process during the initial formation phase. In all cases (V-notched, A-notched and W-notched nozzles), the leading ring initially follows the nozzle's exit shape at early time. However, it tends to relax to the position orthogonal to the nozzle's axis (flattening). After the flattening of *R1*, it exhibit slight bending. The leading ring maintains its curvature as it propagates downward for a relative long period of time. The absence of axis-switching phenomenon in the current work is also consistent with experimental (Troolin and Longmire, 2010) and computational (Le et al., 2011) data of vortex rings from inclined nozzles. Interestingly, recent experimental data of (New and Tsovolos, 2013) has reported similar behavior in continuous inclined elliptic and rectangular jets. This characteristic is distinctively different from planar nozzles where the resulted jets are dominated with the axis-switching dynamics of the formed vortex ring (Gutmark and Grinstein, 2003). The lack of axis-switching dynamics of *R1* here in all cases can be explained by the existence of the circumferential flow in Figure 16, which is created by the non-planar exit. In addition, the formation of additional trailing vortex tubes (*R2*) further interrupts the deformation of the leading ring as shown in Figure 9 and Figure 11. This mechanism is similar to what has been reported in "tabbed" nozzles (Zaman, 1996) in which streamwise structures are shown to inhibit the vortex ring from axis-switching via so call "$\omega_x$-dynamics".

Non-circular nozzles have been widely used in many engineering applications as a mean of passive flow control device. Experimental studies(LONGMIRE et al., 1992b) have shown that the interaction of two flow modes: i) vortex rings and ii) streamwise vortices lead to the superiority of mixing efficiency in the resulted jets. Many experimental studies have confirmed such interaction in variety of nozzle types (Poussou et al., 2007; SHU et al., 2005). Although the existence of two flow modes has been confirmed the generation mechanism of such modes has not been identified. In circular nozzle, streamwise vortices are found to locate randomly at any circumferential location. However experimental data of (New et al., 2005) and computational data of (Cai et al., 2010) show that streamwise vortex pairs are generated only at peaks and troughs in notched nozzles. Therefore, the peaks and troughs play the most significant role in vortex dynamics of notched nozzles.

The entrainment of ambient fluid at troughs as shown in Figure 5 has shown that the entrainment indeed occurs at troughs. The entrainment profile depends on the shape of nozzle exit. The formation of *R2* can be used to explain the generation of streamwise structures (ribs) at trough position in notched nozzles. Recent time-resolved PIV data of (New and Tsovolos, 2009) in A-notched nozzles indicated that there exists footprints of flow entrainments at troughs as in the stream-wise structure (rib-like) visible in the *peak-to-peak* plane. These ribs are connected with the primary ring initially at $\psi=0$ and $\psi=\pi$ and elongated as the primary ring propagates far from the exit. In contrast to the spreading out from the centerline as observed in the conventional nozzle (Liepmann and Gharib, 1992), these ribs move toward the centerline and collide at about z/D=+3. They reported the formation of secondary structures in the *trough-to-trough* plane induced by the interaction of primary rings and the rib structures coincidentally at z/D=+3. In our case, the entrainment of ambient fluid via two trough location is clearly observed. The formation of *R2a-I* and *R2a-II* at troughs during the propagation of *R1* creates the stretching of *R1*'s core as shown in Figure 8b and

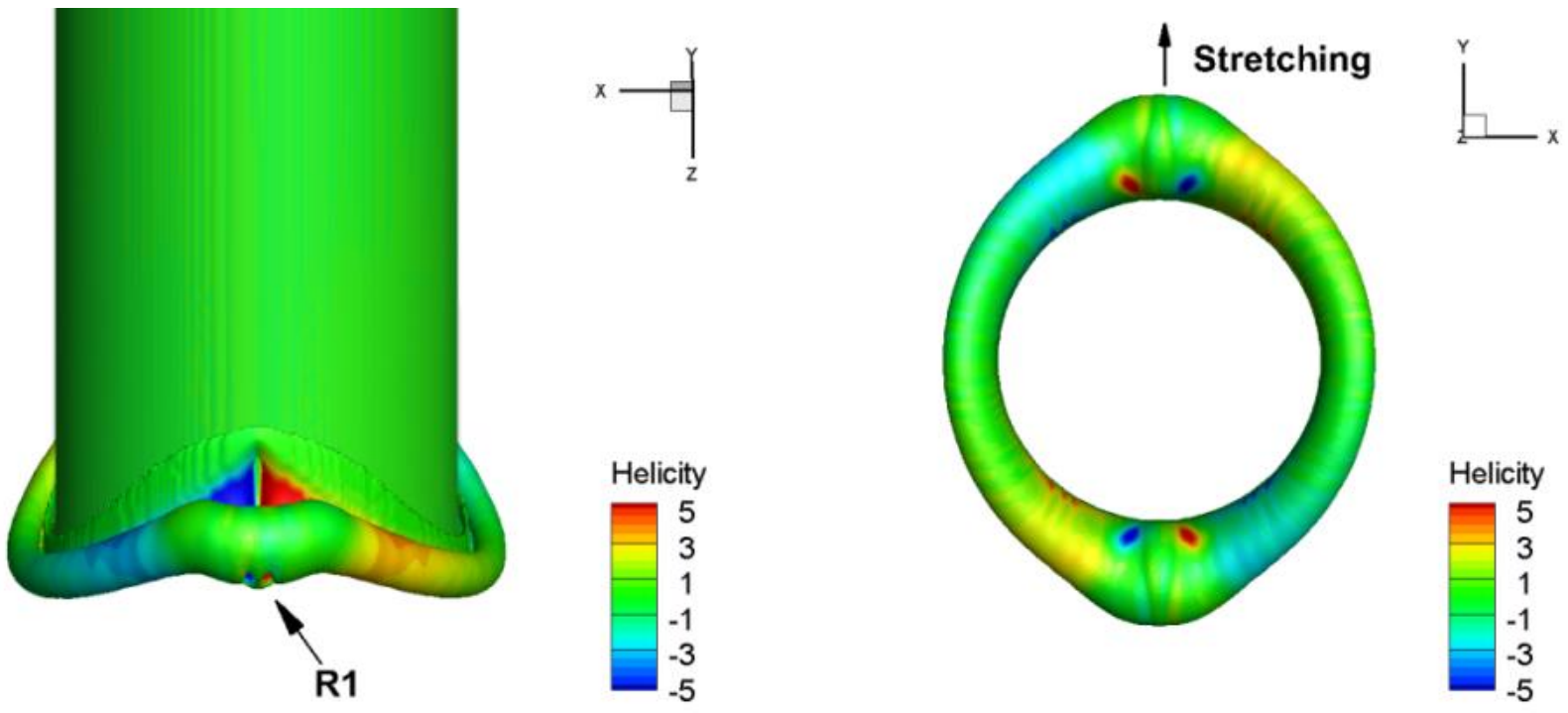

(bb) t*=0.5

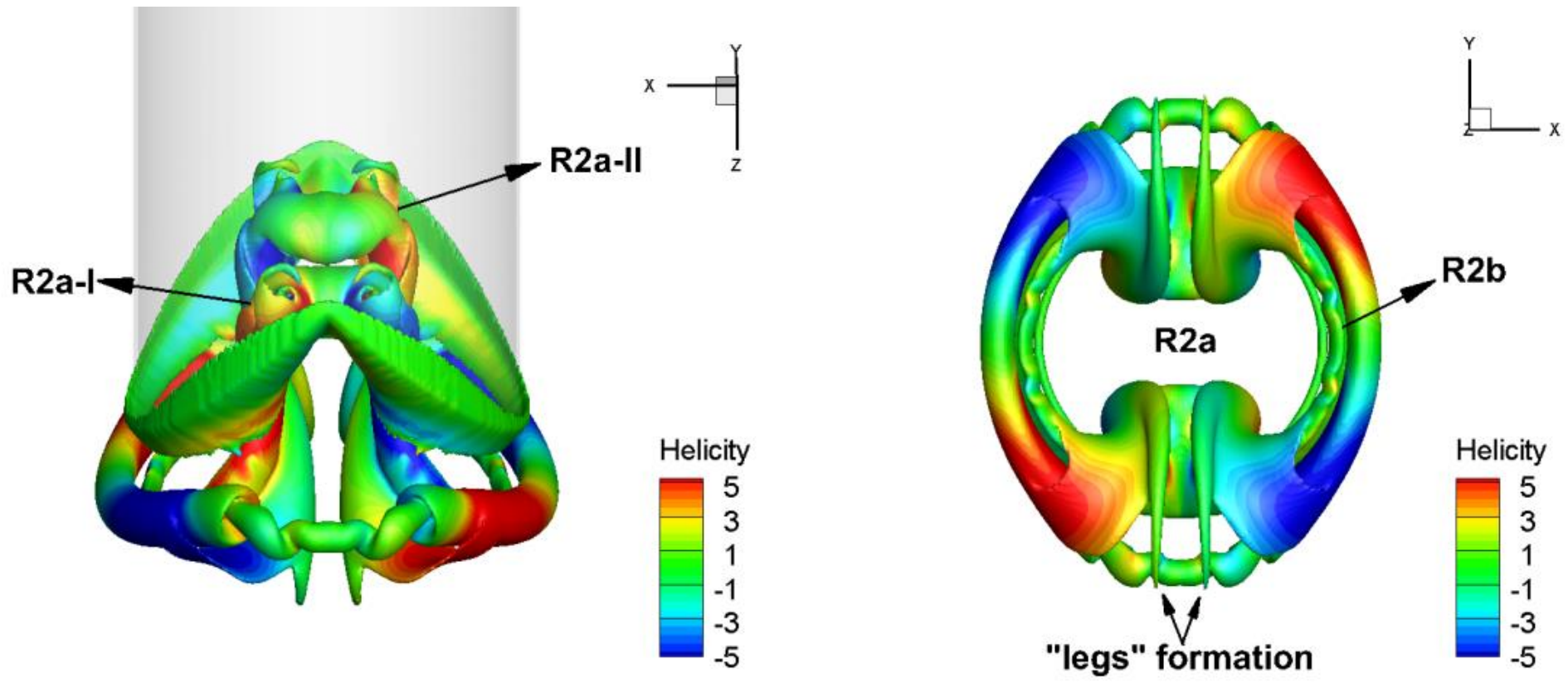

(cc) t* = 1.56

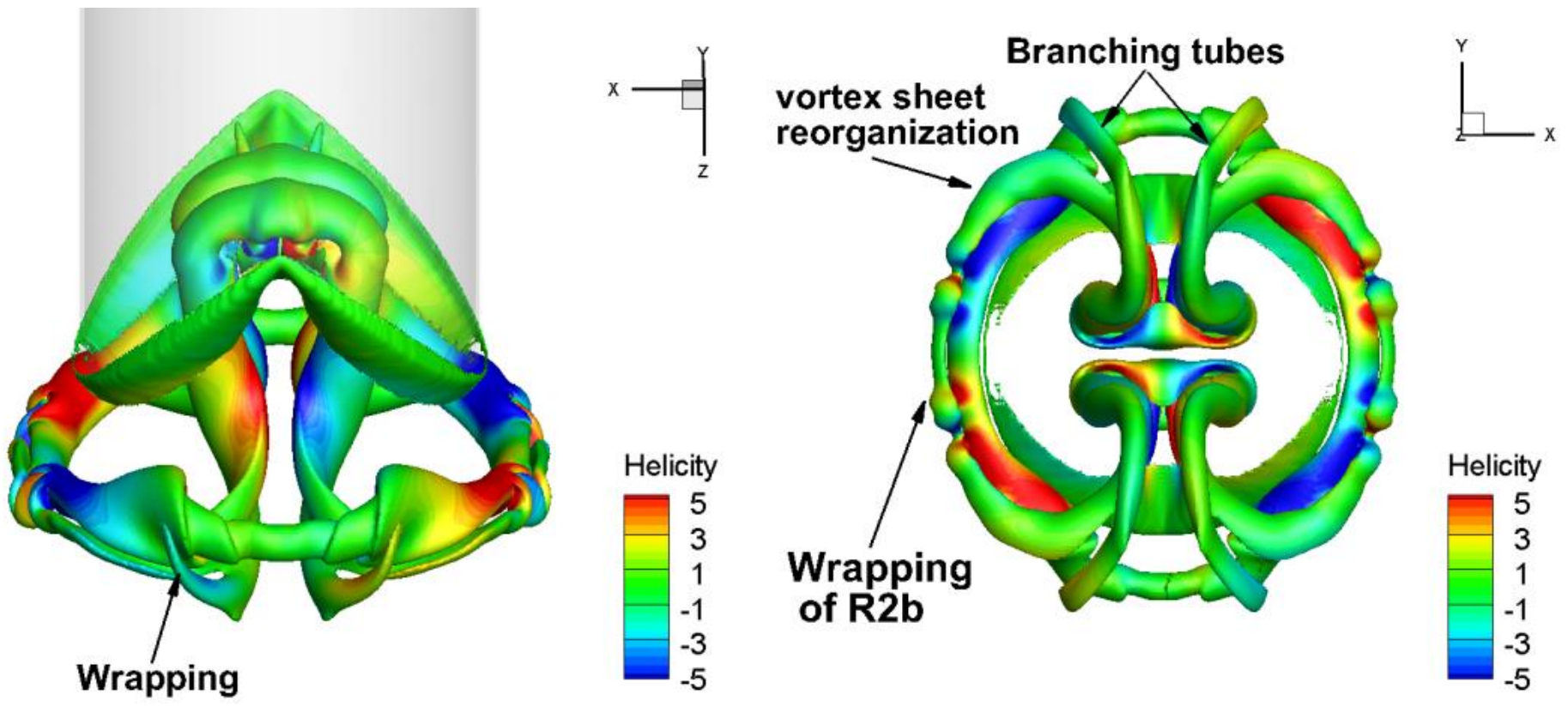

(dd) t* = 2.0

Figure 10b. The leading ring deforms from its original shape to the square-like shape. These results also agree with observation of (Cai et al., 2010; LONGMIRE et al., 1992b; New et al., 2005) on the transformation of initial elliptic cross-section of *R1* into "square-like" shape during its propagation downstream. Our model suggests that the stream-wise structures are formed at troughs, propagate and finally collide at the centerline creating the stream-wise vortex tubes.

The interaction between *R2a* and *R1* for nozzle types possesses similarities with the interaction of two orthogonally-offset tubes as studied by (Zabusky and Melander, 1989). As (Zabusky and Melander, 1989) pointed out, the smaller vortex tube tends to wrap the larger one. In our case, *R2a* is generated locally in a perpendicular direction to *R1* and this wrapping effect ("leg" formation) is expected to occur consistently as seen in Figure 8, Figure 10 and Figure 14. Indeed, *R2c* and *R2d* also exhibit the same wrapping dynamics around *R1* since they are all formed orthogonally to *R1.*

The dynamics of *R2b* is significantly different from other tubes since it is formed nearly parallel to *R1.* As shown in Figure 10b, this small tube first approaches to *R1* with slightly wavy shape. The amplification of this instability is seen in the plane view of Figure 10c. This type of instability's amplification of *R2b* is due to the presence of the larger core *R1*, which has been suggested by (Shelley et al., 1993). Note that the dynamics of *R2b* in V-notched nozzle is similar to one in inclined nozzles (Le et al., 2011).

As observed in Figure 9, Figure 11 and Figure 14, collision among duplications of *R2a* exists in all types of nozzle. It is well known that if two perfect circular rings collide head-on the final stage could result in the breakups of both ring's core and the occurence of vortex reconnection(Kida et al., 1991). In this case, due to the geometrical asymmetry vortex rings *R2a* from the entrainment location have the same rotational strength. The collision induces the core to exhibit self-twist and writhe motion.

From the practical standpoint, the higher number of troughs would increase the entrainment into the jet's core as shown in Figure 5. This entrainment would lead to the formation of the trailing vortex tubes in pulse jet in similar fashion describe in W-notch nozzle. The collision of these vortex tubes at the nozzle's center axis would lead to vortex reconnection and breakdown in the near field region and thus enhance mixing. From the mixing efficiency standpoint, the higher number of troughs would lead to the higher entrainment. This result is consistent to the experimental data of (LONGMIRE et al., 1992a). Note that at the extreme limit, the optimal nozzle might include thin appendages such as needles that are attached to the original nozzle, which is similar to the shape of jellyfish appendages.

Comparing the formation of V-notched and A-notched nozzles in Figure 8 and

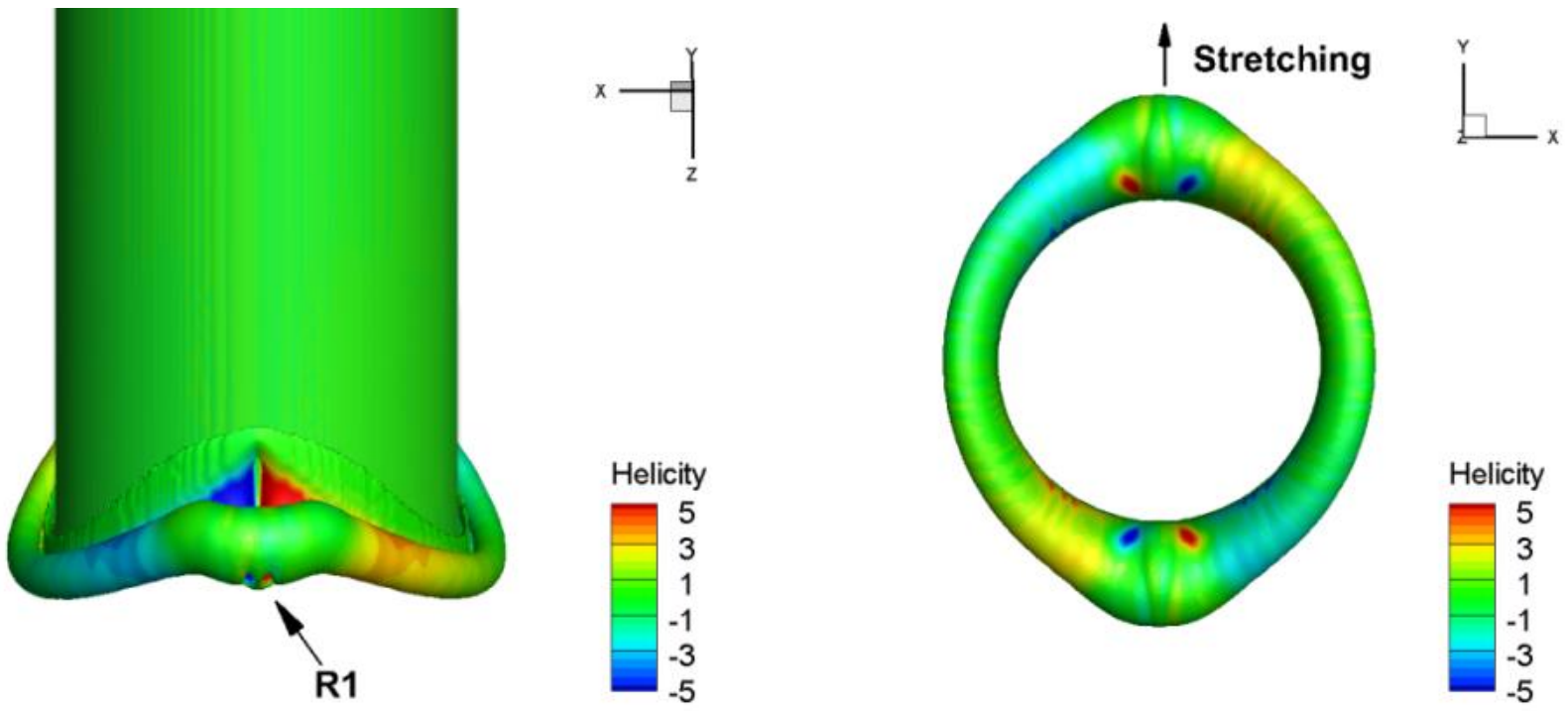

(ee) t*=0.5

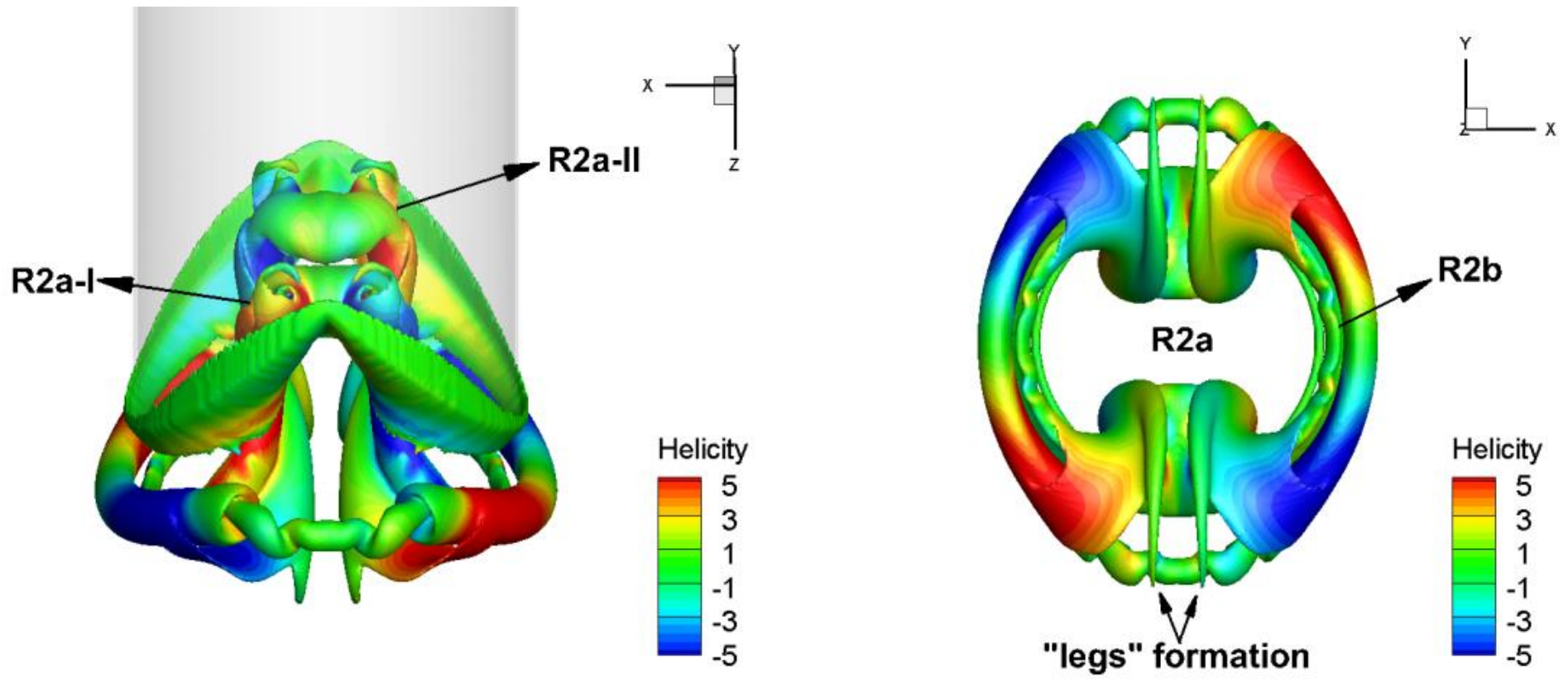

(ff) t* = 1.56

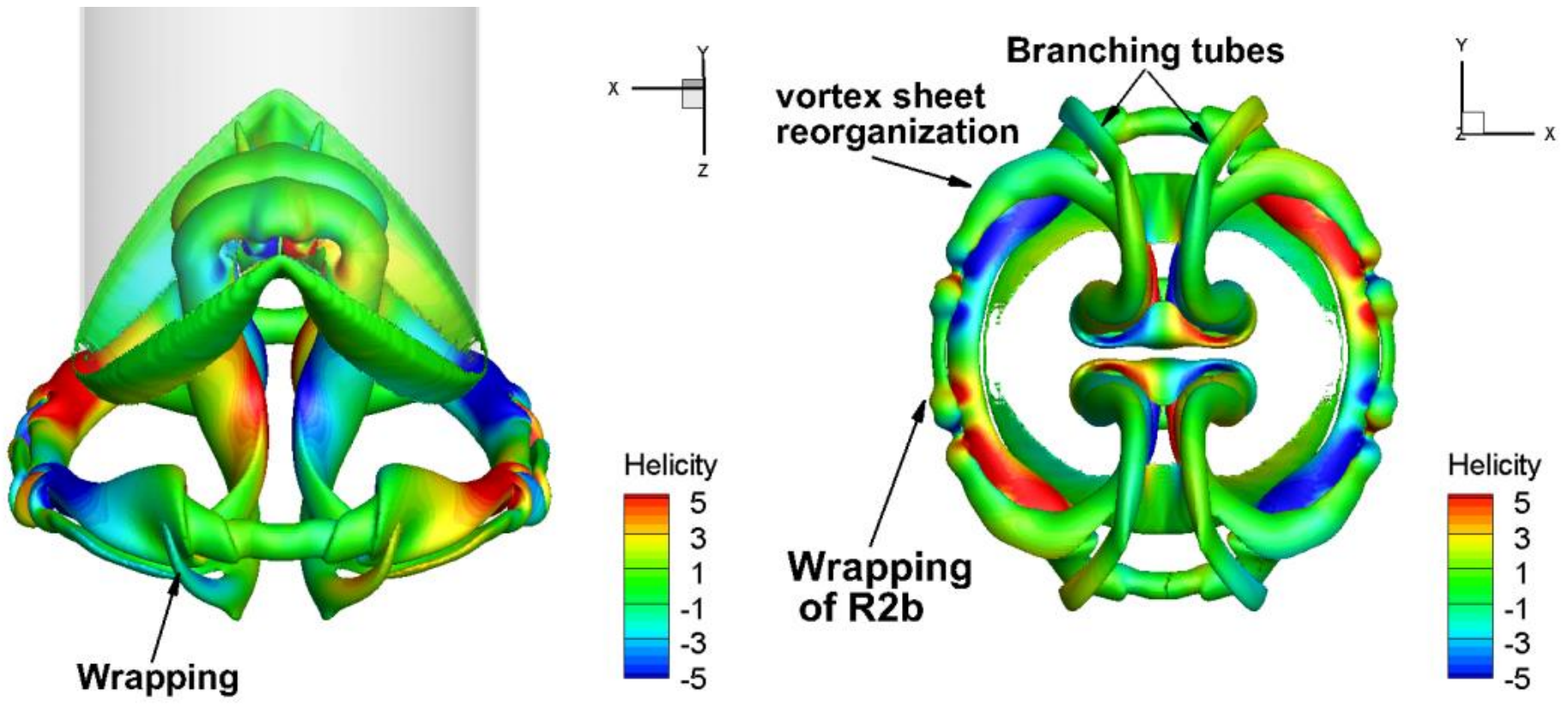

(gg) t* = 2.0

Figure 10 shows the role of exit's shape on the vortex formation process. Since the only difference in these two simulations is the nozzle's exit, it is possible to identify the the nozzle's exit effect. The main difference between the vortex tube structure of A-notched and V-notched nozzles is the absence of the structure *R2b*. In V-notched nozzle, this branch is formed simultaneously with the branch *R2a*. This branch does not exist in A-notched nozzle. This difference can be attributed to the difference in the exit's smoothness. Comparing Figure 8 and

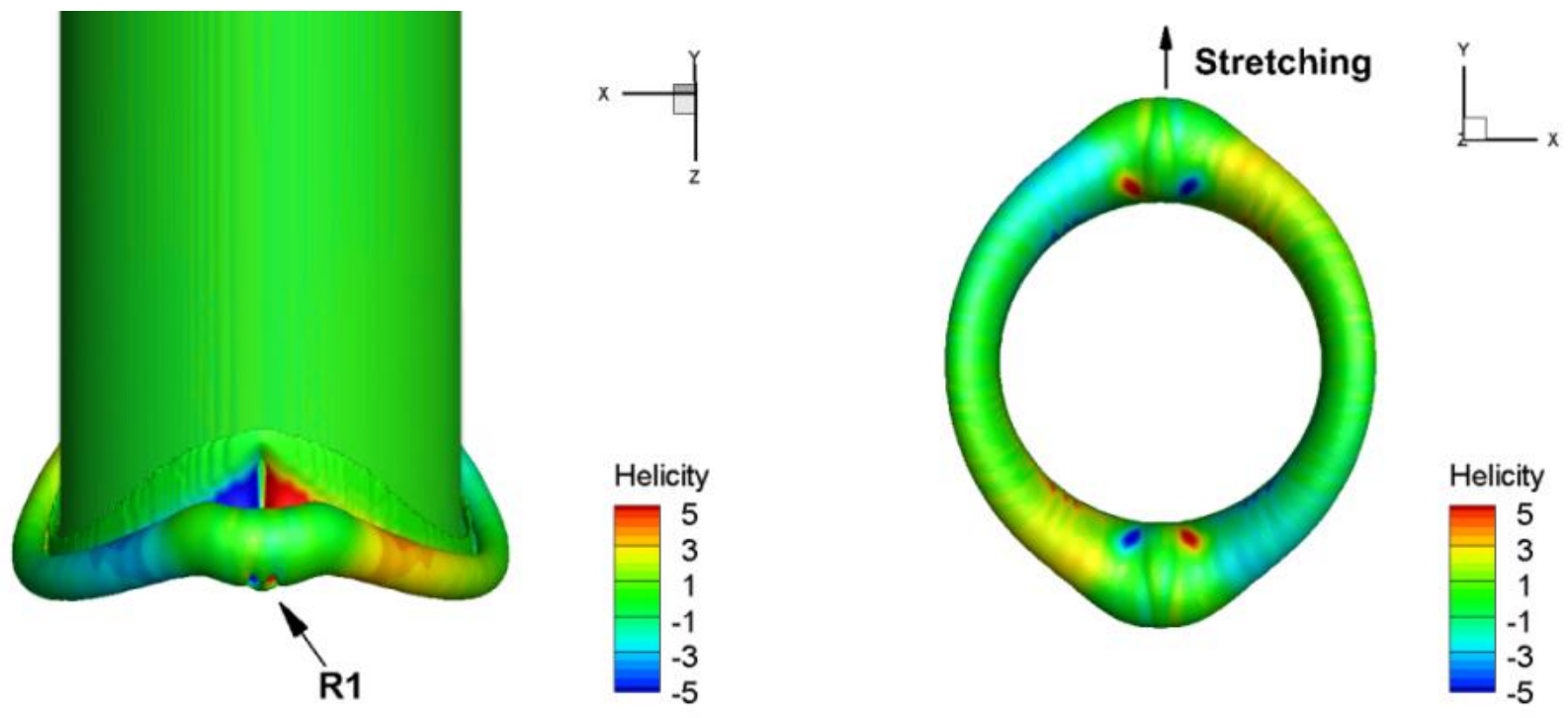

(hh) t*=0.5

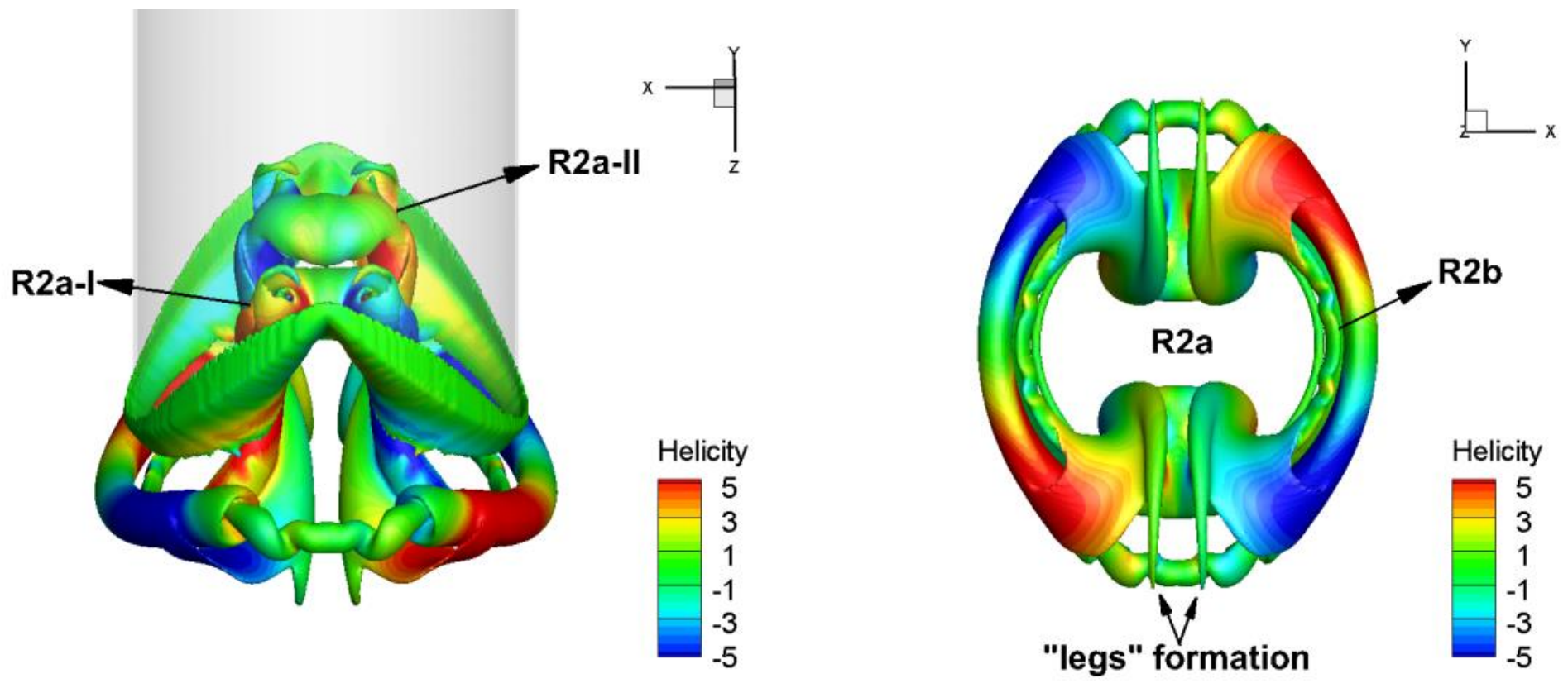

(ii) t* = 1.56

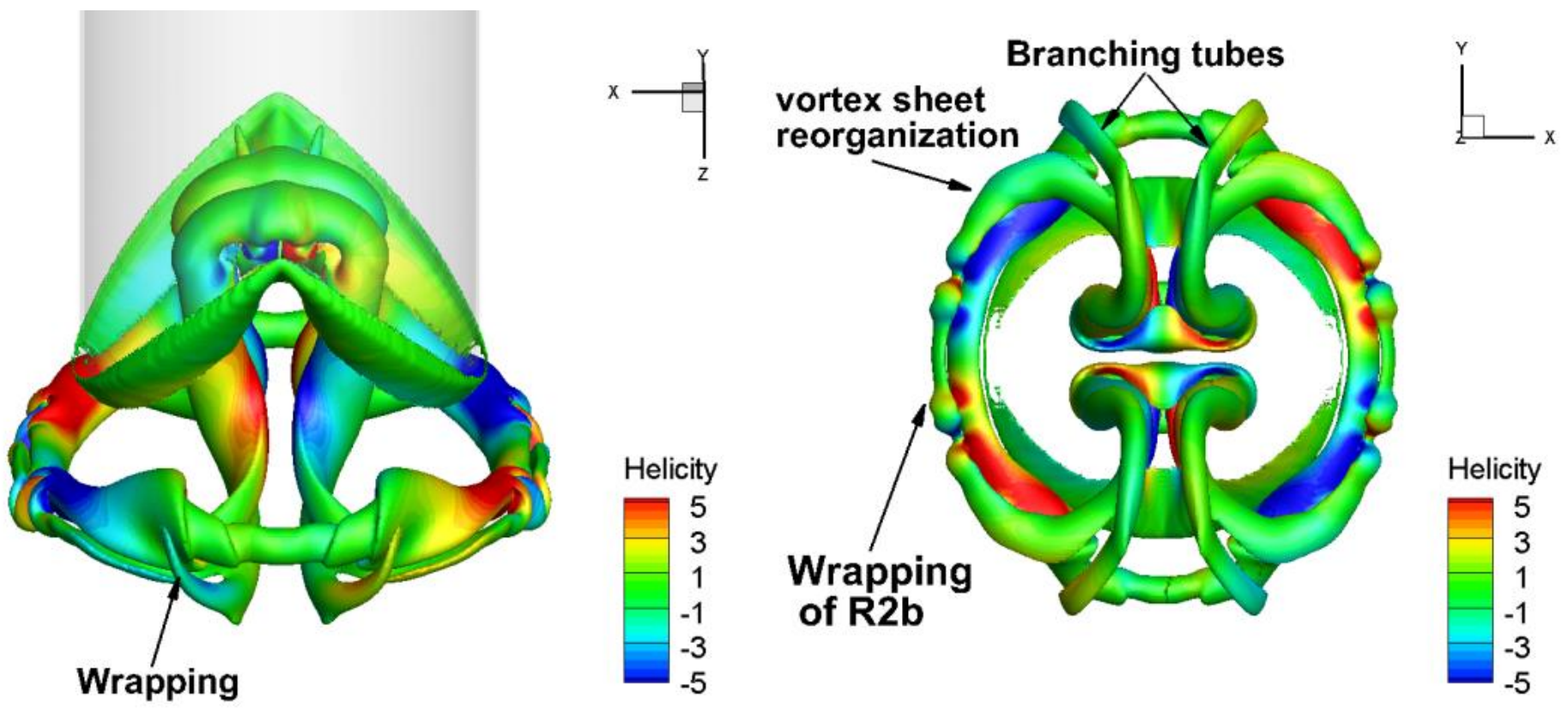

(jj) t* = 2.0

Figure 10 from a-c shows that the discontinuity (sharpness) of the exit at peaks in A-notched nozzles regulates the interruption of the vortex sheet's rolling-up at peaks. The vortex sheet is fully separated from the peaks of A-notched nozzle due to the sharp discontinuity of the peaks thus prevents the formation of the branch *R2b*. This difference leads to the entanglement and complex evolution of *R2b* and faster breakdown of the core at later time in V-notched nozzle.

In biological flows such as in jet-propelled swimmers (Katija and Dabiri, 2009), pistol shrimp (Hess et al., 2013) or left heart (Le et al., 2012) the asymmetrical vortex ring is generated from very complex nozzles. Normally a perfect circular vortex ring has been used to model the flow dynamics. Therefore, one legitimate question has been raised if such approximation is sufficiently accurate (Domenichini, 2011). For elliptic nozzles such question has been partially affirmed by showing the independence of vortex ring formation time on the nozzle exit's shape (O'Farrell and Dabiri, 2014). From previous studies of circular and inclined nozzles (Le et al., 2011; Webster and Longmire, 1998) and the current study in Figure 6, Figure 7 and Figure 13 under the same stroke length L/ D =1, the propagation speed of the leading ring is approximately the same regardless of the nozzle type. This results leads to the suggestion that the formation of the leading ring is independent of the nozzle exit shape. We hypothesize that the formation and propagation of the leading ring can be described equivalently using a circular ring in similar fashion to the work of (Hussain and Husain, 1989; O'Farrell and Dabiri, 2014). This result might be useful for the study of biological flows where complex geometries frequently involve and asymmetrical vortex rings are typically found. In addition, the results on the flow structure from complex nozzles such as W-notched nozzle in Figure 14 can provide physical ground to analyze the formation of vortex rings in cardiovascular flows (Le et al., 2012; Markl et al., 2014), which can well serve the purpose of cardiovascular disease diagnoses and monitoring.

## 6. Conclusions

In piston-cylinder apparatus, direct numerical simulations have been performed for different types of nozzle exits to investigate the dependence of formation process on the nozzle exit's shape. Three types of notched nozzles: i) V-notched; ii) A-notched and iii) W-notched nozzles are examined. The results show that the vortex formation processes from notched nozzles obey the same 3-stage sequence as follow:

a. At the early stage, there is a primary ring formed following the shape of the nozzle exit. The vorticity distribution varies along the circumferential direction creating flow in azimuthal direction. The circumferential flow is split into smaller portions which have the symmetry configuration.

b. There exists a strong dependence of the leading vortex ring shape on the nozzle's exit. However, the average propagation speed of the leading ring does not significantly vary across different types of nozzle exits

c. After the piston stops, the flow entrainment occurs at trough locations creating secondary trailing vortex tubes. These trailing vortex tubes subsequently propagate toward the main nozzle's axis. If there are multiple trough locations, there are multiple pairs of trailing rings formed.

d. At the last stage, they eventually collide at the nozzle's axis. Their interaction induces splitting, merging and reconnection of vortex filaments. The interaction (ring-ring collision or impingement) can lead to vortex reconnection process and induce secondary vortex tubes at peaks

We observe that vortex reconnection occurs in all studied nozzles as the trailing vortex tubes grow and collide head-on toward each other at the nozzle's center line at the later stage of formation. The reconnection results in stretched streamwise structures and small scale separated rings. This suggests that reconnection is a common phenomenon occurring in complex jets, which is possibly the mechanism for energy cascade at higher Reynolds number in complex jet flows.

From our previous work on the inclined nozzles (Le et al., 2011) and the current results, we postulate that the circumferential flow exists in all non-circular nozzles. We hypothesize that the vortex topology reported here might be applicable in a larger set of in-determinate origin nozzles where the exit shape has multiple peaks and troughs.

This work is originally motivated by the need to explain the breakdown mechanism of the mitral vortex ring inside the left ventricle of human heart. In such case, the stroke length ratio $\frac{L}{D}$ is relatively low (see (Gharib et al., 2006), for example). In the current work, only L/D = 1 is examined for all nozzles, and thus vorticity entrainment does not reach the "optimal" value (i.e L/D >4). The "pinch-off" phenomenon therefore does not occur in our simulations. Our conclusions on the flow structure are thus applicable only for small stroke length (L/D <4). In nature, that range of L/D is quite ubiquitous in variety of biological processes (Dabiri, 2008) and (Hess et al., 2013). Thus it is possible to extend the conclusions in the current work for appropriate classes of biological flows where complex nozzle's geometry exists. Further studies with larger stroke length (i.e L/D > 4) might be useful to understand the dynamics of the trailing vortex sheet (crescent vortex) and its interaction with the leading ring during “pinch-off” process.

## 7. Acknowledgements

We greatly appreciate the support of Minnesota Supercomputing Institute for the computational time of this work.